%% file: ms.tex
\documentclass[acmsmall,acmthm=false,nonacm]{acmart}

\pdfoutput=1

\newtoggle{isappendix}
\togglefalse{isappendix}
\newtoggle{EXTND}
\newtoggle{ARXIV}

\toggletrue{EXTND}
\toggletrue{ARXIV}

\iftoggle{EXTND}{
  \settopmatter{printfolios=true,printccs=false,printacmref=false}
  \setcopyright{none}
  \renewcommand{\footnotetextcopyrightpermission}[1]{}
  \renewcommand{\footnotetextauthorsaddresses}[1]{}
}{
  \settopmatter{}
  \setcopyright{rightsretained}
  \acmJournal{PACMPL}  
  \acmYear{2021}
  \acmMonth{8}
  \acmVolume{5}
  \acmNumber{}
  \acmArticle{}
  \acmDOI{}
  \startPage{}
  \acmPrice{}
  \acmISBN{}
  \acmConference[ICFP]{}{2021}{}
}

\usepackage{booktabs}
\usepackage{mathpartir}
\usepackage{amsthm}
\usepackage{amsmath}
\usepackage{MnSymbol}
\usepackage{multirow}
\usepackage{subcaption}
\usepackage[inline]{enumitem}
\usepackage[bbgreekl]{mathbbol}

\usepackage{ellipsis}

\DeclareSymbolFontAlphabet{\mathbb}{AMSb}

\DeclareSymbolFontAlphabet{\mathbbl}{bbold}

\newtheorem{theorem}{Theorem}[section]
\newtheorem{lemma}[theorem]{Lemma}

\newtheoremstyle{straighttext0}%
                {}{}%
                {}{}%
                {\bfseries}{}%
                { }%
                {\thmname{\textbf{#1}}\thmnumber{ #2}: \thmnote{ #3}}
\theoremstyle{straighttext0}
\newtheorem*{bproof}{Proof}

\newtheoremstyle{straighttext}%
                {}{}%
                {}{}%
                {}{}%
                {\newline}%
                {\thmname{\textbf{#1}}\thmnumber{ \textbf{#2}}: \thmnote{ #3}}
\theoremstyle{straighttext}

\newtheorem*{bcase}{Case}

\iftoggle{EXTND}{
\newcommand{\Secref}[1]{Section~\ref{#1}}
\newcommand{\Figref}[1]{Figure~\ref{#1}}
\newcommand{\new}[1]{#1}
}{
  \input{editingmarks}
}
\input{defs.tex}

\input{haskell_style}

\input{formal_dynamics}

\input{formal_typings1}

\begin{document}

\title[]{Efficient Tree-Traversals: Reconciling Parallelism and Dense Data Representations}

\iftoggle{EXTND}{
\subtitle{Extended Version}
\titlenote{Extended version of ``Efficient Tree-Traversals: Reconciling Parallelism and Dense Data Representations,''~\cite{ParallelLoCal-icfp}.}
}{}


\newcommand{\iuaffil}[0]{
  \affiliation{
    \institution{Indiana University}            
    \country{United States}
}}

\newcommand{\fbaffil}[0]{
  \affiliation{
    \institution{Facebook}            
    \country{United States}
}}

\newcommand{\purdueaffil}[0]{
  \affiliation{
    \institution{Purdue University}            
    \country{United States}
}}

\newcommand{\cmuaffil}[0]{
  \affiliation{
    \institution{Carnegie Mellon University}            
    \country{United States}
}}

\newcommand{\kentaffil}[0]{
  \affiliation{
    \department{School of Computing}              
    \institution{University of Kent}            
    \country{United Kingdom}
}}

\author{Chaitanya Koparkar}\iuaffil\email{ckoparka@indiana.edu}
\author{Mike Rainey}\cmuaffil\email{me@mike-rainey.site}
\author{Michael Vollmer}\kentaffil\email{m.vollmer@kent.ac.uk}
\author{Milind Kulkarni}\purdueaffil\email{milind@purdue.edu}
\author{Ryan R. Newton}\purdueaffil\email{rrnewton@purdue.edu}

\begin{abstract}
Recent work showed that compiling functional programs to use dense, serialized
memory representations for recursive algebraic datatypes can yield significant
constant-factor speedups for sequential programs. 
But serializing data in a
{\em maximally} dense format consequently serializes the processing of that
data, yielding a tension between density and parallelism.  
This paper shows that a disciplined, practical compromise is possible.
We present \pargibbon{}, a
compiler that obtains the benefits of dense data formats {\em and} parallelism.
We formalize the semantics of the {\em parallel location calculus}
underpinning this novel implementation strategy, 
and show that it is type-safe.
\pargibbon{} exceeds the parallel performance of existing compilers for purely
functional programs that use recursive algebraic datatypes,
including, notably, abstract-syntax-tree traversals as in compilers.
\end{abstract}


\iftoggle{EXTND}{}{
}

\maketitle

\renewcommand{\shortauthors}{C. Koparkar, M. Rainey, M. Vollmer, M. Kulkarni and R. R. Newton}

\section{Introduction}

Representing tree-like data as \emph{pointer-free, serialized byte arrays} can
be extremely efficient for tree traversals, as it minimizes pointer-chasing and
maximizes locality \cite{goldfarb13sc, meyerovich11hotpar, makino90}. 
Moreover, by using an in-memory representation also suitable for external
transfer and storage \cite{capnproto, cnf-icfp15}, programs can rapidly process
data without the overhead of deserialization.
Traditionally, any such tree-layout optimizations would be implemented manually by
the programmer---for example, in a scientific application with balanced
trees.

Recent work, however, has shown the benefits of {\em automatically} compiling tree
traversals to use denser representations, even for source programs written in a
general-purpose language. 
The Gibbon compiler for a subset of Haskell
exemplifies this approach \cite{Gibbon, LoCal}.
%
%
While the dense data representation strategy works well for sequential
programs, there is an intrinsic tension if we want to parallelize
these tree traversals.  As the name implies, efficiently {\em serialized}
data must often be read serially.  To change that, first, enough {\em indexing} data
must be left in the representation so parallel tasks can
``skip ahead'' and process multiple subtrees in parallel.  Second, the
allocation areas must be bifurcated to allow allocation of outputs in
parallel.

In this paper, we offer a solution to these challenges.
We propose a strategy where form follows function: where data
representation is random-access only insofar as parallelism is needed,
and both data representation and control flow ``bottom out'' to
sequential pieces of work.  That is, granularity-control in the data
mirrors traditional granularity-control in parallel task
scheduling. We demonstrate our solution by extending the Gibbon
compiler with support for parallel computation, introducing {\em \pargibbon}.
We also extend {\em LoCal}, Gibbon's typed intermediate language,
adding parallelism and give an updated formal
semantics~(\Secref{sec:region-parallel-local}).

In addition to tree traversals, we show that \pargibbon{} can efficiently
compile other parallel programs, such as sort and search algorithms
(\Secref{sec:evaluation}) to match or exceed the performance of the best
existing parallel functional compilers.  We choose a {\em functional} focus
for three primary reasons:

\begin{itemize}
\item Many tree traversals have different input and output types---as in
  a compiler pass that converts between intermediate languages---which necessitates 
  {\em out-of-place} traversals even in an imperative language.
\item Even pure programs can use mutable data, via linear
  types. (Gibbon uses these and eschews the {\em IO monad}.)
\item The purely-functional parallel Gibbon programs considered in this work are intrinsically
  {\em data-race free}.
\end{itemize}

The last point is worth emphasizing: every time a language adds both parallel
constructs and mutable data, it enables data-races and must define a memory
model to give them meaning.  In this work, we extend Gibbon with linearly-typed primitives for mutable data\footnote{Leveraging the Linear Haskell~\cite{linear-haskell}
extensions now available in GHC 9}
(\Secref{subsec:parallel-arrays}),
while keeping the language {race-free}.
We claim that linearly-typed mutable data, efficient data representation, and
compiler-supported parallelism are a synergistic combination.
In Parallel Gibbon programs, as in other
purely functional parallel programs, parallelism annotations not only don't
introduce races but also {\em do not affect program semantics},
meaning that these programs are {\em deterministic} as well as data-race free.

Ultimately, we believe that this work shows one path forward for
high-performance, purely-functional programs.  Parallelism in functional
programming has long been regarded as theoretically promising, but has a
spottier track record in practice, due to problems in runtime systems, data
representation, and memory management.
\pargibbon{} directly addresses these sore spots, showing how a
purely-functional program operating on fine-grained, irregular data can also run
fast (sequentially) and parallelize efficiently.
This complements more well-trodden areas of compiler research on parallelism, such
as dense and sparse collective operations on arrays~\cite{julia-parallel-accelerator, tensorflow, pytorch, nesl}.
That is, the approach described in this paper---for general-purpose, recursive
functional programs, including tree traversals---could be combined with targeted EDSLs or libraries implementing additional parallel programming idioms,
such as Haskell's Accelerate~\cite{accelerate}.  Both determinism and
data-race-freedom would be {\em compositional} within the functional-parallelism
setting.
Indeed, we have taken the first steps in this direction, adding a small set of parallel array primitives to Gibbon (\Secref{subsec:parallel-arrays}).

In this paper, we make the following contributions:

\begin{itemize}  
\item We introduce the first compiler that combines parallelism with automatic dense data
  representations for trees. While dense data~\cite{LoCal} and
  efficient parallelism~\cite{MPL,icfp-hierarchical-heaps} have been shown to independently yield large
  speedups on tree-traversing programs, our system is the first to combine
  these sources of speedup, yielding the fastest known performance generated by
  a compiler for this class of programs.

\item We formalize the semantics of a {\em parallel location calculus}
  (\Secref{sec:region-parallel-local}) that underpins the compiler, 
  including a proof of its type-safety (\Secref{subsec:type-safety}).
  To do so, we extend prior work on location calculi~\cite{LoCal},
  which in turn builds on work in region calculi~\cite{regioncalcs}.

\item We evaluate our implementation (\Secref{sec:implementation})
  on several benchmarks
  from the literature (\Secref{sec:evaluation}).
  On a single thread, our implementation is
  $1.93\times$, $2.53\times$, and $2.14\times$
  faster than \MPL{}~\cite{MPL} (an extension of MLton),
  OCaml, and GHC, respectively.
  When utilizing 48 threads, our geomean speedup is
  $1.92\times$, $3.73\times$ and $4.01\times$,
  meaning that the use of dense
  representations to improve sequential processing performance
  {\em coexists with scalable parallelism}.
  Most notably, the speedup on a five-pass compiler drawn from a university
  compiler course was $1.02\times$, $2.2\times$ and $10.7\times$ over those alternative
  languages.
  
\end{itemize}


\section{Overview}
\label{sec:overview}

We give a high-level overview of the ideas presented in this paper
using a program given in \Figref{fig:example1-haskell}
\iftoggle{EXTND}{
  (with a larger sample program given in Appendix \ref{sec:appendix-code-snippet}).
}{
  (with a larger sample program given in the Appendix of the extended version~\cite{ParallelLoCal-tr}).
}
Because the techniques we present for compiling tree-traversals are
directly applicable to {\em compilers themselves}, we use 
a miniature compiler pass as our example.
The example defines a datatype \il{Exp} which represents the abstract syntax of a language that supports
integer arithmetic, and a function \il{constFold} that implements
constant folding for this language.
Constant folding is a common compiler optimization in which expressions with
constant operands are evaluated at compile time, thus improving the run-time performance.
But here we are trying to optimize the performance of {\em the constant folding pass itself}
rather than the performance of the program produced by constant folding.
%
\il{constFold} walks over the abstract-syntax-tree, and substitutes all
expressions of the form \il{(Plus (Lit i) (Lit j))} with \il{(Lit (i+j))}.
We only show a simplified \il{constFold} ---
for example it doesn't recur on the children of \il{Plus} before checking if they're literals ---
to keep it simple enough to serve as a running example.

The program in \Figref{fig:example1-haskell} is written using the
front-end language for Gibbon, a polymorphic, higher-order subset of Haskell,
with strict rather than lazy evaluation.
The \il{($\partup$)} operator used on line~\ref{line:parallel-tuple} denotes a parallel tuple --- it marks its operands to evaluate in parallel with each other.
But with a purely functional source language, it is semantically equivalent to a sequential
tuple, i.e., replacing all ``\il{$\partup$}'' occurences with ``\il{,}'' yields a valid program.
%
We will return to this topic in \Secref{sec:implementation}.

Gibbon uses LoCal (short for location calculus) as an intermediate
representation (IR) with explicit byte-addressed, mostly-serialized data layout.
To go from the vanilla Haskell front-end language to \seqcalc{},
it performs {\em location inference},
a variant of region inference~\cite{regioncalcs,mlkit-retrospective},
on the input programs.
%
%
The LoCal IR code generated by Gibbon for the \il{constFold} function
is shown in \Figref{fig:example1}.
In the following, we use it to sketch out how LoCal works.

{
\small
\input{fig1.tex}
}

\subsection{A Primer on Location-Calculus}
\label{subsec:local-primer}

LoCal is a type-safe IR that represents programs operating on
densely encoded (serialized) data.  All serialized values live in regions,
which are unbounded memory buffers \new{that never overlap and} that store the raw data.
All programs make explicit not only the region to which a value belongs to, but
also a \textit{location} at which that value is written.
In our notation, a location $\locreg{l}{r}$ resides in region $r$.
Locations are fine-grained indices into a region, but unlike pointers in
languages like C, arbitrary arithmetic on locations is not allowed.
Locations are only introduced relative to other locations,
\new{and they can be written to only once.}
\new{
Once allocated at a particular location, a value cannot be {\em shared} with another location
(within the same region or across regions), and it has to be {\em copied}
to allocate it at a different location.
(In practice, the Gibbon compiler supports sharing using pointers, which
we discuss in \Secref{subsec:indirection-pointers}.)
}

A new location is either: at the start of a region, one unit past
an existing location, or {\em after} all elements of a value rooted at an
existing location.
In the program given in \Figref{fig:example1},
the location $\locreg{\loc_3}{\reg_2}$ is one past the
location $\locreg{\loc_2}{\reg_2}$ (line~\ref{line:letloctag})
and $\locreg{\loc_4}{\reg_2}$ is after
every element of the value rooted at location $\locreg{\loc_3}{\reg_2}$
(line~\ref{line:letlocafter}).
Any expression that allocates takes an extra argument:
a location-region pair that specifies where the allocation should happen.
The types of such expressions are decorated with these location-region pairs.
For example, the \il{(Lit $\;\locreg{\loc_2}{\reg_2}\;$ i)} data constructor
(line~\ref{line:datacon})
allocates a tag at location $\loc_2$ in region $\reg_2$, and has type \il{(Exp$\tyatlocreg{}{\loc_2}{\reg_2}\;$)}.
Any scalar arguments passed to a data constructor, such as the
unpacked integer \il{i} in this case, are allocated immediately after
the tag.
Functions may be polymorphic over any of their input or output locations,
and these locations are provided at call-sites.
In the example, the function \il{constFold} is polymorphic over an
input location $\locreg{\loc_1}{\reg_1}$ and an output location $\locreg{\loc_2}{\reg_2}$,
and values for these are given at all call-sites.
\new{
In spite of the forall quantifier in its type signature,
the input and output regions given at its call-site must be distinct ($\reg_1 \neq \reg_2$) to prevent overwrites.
This property is checked by LoCal's type-system (described in \Secref{subsec:type-system}),
which makes multiple writes to any location illegal---with the use of a nursery environment---
ensuring that function calls like
(\il{constFold} [$\locreg{\loc_x}{\reg_x}$ $\locreg{\loc_x}{\reg_x}$] \il{x})
don't type check.
}

\subsubsection{Sequential Execution Model}
\label{subsec:local-primer-eval}

LoCal has a dynamic semantics which can run programs sequentially~\cite{LoCal}.
In this model, regions are represented as serialized heaps, where each heap
is an array of cells that can store primitive values
(data constructor tags, numbers, etc).
A write operation, such as the application of a data constructor, allocates to a fresh cell
on the heap, and a read operation reads the contents of a cell.
Performing multiple reads on a single cell is safe, and the type-system ensures
that each cell (location) is written to only once.
At run time, locations in the source language translate to heap indices,
which are the concrete addresses of the cells where reads/writes happen.
Computing addresses of locations which are at the start of a region,
or one past another location is straightforward
--- the addresses get initialized to \il{0} and \il{(prev + 1)} respectively.
But evaluating an $\gramwd{after}$ expression,
to get an address one past the end of another, variable-sized value,
requires more work.

A naive computational interpretation of this $\gramwd{after}$ is to simply scan over
a value to compute its end.
In LoCal's formal model, this is referred to as the {\em end-witness judgment}.
Locations computed via $\gramwd{after}$ are used during both read and write operations.
For example,
when a LoCal \il{case} expression pattern matches on \il{(Plus e1 e2)},
it has to scan past \il{e1} in order to know the starting address of \il{e2},
which adds $O(n)$ amount of extra work in a {\em fully} serialized representation!
In practice, if values are read in the same order in which they were serialized,
a linear scan can be avoided by {\em tracking end-witnesses} that are
naturally computed in the evaluation of the program,
for example, by having every read return the address of the cell after it.
Intuitively, we can imagine there being a single read pointer that
is used to perform all reads in the program.
It always points to the next cell to be read on heap, and each read
advances it by one.
When the program starts executing, the read pointer starts at the
beginning of the heap and it chugs along in a continuous fashion.
Allocating a serialized value can  be thought of in a similar way ---
that there is a single allocation pointer that starts at the
beginning of the heap, and moves along its length performing writes,
as illustrated in \Figref{fig:step-by-step-dynamics-sequential-right}.
To avoid changing the asymptotic complexity of programs which read values out-of-order,
the Gibbon compiler by default inserts some offset information ---
such as pointers to some fields of a data constructor ---  back into the representation.  
But it doesn't allow {\em out-of-order allocations},
which will be needed as we add parallelism to LoCal (\Secref{subsec:parallelism-in-local}).

\subsubsection{Sequential Execution Model, Example}

{
\begin{figure*}
\hspace{0.07\linewidth}
\begin{subfigure}[t]{0.35\linewidth}
  \vspace{0pt}
  \includegraphics[scale=0.36]{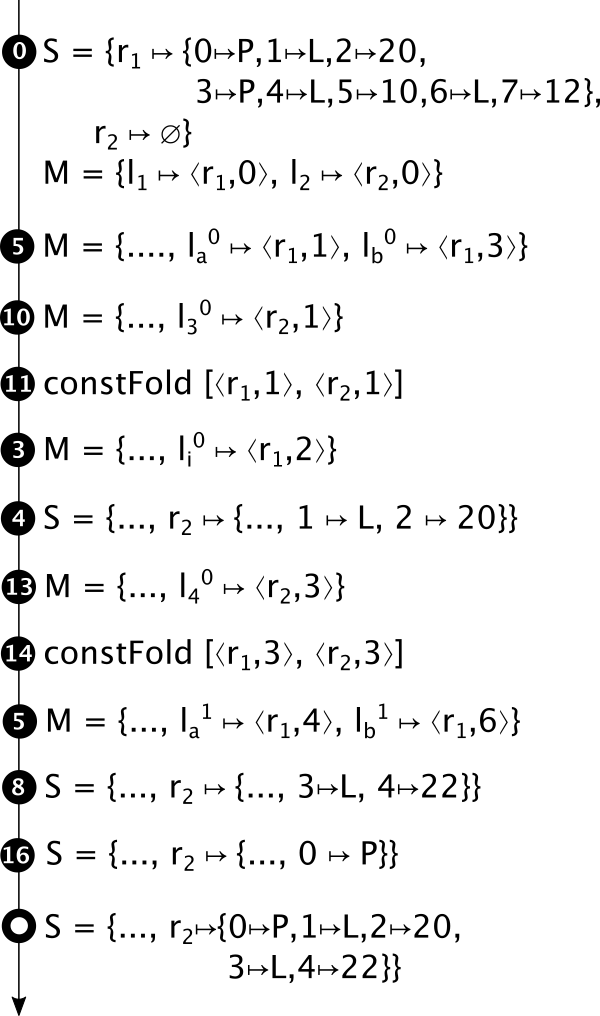}
  \caption{}
  \label{fig:step-by-step-dynamics-sequential-left}
\end{subfigure}
\hspace{0.16\linewidth}
\begin{subfigure}[t]{0.35\linewidth}
  \vspace{0pt}
  \includegraphics[scale=0.36]{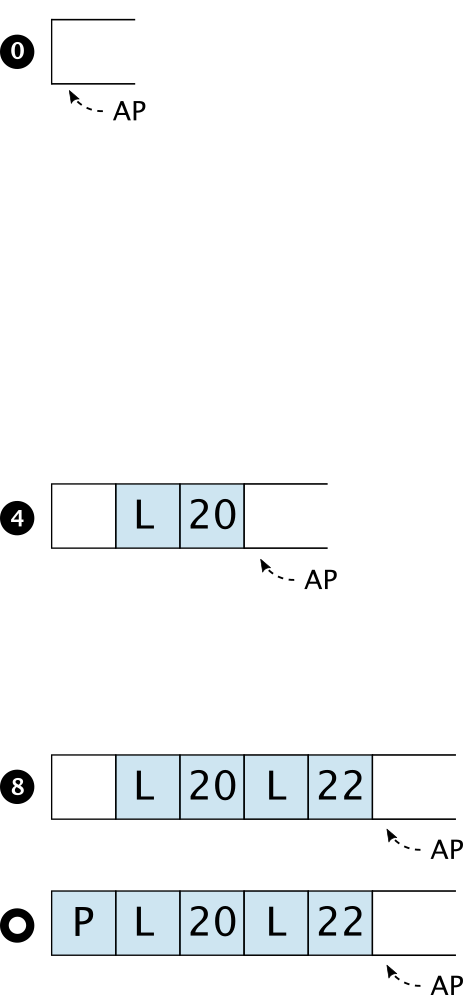}
  \caption{}
  \label{fig:step-by-step-dynamics-sequential-right}
\end{subfigure}
\caption{
  (a) Sequential, step-by-step execution of the program from
  \Figref{fig:example1},
  and (b) \new{the heap operations corresponding to the output region $\reg_2$.}
  Each step is named after its line number in the program and only
  shows the changes relative to the previous step.
  \textit{AP} is the allocation pointer.
  \il{P} is short for \il{Plus}, and \il{L} is short for \il{Lit}.
  %
}
\label{fig:step-by-step-dynamics-sequential}
\end{figure*}
}

To make this execution model concrete, let us go over
a step-by-step trace of the semantics executing \il{constFold} on
\il{(Plus (Lit 20) (Plus (Lit 10) (Lit 12)))}.
%
%
The execution trace is given in \Figref{fig:step-by-step-dynamics-sequential}.
The store $\STOR$ maps regions to their corresponding heaps, and
the location map $\MENV$ maps symbolic locations to their corresponding
heap indices.
The evaluation starts at
\il{(constFold [$\locreg{l_1}{r_1}\;\locreg{l_2}{r_2}$] e)},
and is given a store containing a fully allocated input region $r_1$
and an empty region $r_2$ to allocate the output,
along with a location map
containing the locations $\locreg{l_1}{r_1}$ and $\locreg{l_2}{r_2}$
initialized to the starting addresses of these regions.
Since the input region has a \il{Plus} at the top, 
execution continues at line~5.
The pattern match binds the locations $l_{a}^0$ and $l_{b}^0$ to
the addresses of the sub-expressions \il{(Lit 20)} and \il{(Plus (Lit 10) (Lit 12))}
respectively .
Since both the sub-expressions are not constants, execution continues at line~10.
Then, the output location of the first sub-expression, $\locreg{l_3}{r_2}$, is defined to
be one past $\locreg{l_2}{r_2}$, and \il{constFold} is invoked recursively
on this sub-expression.
%
%
Step 4 \new{{\em copies}\footnote{\new{This value is copied because line~\ref{line:datacon} in \Figref{fig:example1} has a data constructor \il{(Lit $\;\locreg{l_2}{r_2}\;$ i)} on the right hand side of the case alternative. If we update the program to return the input expression \il{exp} directly, Gibbon would allocate a pointer and the value \il{(Lit 20)} would be {\em shared} between the input and the output regions. We discuss how sharing works in \Secref{subsec:indirection-pointers}.}}}
the first sub-expression by writing a tag $L$ (short for \il{Lit}),
followed by the integer 20 on the heap.
Then, the output location of the second sub-expression, $\locreg{l_4}{r_2}$, is defined to be one past
every element of the first sub-expression, which occupies two cells after the 0$^{th}$ cell.
Thus, $\locreg{l_4}{r_2}$ gets initialized with the address of the 3$^{rd}$ cell.
\il{constFold} is now invoked recursively
for the second sub-expression.
%
%
Following similar steps, the second sub-expression is allocated at $\locreg{l_4}{r_2}$.
Since the second sub-expression is a \il{Plus} with constant operands,
it is transformed to \il{(Lit 22)}.
Finally, Step 16 writes the tag $P$ (short for \il{Plus}) which
completes the construction of the full expression,
\il{(Plus (Lit 20) (Lit 22))}.

\subsection{Parallelism in Location-Calculus}
\label{subsec:parallelism-in-local}

In this section, we outline various {\em latent} opportunities for parallelism
that exist in \seqcalc{} programs (irrespective of annotation with ``\il{$\partup$}'').
The first kind of parallelism is available when programs access
the store in a read-only fashion,
such as in an interpreter, for example.
\vspace{1mm}
\begin{code}
interp : forall @\locreg{l}{r}@ . @\tyatlocreg{Exp}{l}{r}@ -> Int
interp [@\locreg{l}{r}@] t = case t of
  Lit (i : @\tyatlocreg{Int}{l_i}{r}@) -> i
  Plus (e1 : @\tyatlocreg{Exp}{l_a}{r}@) (e2 : @\tyatlocreg{Exp}{l_b}{r}@) ->
    (interp [@\locreg{l_a}{r}@] a) + (interp [@\locreg{l_b}{r}@] b)
  ...
\end{code}
Even though the recursive calls in the \il{Plus} case can
safely evaluate in parallel, there is a subtlety: parallel evaluation
is efficient only if the \il{Plus} constructor stores offset
information for its right child node.
If it does, then the address of \il{e2} can be calculated in constant
time, thereby allowing the calls to proceed immediately in parallel.
If there is no offset information, then the overall tree traversal is
necessarily sequential, because the starting address of \il{e2} can be
obtained only after a full traversal of \il{e1}.
As such, there is a tradeoff between space and time, that is, the cost
of the space to store the offsets, 
versus the time of the sequential traversal 
forced by the absence of offsets.

Programs that write to the store also provide opportunities for
parallelism.
The most immediate such opportunity exists when the program performs
writes that affect different regions.
Such writes can happen in parallel because different regions cannot overlap
in memory.
%
%
There is another kind of parallelism that is more challenging
to exploit, but is at least as important as the others:
intra-region parallelism that can be realized by allowing different fields
of the same constructor to be filled in parallel.
This is crucial in \seqcalc{} programs, where large, serialized data
(large trees or DAGs) frequently occupy only a small number of regions, and yet there are
opportunities to exploit parallelism in their construction.

Consider the case in \Figref{fig:example1} which recursively calls
\il{constFold} on the sub-expressions of \il{Plus}.
%
If we want to access the parallelism between the recursive calls,
we need to break the data dependency that the right branch has on
the left.
The starting address of the right branch, namely $\locreg{l_4}{r_2}$, is
assigned to be end witness of the left branch by the \il{after}
expression.
But the end witness of the left branch is, in general, known only
after the left branch is completely filled, which would effectively
sequentialize the computation.
One non-starter would be to ask the programmer to specify the
size of the left branch up front, which would make it possible to
calculate the starting address of the right branch.
Unfortunately, this approach would introduce safety issues, such as
incorrect size information, of exactly the kind that \seqcalc{}
is designed to prevent.
Instead, we explore an approach that is safe-by-construction and
efficient, as we explain next.

\section{Region-Parallel LoCal}
\label{sec:region-parallel-local}

To address the challenges of parallel evaluation---in concert with dense,
mostly-serialized data representations---we start by
presenting an execution model, \rparcalc{},
which can utilize {\em all} potential parallelism in \seqcalc{} programs.
Parallelism in this formal model is generated {\em implicitly},
by allowing every let-bound expression to potentially evaluate in parallel with the body.
Accordingly, the language omits explicit parallelism ``hints'' (\il{$\partup$}).
That is, you'll see in the next sections that implicitly parallel \il{let} has both a sequential and parallel evaluation rule.
%
%
%
By modeling every possible parallelization, the formal model is general ---
it formalizes all possible valid parallel schedules, and all valid heap layouts.
We return to the pragmatic issue of selecting {\em efficient} parallelizations, i.e. granularity control, in \Secref{subsec:granularity}.
%

\subsection{Region Memory and Parallel Tasks}
\label{sec:indirection-semantics-intro}

{
\begin{figure*}
\hspace{-5mm}
\begin{subfigure}[t]{0.45\linewidth}
  \vspace{0pt}
  \includegraphics[scale=0.34]{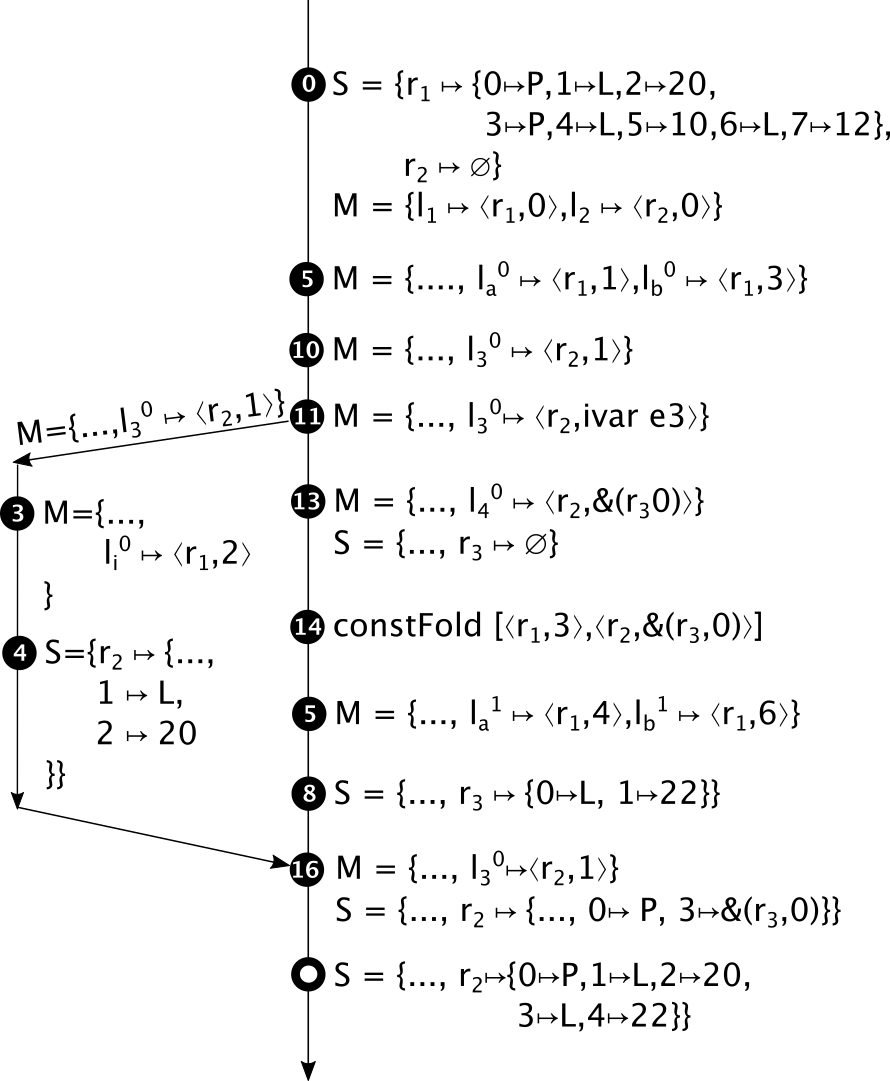}
  \caption{}
  \label{fig:step-by-step-dynamics-region-parallel-left}
\end{subfigure}
\hspace{0.12\linewidth}
\begin{subfigure}[t]{0.4\linewidth}
  \vspace{0pt}
  \includegraphics[scale=0.34]{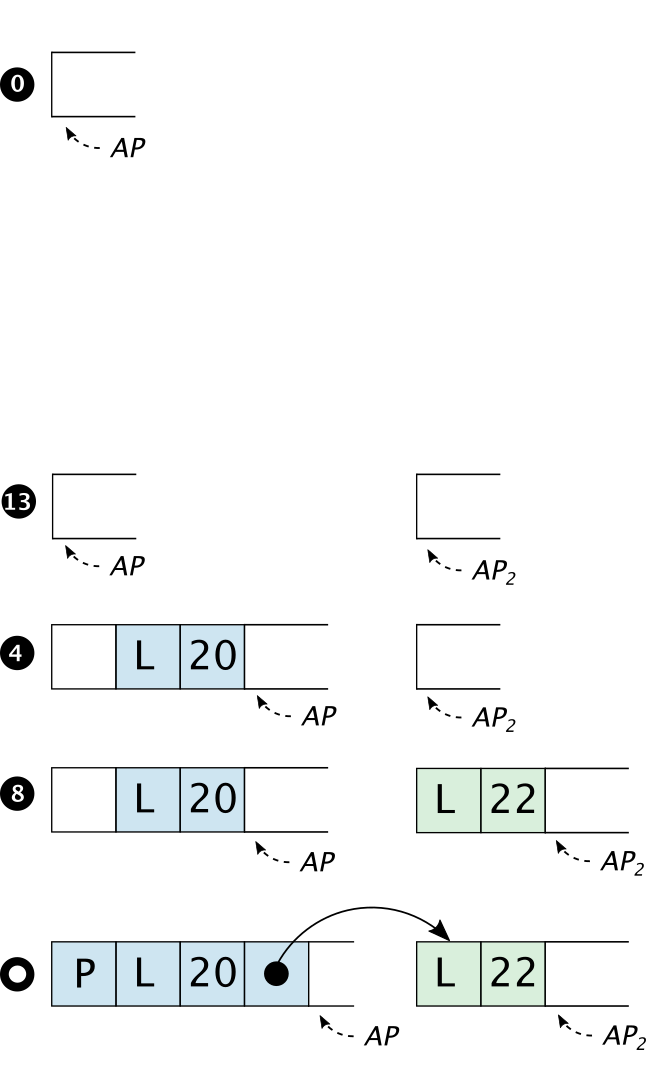}
  \caption{}
  \label{fig:step-by-step-dynamics-region-parallel-right}
\end{subfigure}
\caption{
  (a) Parallel, step-by-step execution of the program from \Figref{fig:example1} such that
  parallel allocations happen only in separate regions,
  and (b) \new{the heap operations corresponding to the output region $\reg_2$.}
  Each step is named after its line number in the program and only
  shows the changes relative to the previous step.
  \il{P} is short for \il{Plus}, and \il{L} is short for \il{Lit}.
}
\label{fig:step-by-step-dynamics-region-parallel}
\end{figure*}
}


In the formal model, while parallelism is implicit, there is
still a restriction that at
most one task allocates in a given region at a time.
To realize intra-region parallelism, the model introduces fresh,
intermediate regions as needed, that is, when
the schedule takes a parallel evaluation step for a given let-bound expression,
and the body tries to allocate in the same region.
To demonstrate this, let us consider a trace of the
region-parallel evaluation of the program from \Figref{fig:example1},
%
corresponding to the schedule
shown in \Figref{fig:step-by-step-dynamics-region-parallel},
where the recursive calls to \il{constFold} on lines~12 and 14 run in parallel
with each other.
The parallel fork point for the first recursive call
occurs on the $11^{th}$ step of the trace.
At this point, the evaluation of the let-bound expression results in
the creation of a new child task, and the
continuation of the body of the let expression in the parent task.

Each task has its own private view of memory, which is realized by
giving the child and parent task copies of the store $\STOR$ and
location map $\MENV$.
These copies differ in one way, however: each sees a different mapping
for the starting location of \il{e3}, namely $\locreg{l_3}{r_2}$.
The child task sees the mapping
$\locreg{l_3}{r_2} \mapsto \concreteloc{r_2}{1}{}$, which
is the ultimate starting address of \il{e3} in the heap.
The parent task sees a different mapping for $\locreg{l_3}{r_2}$,
namely $\concreteloc{r_2}{\indivar{e3}}{}$.
This address is an \emph{ivar}: it behaves exactly like an I-Var~\cite{IStructures},
and, in our example, stands in for the completion of the memory being
filled for \il{e3}, by the child task.
Any expression in the body of the let expression that tries to read
from this location blocks on the completion of the child task.

The only exception to this blocking-rule is a letloc $\gramwd{after}$ expression, which is handled differently.
Such an expression occurs at line~\ref{line:letlocafter},
just after the parent continues after the fork point.
At this step, the parent task uses an \il{after} expression to
assign an appropriate location for the starting address of \il{e4},
one past every byte occupied by \il{e3}.
If we synchronize with the child task here, the computation will effectively be
sequential.
In order to avoid that,
the starting address of \il{e4} is assigned to be
$\locreg{l_4}{r_2} \mapsto \concreteloc{r_2}{\indirection{r_3}{0}}{}$.
This address is an \emph{indirection} pointing to the start of fresh region $r_3$, and
causes the parent task to allocate \il{e4} in the region $r_3$ instead of $r_2$,
which is being allocated to by the child task,
thus maintaining the single-threaded-per-region allocation invariant.
The parent and child tasks have, in effect, two different
allocation pointers for 
what will functionally be the same region (after joining).
The use of \il{e3} on line~\ref{line:joinpoint} forces the parent task
to join with its child task.
In particular,
$\concreteloc{r_2}{\indivar{e3}}{}$ is
substituted by $\concreteloc{r_2}{1}{}$, the starting address of
\il{e3}, in the expression and the location map $\MENV$.
Also, all the new entries in the location map $\MENV$ and store
$\STOR$ of the child are merged into the corresponding environments in
the parent.
Finally, the regions $r_2$ and $r_3$ are linked with a pointer,
corresponding to the indirection pointer that was added for the starting address of \il{e4}.

\subsection{Syntax and Operational Semantics}
\label{subsec:region-parallel-semantics}

\begin{figure}
  \small
  \input{formal_grammar}
  \input{formal_dynamics_grammar}
  \normalsize
  \caption{Grammar of \rparcalc{}.}  
  \label{fig:opergram}
\end{figure}

In this section, we present the formal semantics of our parallel
location calculus, \rparcalc{}. {This semantics has also been mechanically tested in PLT
  Redex~\cite{redex}.
The grammar for the language is given in \Figref{fig:opergram}.
Again, all parallelism in this model language is introduced implicitly, by
evaluating \il{let} expressions.
There is no explicit syntax for introducing parallelism in our
language, and consequently the language is,
from the perspective of a client, exactly same as the
sequential language~\cite{LoCal}.

The parallel operational semantics does, however, differ from the
sequential semantics, most notably from the introduction of a richer
form of indexing in regions.
Whereas in sequential \seqcalc{} a region index consists simply of a non-negative integer,
it is enriched to an extended region index $\indbef{\ind}$ in \rparcalc{}.
It consists of either a concrete index
$\concreteind{\ind}$, an $\indivar{\ivarid}$, or an indirection pointer
$\indirection{\reg}{\concreteind{\ind}}$.
A concrete index is a non-negative integer that specifies the
final position of a value in a region.
An ivar is a synchronization variable that is used to coordinate between parallel tasks.
For example, the $\indivar{e3}$ in the sample
trace in \Figref{fig:step-by-step-dynamics-region-parallel}, is used
to synchronize with the child task that is allocating \il{e3}.
An indirection $\indirection{\reg}{\concreteind{\ind}}$ points
to the address $\concreteind{\ind}$ in the region $\reg$,
and is used to link together different \emph{chunks} of the same logical region,
which may have been introduced to enable intra-region parallel allocation.
For example, in the sample
trace in \Figref{fig:step-by-step-dynamics-region-parallel}, a pointer
$\indirection{\reg_3}{0}$ written at the end of the value \il{e3}
links it with the value \il{e4}, which is allocated to a separate region $\reg_3$.
And a concrete location $\concretelocvar$ is enriched to a pair
$\concreteloc{\reg}{\indbef{\ind}}{}$,
of a region $\reg$,
and an extended region index $\indbef{\ind}$.
The state configurations of \rparcalc{} appear at the bottom of
\Figref{fig:opergram}.
Just like in sequential \seqcalc{}, a sequential state of
\rparcalc{}, $\SEQSTATE$, contains a store $\STOR$, location map $\MENV$, and an
expression $\EXPR$.
But by using enriched concrete locations,
the location map also has the ability to contain indirection pointers.
A value that can be written to a heap, $\heapval$, is
similarly enriched to allow indirection pointers.

\subsubsection{Sequential Transitions}
\label{subsubsec:sequential-semantics}

{
\begin{figure}
  \small
  \normalsize
  \begin{mathpar}
    \mprset{flushleft}
    \rdregionpardatacon{}\\
    \rdregparletlocafterseq{}\\
    \rdregparletlocafternewreg{}\\    
    \rregpardcase{}\\
    \rdregparletexp{}\\
    \rdregparletval{}
  \end{mathpar}
  \caption{Selected dynamic semantics rules (sequential transitions).}
  \label{fig:dynamic-sequential}
\end{figure}
}

A subset of the sequential transition rules are given in
\Figref{fig:dynamic-sequential}.
The rules are close to the original sequential rules,
except for some minor differences.
For the rule \ddatacon{}, we need to handle the case
where an indirection is assigned to the source symbolic location
$\locreg{\loc}{\reg}$.
For this purpose, we use a metafunction $\hat{\MENV}$
\iftoggle{EXTND}{
  (formally defined in Appendix~\ref{subsec:maplookup})
}{
  (formally defined in the Appendix of the extended version~\cite{ParallelLoCal-tr})
}
that can dereference indirection pointers when looking up its address in the location map $\MENV$.
With respect to the rule \dregparletlocafternewreg{}, we now
allow the concrete location assigned to the source location
$\locreg{\loc_0}{\reg}$ to hold an ivar.
The purpose of this relaxation is to allow an expression downstream
from a parallelized \il{let} binding to continue evaluating in
parallel with the task that is evaluating the let-bound expression.
The task evaluating the \il{after} expression continues by using an
indirection pointing to the start of a fresh region $\reg'$.
%
%
The effect is to make $\concreteloc{\reg'}{0}{}$ the setting for the allocation
pointer for the task.
If the source location $\locreg{\loc_0}{\reg}$ is assigned to hold a concrete index
$\concreteind{\ind}$, the rule \dletlocafter{} yields an address by using the
the end-witness judgment.
The remaining rules are similar to sequential \seqcalc{},
\iftoggle{EXTND}{
  and are available in Appendix~\ref{subsec:appendix-dynamic-semantics}.
}{
  and are available in the Appendix of the extended version~\cite{ParallelLoCal-tr}.
}

\subsubsection{Parallel Transitions}
\label{subsubsec:parallel-semantics}

{
\begin{figure}
  \small
  \normalsize
  \begin{mathpar}
    \mprset{flushleft}
    \rdregparsteptask{}\\
    \rdregparletexppar{}\\
    \rdregparcasejoin{}\\
    \rdregpardataconjoin{}
  \end{mathpar}
  \vspace{-6mm}
  \caption{Dynamic semantics rules (parallel transitions).}
  \label{fig:dynamic-parallel}
\end{figure}
}

We generalize a sequential state to a parallel task $\TASKVAR$ by
adding two more fields: a located type and a concrete location, which
together describe the type and location of the final result allocated
by the task.
A parallel transition in \rparcalc{} takes the form of the following
rule, where any number of tasks
in a task set $\TASKSET$ may step together.
\begin{displaymath}
\TASKSET \rparstepsto  \TASKSET'
\end{displaymath}
In each step, a given task may make a sequential transition, it may
fork a new parallel task, it may join with another parallel task,
or it may remain unchanged.

The parallel transition rules are given in
\Figref{fig:dynamic-parallel}.
In these rules, we model parallelism by an interleaving semantics.
Any of the tasks that are ready to take a sequential step may make a
transition in rule \dregparsteptask{}.
A parallel task can be spawned by the \dletexppar{} rule, from which an
in-flight \il{let} expression breaks into two tasks.
The child task handles the evaluation of the \il{let}-bound expression
$\EXPR_1$, and the parent task handles the body $\EXPR_2$.
To represent the future location of the let-bound expression,
and to create a data dependency on it,
the rule creates a fresh ivar, which is
passed to the body of the \il{let} expression.
This same ivar is also the target concrete location of the
child task, thereby indicating that it produces this value.

%
A task can satisfy a data dependency in a rule such as
\dregparcasejoin{}, where a \il{case} expression is blocked
on the value
located at $\indivar{\ivarid_c}$, by joining with the task producing the value.
Because each task has a private copy of the store and location map,
the process of joining two tasks involves merging environments.
The merging of the task memories is performed by the metafunctions $\merges$ and $\mergem$,
\iftoggle{EXTND}{
  defined formally in Appendix~\ref{subsec:appendix-merging-memories-region-parallel}.
}{
  defined formally in the Appendix of the extended version~\cite{ParallelLoCal-tr}.
}
We merge two stores by merging the heaps of all the regions that are
shared in common by the two stores, and then by combining with all
regions that are not shared.
We merge two heaps by taking the set of all the heap values at
indices that are equal, and all the heap values at
indices in only the first and only the second heap.
The merging of location maps follows a similar pattern, but is slightly
complicated by its handling of locations that map to ivars.
In particular, for any location where one of the two location maps
holds an ivar and the other one holds a concrete index, we assign
to the resulting location map the concrete index, because the concrete
index contains the more recent information.
After merging the environments, all occurrences of
$\indivar{\ivarid}_c$ are eliminated in the continuation,
and are replaced by the index $\ind_p$,
that represents the starting index of the value produced by the task $\TASKVAR_p$.
Join points in \rparcalc{} are, in general, deterministic, because they only
{\em increase} the information held by the parent task.

The rule \dregpardataconjoin{} handles the case where a data constructor
is blocked on the value of its $j^{th}$ field, and it joins with the task
producing that value.
It is similar to \dregparcasejoin{}, and also requires merging environments.
But depending on the schedule of execution,
if the $(j+1)^{th}$ field of this constructor was computed in parallel
with the $j^{th}$ field, they will both be allocated to separate regions,
due to the way the rule \dregparletlocafternewreg{} works.
These fields have to be reconciled to simulate a single region.
For this purpose, we use a metafunction $\linkfields$
\iftoggle{EXTND}{
  --- defined formally in Appendix~\ref{subsec:appendix-linking-fields} ---
}{
  --- defined formally in the Appendix of the extended version~\cite{ParallelLoCal-tr} ---
}
which stitches together these fields
by writing an indirection pointing to the start of region containing the $(j+1)^{th}$ field
at an address one past the end of the $j^{th}$ field.
Thus, when all fields of a data constructor are synchronized with, all fields allocated
to different regions are linked together by indirection pointers,
forming a linked-list.


\subsection{Type System}
\label{subsec:type-system}

{
\begin{figure}
\begin{mathpar}
\rttask{}
\quad
\rttasksetempty{} \\
\rttaskset{}
\end{mathpar}
\caption{Typing rules for a parallel task $\TASKVAR$, and a set of parallel tasks $\TASKSET$.}
\label{fig:static-semantics-parallel}
\end{figure}
}

Our type system for \rparcalc{} requires some substantial extensions
to the original type system given by \citet{LoCal}.
These extensions address the need to handle multi-task configurations,
which require a number of new typing environments and rules.
Before we present these extensions, we recall the typing rule
for the configuration of a single task, which is mostly unchanged from
the original.
\[ \TENV;\SENV;\CENV;\AENV;\NENV \vdash \AENV'; \NENV'; \EXPR : \hTYP \]
The context for this judgment includes five different environments.
First, $\TENV$ is a standard typing environment.
Second, $\SENV$ is a store-typing environment, mapping 
\emph{materialized} symbolic locations to their types. That is, every location in $\SENV$ {\em has been written} and contains a value of type $\SENV(l^r)$.
Third, $\CENV$ is a constraint environment, keeping
track of how symbolic locations relate to each other.
Fourth, $\AENV$ maps each region in scope to a location, and is used to symbolically
track the allocation and incremental construction of data structures;
Finally, $\NENV$ is a nursery of all symbolic locations that have been allocated,
but not yet written to.
\new{Locations are removed from $\NENV$ upon being written to,
as the purpose is to prevent multiple writes to a location.}
Both $\AENV$ and $\NENV$ are threaded through the typing
rules, also occurring in the output of the judgment, to the right of the
turnstile.

To generalize our typing rules to handle multi-task configurations,
we introduce new environments for variables $\TENVMAP$, store
typing $\SENVMAP$, allocation constraints $\CENVMAP$, allocation
pointers $\AENVMAP$, and nurseries $\NENVMAP$.
These environments extend their counterparts in the sequential
\seqcalc{} type system, and are needed to track state on a per-task
basis.
\Figref{fig:static-semantics-parallel} gives 
the precise typing rules to type check a parallel task $\TASKVAR$, and
a set of parallel tasks $\TASKSET$.
A parallel task $\TASKVAR$ is well-typed if its target expression
$\EXPR$ is well-typed, using the original \seqcalc{} typing rules, and
a task set $\TASKSET$ is well-typed if all tasks in it are
well-typed.
\iftoggle{EXTND}{
  \rparcalc{}'s complete typing rules are given in Appendix~\ref{subsec:appendix-typing-rules}.
}{
  \rparcalc{}'s complete typing rules are given in the Appendix
  of the extended version~\cite{ParallelLoCal-tr}.
}

\subsection{Type Safety}
\label{subsec:type-safety}

Compared to the original type-safety result proved for single-task
\seqcalc{}, ours generalizes to parallel evaluation by requiring that,
for any given multi-task configuration, either the program has fully
evaluated or at least one task can take a step.
As usual, we prove this theorem by showing progress and preservation.
The main complication relates to the property that parts of the
overall store are now spread across the individual stores of the
tasks, whereas in the original proof there is only one store.
In particular, our proof must establish that the store of each task
remains well formed, even while that task waits on a data dependency,
and moreover after the task joins with another task and their stores
are merged.
The complete proof is available
\iftoggle{EXTND}{
  in Appendix~\ref{subsec:appendix-type-safety}.
}{
  in the Appendix of the extended version~\cite{ParallelLoCal-tr}.
}
Here we summarize key invariants.

Many such invariants are specified by our well-formedness rule, which
applies to a set of tasks executing in the parallel machine.
\iftoggle{EXTND}{
  We give the full rule in Appendix~\ref{sec:well-formedness-tasks}.
}{
  The full rule is given in the Appendix of the extended version\cite{ParallelLoCal-tr}.
}
\begin{displaymath}
\SENVMAP;\CENVMAP;\AENVMAP;\NENVMAP \storewftasks{\TASKSET}
\end{displaymath}
This judgment specifies two new invariants that must hold for all
tasks $\TASKVAR \in \TASKSET$.
The first enforces that all ivars get filled with an appropriate value.
In particular, if an expression being evaluated by a task references an ivar,
then there must be exactly one other task in the task set which supplies a
{\em well-typed} value for it.
The second invariant consists of the well-formedness judgment that
verifies certain properties hold for each store of a given task.
This judgment generalizes a similar rule used in the original proof
by its use of the overall task set $\TASKSET$.
We discuss in detail the necessary extensions in the sequel.

With these new typing judgments in hand, we can now state the type-safety
theorem, shown below.
This theorem states that, if a given task set $\TASKSET$ is well typed
and its overall store is well formed, and if $\TASKSET$
makes a transition to some task set $\TASKSET'$ in $n$ steps,
then either all tasks in
$\TASKSET'$ are fully evaluated or $\TASKSET'$ can take
a step to some task set $\TASKSET''$.
\begin{theorem}[Type Safety]
  \label{theorem:type-safety}
  \begin{displaymath}    
  \begin{aligned}
  \text{If} \;\; & \emptyset;\SENVMAP;\CENVMAP;\AENVMAP;\NENVMAP \vdash_{taskset} \AENVMAP' ; \NENVMAP' ; \TASKSET \; \wedge \; \SENVMAP;\CENVMAP;\AENVMAP;\NENVMAP \storewftasks{\TASKSET} \\
  \text{and} \;\; & \TASKSET \rparstepsto^{n} \TASKSET' \\
  \text{then, either} \;\; & \forall \TASKVAR \in \TASKSET'. \; \taskcomplete{\TASKVAR} \\
  \text{or} \;\; & \exists \; \TASKSET''. \; \TASKSET' \rparstepsto \TASKSET''.
  \end{aligned}
  \end{displaymath}
\end{theorem}


\subsubsection{Well formedness of the store}
  Our store well-formedness judgment extends the judgment of the
  sequential \seqcalc{} typing system to establish a global criterion
  for well-formedness, checking in particular that parts of regions
  that are distributed across each task-private store, given by
  $\MENV; \STOR$, are well-formed:
\begin{displaymath}
\storewf{\SENV}{\CENV}{\AENV}{\NENV}{\TASKSET}{\MENV}{\STOR}
\end{displaymath}
\iftoggle{EXTND}{
  This judgment, defined formally in Appendix~\ref{sec:well-formedness},
}{
  This judgment, defined formally in the Appendix of
  the extended version~\cite{ParallelLoCal-tr},
}
is one of the most challenging parts of our extension, because it must be
strong enough to ensure safe merging of stores when tasks meet at join points.
Like in sequential LoCal, it specifies three categories of invariants.

The first category enforces that allocations occurring across the
task-private stores are accounted for.
In particular, for each symbolic location in the store-typing environment,
$(\locreg{\loc}{\reg} \mapsto \TYP) \in \SENV$,
a value must be allocated to the appropriate store.
There are two possible ways in which this allocation may occur:
(1) sequentially, in the current task, or (2) in parallel, in a different task.
In the sequential case, $\locreg{\loc}{\reg}$'s address in the location map $\MENV$ must be a
concrete index, and it must have an end-witness.
This technical point ensures that the store never contains partially allocated values.
In the parallel case, $\locreg{\loc}{\reg}$'s address in $\MENV$ must be an ivar,
and there must be exactly one other task, $\TASKVAR_{oth} \in \TASKSET$,
that supplies a well-typed value for it.
Moreover, $\locreg{\loc}{\reg}$'s address in $\TASKVAR_{oth}$'s location map must be a concrete index,
and if $\TASKVAR_{oth}$ has finished evaluating, this value must have an end-witness.
This property ensures that when these tasks merge, the resulting store has {\em complete} values
allocated at the expected addresses and the expected types.

The second category enforces that allocations occur in the sequence specified by the constraint
environment $\CENV$.
In particular, if there is some location $\locreg{\loc}{}$ in the domain of $\CENV$,
then the location map and store must have the expected allocations at the expected types.
The most interesting rule here is that for the \gramwd{after} constraint,
since it involves potential parallel allocations.
For instance, if $(\loc \mapsto (\afterl{\tyatlocreg{\TYP}{\loc'}{}})) \in \CENV$,
then the values at locations $\loc$ and $\loc'$ may be allocated sequentially,
in the same task, or in parallel, in different tasks.
The sequential case is straightforward.
For the parallel case, there are two possibilities --- (1) the task allocating the value at
location $\loc'$ may be still in-flight,
or (2) it may have already synchronized with the current (parent) task.
In the first case, we ensure the presence of an appropriate indirection in the location map
($\loc \mapsto \concreteloc{\reg}{\indirection{\reg_{\text{fresh}}}{0}}{}$)
and of a fresh region in the store ($\reg_{\text{fresh}} \in \STOR$).
Otherwise, we ensure that a link between the values at locations $\loc'$ and $\loc$ exists,
which is accomplished by the metafunction $\linkfields$.

The final category enforces that each location is written to only once.
This is done by checking that the domain of the store-typing environment and the nursery
are disjoint: $dom(\SENV) \cap \NENV = \emptyset$.

\subsection{Controlling Fragmentation}
\label{subsec:fragmentation}

A consequence of \rparcalc{} introducing fresh regions
is that the schedule of evaluation dictates the way a value is laid out on the heap,
as shown in \Figref{fig:datacon-layouts}.
Every choice to parallelize an intra-region allocation implies the creation
of a new region and a new indirection, thereby introducing fragmentation.
Thus, in addition to the usual task-scheduling overheads,
in our system, a schedule that parallelizes too many allocations also leads
to fragmentation.
Conversely, effort at amortizing the overhead of parallelism
simultaneously amortizes the overhead of indirections and region
fragmentation.
We return to this topic and address fragmentation along with parallelism granularity management in \Secref{subsec:granularity}.

{
\begin{figure}
\par\medskip
\begin{subfigure}{0.45\textwidth}
  \par\medskip
  \includegraphics[scale=0.3]{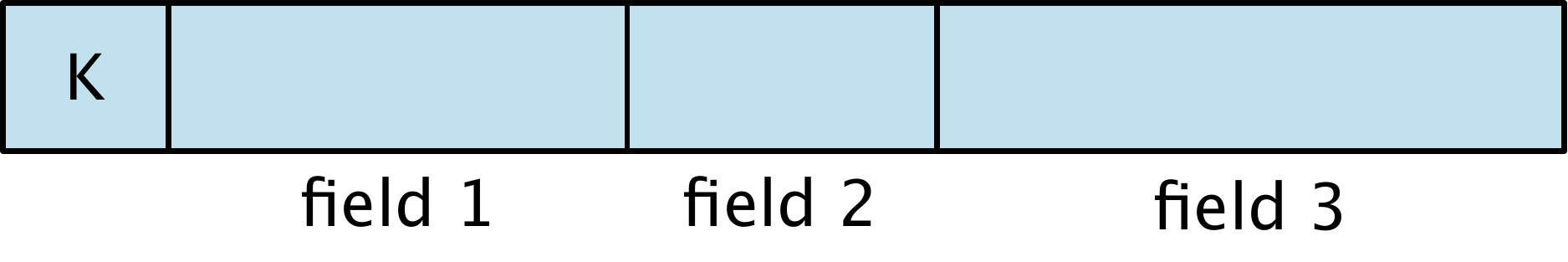}
  \caption{}
  \label{fig:datacon-layout-1}
  \par\bigskip
  \includegraphics[scale=0.3]{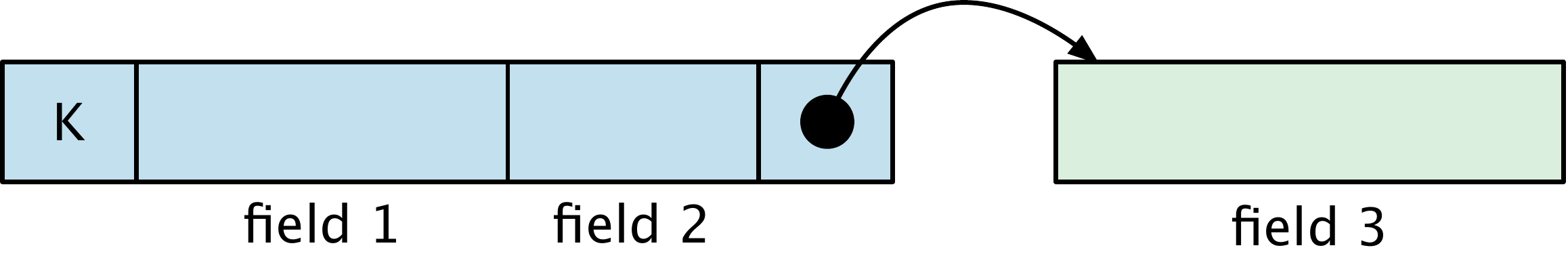}
  \caption{}
  \label{fig:datacon-layout-2}
  \par\bigskip
  \includegraphics[scale=0.3]{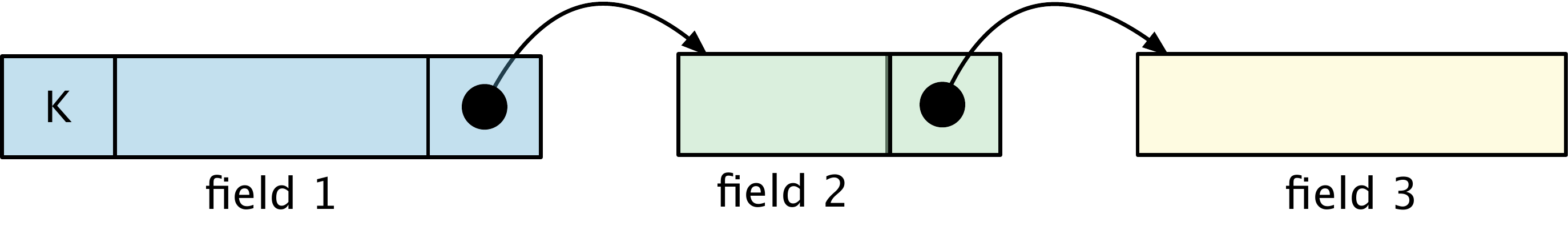}
  \caption{}
  \label{fig:datacon-layout-3}
\end{subfigure}
\caption{
  The heap layout for a data constructor if:
  (a) all fields are allocated sequentially,
  (b) only the second and third fields are allocated in parallel with each other, and
  (c) all fields are allocated in parallel.  
}
\label{fig:datacon-layouts}
\end{figure}
}

\section{Implementation}
\label{sec:implementation}

Gibbon is a whole-program\footnote{\new{Gibbon’s automatic selection of data representation works best if it can see the whole program, much like the data-representation optimizations in MLton. One way to get around this issue would be to make the programmer responsible for choosing the representation, by using appropriate annotations in datatype definitions. Another option is to conservatively insert random-access information in all datatypes that flow into code within other compilation units. Our current implementation does not offer these options, and only supports whole program compilation.}} micropass compiler that compiles a
polymorphic, higher-order subset of (strict) Haskell\footnote{Note
  that we are not the first to propose a strict variant of Haskell, not only do many of its cousins like Idris take a strict approach, but GHC itself supports a module-level strict mode.}.
The Gibbon front-end uses standard whole-program compilation and monomorphization
techniques~\cite{urweb-icfp} to lower input programs into a
first-order, monomorphic representation.
On this representation, Gibbon performs \textit{location inference} 
to convert it into a LoCal program, which has region and location annotations.
Then a big middle section of the compiler is a series of LoCal->LoCal compiler passes
that perform various transformations.
Finally, it generates C code.

Our parallelism extension operates in the middle end,
with minor additions to the backend code generator and the runtime system.
We add a collection of LoCal->LoCal compiler passes that transform the program
so that reads and allocations can run in parallel.
At run time, we make use of the Intel Cilk Plus language extension~\cite{blumofe1996cilk}
(and its work-stealing scheduler) to realize parallel execution.
Our implementation closely follows the formal model described in \Secref{subsec:region-parallel-semantics},
but with explicit parallelism annotations.
%


\subsection{Granularity Control}
\label{subsec:granularity}

Before going further into the details of what we implemented, we first
explain our choices in what we do and {\em do not} implement.
As we saw in \Figref{fig:example1-haskell}, we use manual annotations
for the programmer to mark parallelism opportunities.  This is the
norm in both current and past parallel programming practice: from
MultiLisp~\cite{multilisp} to OpenMP~\cite{openmp}, Cilk~\cite{blumofe1996cilk},
Java fork-join~\cite{lea2000java}, etc.
Recall also that with a purely functional source language,
parallel-tuple annotations change {\em performance only}, not program
semantics, so a programmer need not worry about safety when inserting
annotations\footnote{This same property holds for inserting parallel
  annotations in pure GHC Haskell code, which has been used to modest
  benefit in past experimental work \cite{ghc-insert-parallelism}, but
  is not commonplace practice.}.
Task granularity thresholds can be fine-tuned by using the same reasoning
as in other parallel systems --- switch to sequential for small
problem sizes.
But there's also the issue of fragmentation
(\Secref{subsec:fragmentation}), i.e. amortizing the overhead of
pointers in the representation as well as parallel tasks in the
control flow. One might wonder how these interact.

\subsubsection{How to optimize granularity in Gibbon?}

%
Relatively small chunks of sequential data can effectively amortize the
cost of creating regions and indirections.
For instance, serializing just the bottom two levels in our binary tree examples eliminates 75\% of pointers and ensures pointers use only 11\% of memory.
The task size to amortize parallel scheduling overheads, on the other
hand, is usually much larger.  Therefore, it's best the Parallel
Gibbon programmer thinks about parallelism granularity exclusively,
and the data representation can comfortably follow from that.
%
%
In other words, a manual (or automatic) solution to task granularity, also gives ``for free'', an efficient, mostly-serialized data representation with amortized indirections.

\subsubsection{Why not {\em automatic} granularity?}

{\em Automatic} task-parallel granularity control is an active research area
\cite{Heartbeat_17,OracleGuided_19}.
Combining Parallel Gibbon's automatic
control over data representation, {\em together} with automatic
granularity control, is a promising avenue for future work.  This goes
doubly so if approaches like {\em Heartbeat scheduling}
\cite{Heartbeat_17} mature to the point of offering robust backends
and runtime systems that compilers like Parallel Gibbon may target,
and very recent work offers a step in that direction \cite{TPAL_21}.

For this paper, however, it would be confounding to address automating
granularity {\em simultaneously} with compacting data representation and
assigning regions.
In our experiments (\Secref{sec:evaluation}), we hold task granularity
constant across different implementations of the same benchmarks,
focusing only on the impact of each compiler's code generation and
data representation choices.
Parallel Gibbon, as well as all of its competitors, use explicit
parallelism annotations, and schedule the same set of tasks at runtime
for the same program inputs, unless mentioned otherwise.

\subsection{Desugaring Parallel Tuples}

As shown in \Figref{fig:example1-haskell},
in the front-end language, we use the standard parallel tuples to express parallelism,
like other eager, parallel functional languages~\cite{manticore,MPL}.
A parallel tuple \il{(e1 $\;\partup\;$ e2)}
marks the expressions \il{e1} and \il{e2} to evaluate in parallel with each other.
To more closely match Cilk, we desugar these parallel tuples into a spawn/sync representation in the compiler IR:
\begin{code}
let x = spawn e1 in
let y = e2 in
let _ = sync in
(x,y)
\end{code}
Using this representation simplifies the subsequent conversion to LoCal,
in which additional steps like allocating regions or binding locations
may be required before getting to \il{e1} or \il{e2}.
Generating the corresponding \il{letregion}/\il{letloc} bindings, such that they have
the correct scope, is easier with a spawn/sync representation.
Also, we preserve these parallelism annotations in the LoCal code we
generate.  In contrast with the formal model (with implicit
parallelism, \Secref{sec:region-parallel-local}), \il{let y = e2}
is always sequential, whereas \il{let y = spawn e2} essentially
corresponds to a {\em potentially} parallel let binding, though the decision
is ultimately dynamic.

We do not support first-class futures, or tasks that communicate through channels
or other mutable data structures, and thus the task-parallelism opportunities available in
Parallel Gibbon remain effectively {\em series-parallel}.
But this is sufficiently expressive for writing a large number of parallel algorithms.
Note that the formal model can express some {\em local}, non-escaping futures that are not strictly series-parallel by using parallel lets that are
forced out of order. This pattern of parallelism does not provide much additional expressive power over that provided by parallel tuples, so we do not give up much by not exposing this capability in the front-end language.

\subsection{Indirection Pointers}
\label{subsec:indirection-pointers}

In the implementation, we need a runtime representation of optional
pointers to include in the data that corresponds to the indirections in the
semantics (\Secref{sec:indirection-semantics-intro}).
Fortunately, in the Gibbon compiler there is already a
pointer mechanism that is sufficient for our purposes.
This exists because of how Gibbon's regions grow---rather than copying
data into a larger buffer, Gibbon accumulates a linked list of contiguous
chunks, doubling the size on each extension. The last filled cell in a
chunk is an indirection into the next chunk.
\new{
Indirections also enable Gibbon to allocate a value that is {\em shared}
between multiple locations (within the same region or across regions)
without requiring a full copy,
which is crucial for ensuring asymptotic complexity conservation of programs.
Variable aliases indicate this sharing to the compiler.
For example, the aliased variable \il{x} in the expression
\il{(let x = mkBigExpr in Plus x x)} indicates that a single, shared value
must be allocated for the left and right subtrees of \il{Plus}.
Gibbon rewrites this expression to \il{(let x = mkBigExpr in Plus x (IndPtr x))},
where the right subtree is an indirection pointing to the data allocated for
the left.
Similarly, the identity function \il{(id x = x)} becomes \il{(id x = IndPtr x)}.
However, shared scalar values such as numbers and booleans are always copied,
because it is more efficient to do so.
}
In \pargibbon{}, we reuse this indirection pointer mechanism to implement
intra-region parallel allocations.

\subsection{Parallel Reads}

Using static analysis, Gibbon can infer if a dataype requires offsets,
and it can transform the program to add offsets to datatypes that need them.
In sequential programs, these are used to preserve asymptotic complexity of
certain functions.
%
In \pargibbon{}, we use these offsets to enable parallel reads.
We update that static analysis, and add offsets if a program
performs parallel reads, i.e. via a clause in a case expression that accesses a
data-constructor's fields in parallel.

\subsection{Parallel Allocations}
\label{subsec:parallel-allocations}


The implementation of intra-region parallel allocations closely follows the
design described in \Secref{sec:region-parallel-local}.
A program transformation pass generates code that allocates fresh regions and writes
indirection pointers at appropriate places.
\new{
But the metafunctions $\merges$ and $\mergem$ which merge task memories at join points have
a different run time behavior compared to their formal definition.
%
The implementation does not have a direct notion of a store.
At run time, a region variable $\reg$ is only a structure containing a
pointer to the start of a memory buffer
and some metadata necessary for garbage collection.
%
Two memory buffers are merged (linked) simply by writing a single indirection pointer
in one of them.
This operation is relatively cheap compared to the set union used
in the formal definition.
Similarly, the implementation does not have a direct notion of a location map,
and therefore there is no run time operation equivalent to $\mergem$.
At run time, all location variables become absolute pointers into the heap.
}

%
But there still exists an issue with fragmentation.
With granularity control --- in the form of judicious use of parallel tuples ---
we can restrict excessive creation of fresh regions, but the number of regions created will still
always be equal to the number of parallel tasks spawned by the program.
This can still cause fragmentation because all spawned tasks
might not actually {\em run} in parallel.

The key insight
is to make the number of fresh region allocations equal to the number of
{\em steals}, not spawns.
That is, because our implementation uses a work-stealing scheduler,
but the general idea applies to other schedulers as well: fragmentation
should be proportional to parallelism in the dynamic schedule, not the static
potential for parallelism.
Our implementation creates fresh regions for intra-region parallel allocations
only if they really run in parallel.
We accomplish this by using the Cilk Plus API to implement a hook to detect when steals occur.
Before reaching a parallel fork point (spawn), the runtime system stores the ID of the worker
executing the current code.
Next, the corresponding ID is immediately fetched in the continuation of the fork point.
If the IDs match, it indicates that a steal did not occur.
This optimization enables parallel allocations with minimal fragmentation.

\subsection{Parallel Arrays}
\label{subsec:parallel-arrays}

Programs need arrays as well as trees.
We extend Gibbon with array primitives such as
\il{alloc}, \il{length}, \il{nth}, \il{slice}, and \il{inplaceUpdate},
and use them to build a small library of parallel array operations with
good work and span bounds.
The ability to {\em safely} mutate an array in-place allows us to implement optimizations
that go beyond what is commonly allowed in a purely functional language.
This is enforced using the new Linear Haskell extensions~\cite{linear-haskell}, for example, the signature of an $O(1)$ array mutation is:
\begin{code}
inplaceUpdate :: Int $\rightarrow$ a $\rightarrow$ Array a $\multimap$ Array a  
\end{code}
Using these primitive operations, collective operations on arrays are implemented as recursive
  divide-and-conquer functions in Parallel Gibbon that use parallelism annotations\footnote{These combinators offer variants to explicitly control sequential chunk size, or to use the common heuristic of splitting into a number of tasks that is a multiple of the number of cores (provided as a global constant).}.
For example, our parallel \il{map} first allocates an array to store the output,
and then updates it in parallel with \il{inplaceUpdate}.
But all such potentially-racy operations are hidden behind a pure interface.
Also, an \il{Array} in \pargibbon{} can only store primitive values such as
numbers, booleans, and n-ary tuples of such values.
In the future, we plan to explore ways to support data-parallel operations
on serialized algebraic data.

\subsection{Memory Management}

In the formalism,
regions are modeled as {\em unbounded} memory buffers that offer a byte-indexed storage
for primitive values (data constructor tags, numbers, etc.).
Practically, we start by allocating a single contiguous {\em chunk} of memory of {\em bounded} size.
When this chunk is exhausted, a new one which is double in size is allocated
and linked with the previous one using a pointer.
\new{This policy is used up to an upper bound (1GB)
after which constant sized (1GB) new chunks are allocated.}
Thus, a single region is really a linked-list of chunks:
a small initial chunk, with subsequent ones doubling in size.
\new{
The small initial chunk is beneficial when a region contains a small value,
and the doubling policy reduces the overall \il{malloc} overhead when
allocating large values.
}

Our garbage collection strategy is based on regions and lifetimes, and also reference counting.
In classic region calculi, regions can be immediately deallocated at the end of their lexical scope.
However, we allow indirection pointers to point across {\em different} regions (chunks),
which is crucial to support parallelism, (and also to maintain the asymptotic complexity
of certain sequential programs).
Thus, a region can stay alive beyond its lexical scope, for example if a pointer to
it is captured by another region which is still in scope.

We use reference counting to deallocate such regions.
When a region is initialized---with a \il{letregion}---its reference count is set to 1,
and it is decremented when the region goes out of scope.
At this stage, if its reference count hits zero, it is deallocated by freeing
all of its chunks.
When a chunk is freed, the reference counts of the regions it points to are also decremented,
which may cause some of these regions to be freed as well.
But these reference counts are {\em per-region}, rather than {\em per-chunk}.
Hence, even if a single chunk in a region is truly alive, all of its other chunks are also
considered alive, and cannot be freed.
%
%
Also, there isn't a supplemental garbage-collector for long-lived regions at this time,
as in later versions of MLKit~\cite{mlkit-retrospective}.
In the future, we plan to make improvements in this area.
Note, however, that it is impossible to create heap cycles in Gibbon, because it's a pure strict language. (In-place modification through linear types doesn't change this, and besides, Parallel Gibbon's arrays do not contain pointers or sum types.)

\section{Evaluation}
\label{sec:evaluation}

In this section, we evaluate our implementation using a variety of
benchmarks from the existing literature, as well as a new compiler benchmark.
To measure the latent overheads of adding parallelism,
we compare our single-thread performance
against the original, sequential LoCal, as implemented by the Gibbon compiler.
Sequential Gibbon is also a good baseline for performing speedup calculations
since its programs operate on serialized heaps, and
as shown in prior work, are significantly faster than their pointer-based counterparts.
Note that prior work \cite{Gibbon} compared sequential constant factor performance
against a range of compilers including GCC and Java.  Since Sequential Gibbon
outperformed those compilers in sequential tree-traversal workloads, we focus
here on comparing against Sequential Gibbon for sequential performance.

We also compare the performance of our implementation
to other languages and systems
that support efficient parallelism for recursive, functional programs
--- \MPL{} \cite{MPL},
Multicore OCaml \cite{OCaml,multicore-ocaml}, and GHC.
\MPL{} (an extension of MLton) 
is a whole-program, optimizing compiler for Standard ML \cite{StandardML};
it supports nested fork/join parallelism, and generates extremely efficient code.
%


The experiments in this section are performed on a 48 core machine
(96 hyper-threads) made up of
2 $\times$ 2.9 GHz 24 core Intel Xeon Platinum 8268 processors, with 1.5TB of memory,
and running Ubuntu 18.04.
The shared memory on this machine is divided into two NUMA nodes
such that CPUs \il{0-23} and \il{48-71} use \il{node-0} as their local memory node,
and \il{24-47} and \il{72-95} use \il{node-1}.
In our experiments we only use 48 threads (no SMT), evenly distributed across
both NUMA nodes (\il{numactl $\;--$physcpubind=48-95}).
All experiments are performed using the default memory allocation policy
which always allocates memory on the current NUMA node.
We observed that using a round-robin memory allocation policy
(option \il{$--$interleave=0,1}) did not affect performance,
and therefore we do not report those results.

Each benchmark sample is the median of 9 runs.
To compile the C programs generated by our implementation we use
GCC 7.4.0 with all optimizations enabled (option \il{-O3}), and
the Intel Cilk Plus language extension~\cite{blumofe1996cilk}
(option \il{-fcilkplus}) to realize parallelism.
To compile sequential LoCal programs we use the open-source Sequential Gibbon compiler, \new{but we modify it to include arrays with in-place mutation using linear types, just like Parallel Gibbon.}
%
For \MPL{}, we use version \il{20200220.150446-g16af66d05} compiled from its source code.
%
For OCaml, we use the Multicore OCaml compiler \cite{multicore-ocaml}
(version 4.10 with options \il{-O3}),
along with the \il{domainslib} \footnote{https://github.com/ocaml-multicore/domainslib}
library for parallelism.
We use GHC 8.6.5, with options \il{-threaded -O2}, along with
the monad-par~\cite{monad-par} library for parallelism.

\subsection{Benchmarks}
\label{subsec:benchmarks}

We use the following set of 10 benchmarks.
For GHC, we use {\em strict datatypes} in benchmarks, which generally offers the
same or better performance, and avoids problematic interactions between
laziness and parallelism.
All programs use the same algorithms and datatypes
(including mutable arrays, which are provably race-free in Gibbon and GHC),
have identical granularity control thresholds, and are run with the same inputs.
This way, each pairing of program and input creates a deterministic task graph ---
which does not change when varying the number of threads ---
and the evaluation focuses on data representation and code generation,
rather than on decomposing and scheduling parallel tasks.

\begin{itemize}

\item \textbf{fib}:
  Compute the 48th fibonacci number with a sequential cutoff after \new{depth=18}: a simple baseline for scaling.

\item \textbf{\buildfib{}}:
  This is an artificial benchmark that is included here to
  measure parallel allocation under ideal conditions.
  It constructs a balanced binary tree of height 18, and computes
  the 20th fibonacci number at each leaf, with sequential cutoff after depth=12.


\item \textbf{buildKdTree} and \textbf{countCorrelation} and \textbf{allNearest}:
  \il{buildKDTree} constructs a kd-tree \cite{kdtree}
  containing 1M 3-d points
  in the Plummer distribution.
  The sequential cutoff is at a node containing less than 32K points.
  %
  %
  \il{countCorrelation} takes as input a kd-tree and a list of 100 3-d points,
  and counts the number of points which are correlated to each one.
  The chunk-size for the parallel-map is 4, and the sequential cutoff
  for \il{countCorrelation} is at a node containing less than 8K points.
  %
  \il{allNearest} computes the nearest neighbor of all 1M 3-d points.
  %
  The chunk-size for the parallel-map is 1024.

\item \textbf{barnesHut}:
  Use a quad tree to run an nbody simulation over
  1M 2-d point-masses distributed uniformly within a square.
  The chunk-size for the parallel-map is 4096.

\item \textbf{coins}
  This benchmark is taken from GHC's NoFib \footnote{https://gitlab.haskell.org/ghc/nofib}
  benchmark suite. It is a combinatorial search problem that computes
  the number of ways in which a certain amount of money can be paid by
  using the given set of coins.
  %
  The input set of coins and their quantities are
  \il{[(250,55),(100,88),(25,88),(10,99),(5,122),(1,177)]},
  and the amount to be paid is 999.
  %
  The sequential cutoff is after depth=3.

\item \textbf{countNodes}: This operates on ASTs gathered from the Racket compiler when processing
  large, real programs. The benchmark simply counts the number of
  nodes in a tree.
  %
  %
  For our implementation, we store the ASTs on disk in a serialized format
  which is read using a single \il{mmap} call.
  All others parse the text files before operating on them.
  To ensure an apples-to-apples comparison, we do not measure the time required
  to parse the text files.
  %
  The size of the text file is 1.2G, and that same file
  when serialized for our implementation is 356M.
  The AST has around 100M nodes in it.
  %
  %
  The sequential cutoff is after depth=9.

\item \textbf{constFold}:
  Run the \il{constFold} function shown in
  \Figref{fig:example1-both} on an artificially generated syntax-tree,
  which is a balanced binary tree of \il{Plus} expressions,
  with a \il{Lit} as a leaf.
  The height of the syntax-tree is 26, the sequential cutoff is after depth=8.

\item \textbf{mergeSort}:
  %
  An in-place parallel merge sort,
  which bottoms out to a sequential quick sort
  when the array contains less than 8192 elements.
  For our implementation, we use the \il{qsort} function from the C standard library to
  sort small arrays.
  The Haskell implementation is taken from Kuper et al's artifact
  accompanying their paper~\cite{lvish_2014},
  and it makes an FFI call to a sequential quick sort written in C.
  \MPL{} and OCaml bottom out to a sequential quick sort implemented in their source language.
  The input array contains 8M randomly generated floating point numbers.

\end{itemize}

\subsection{Results: Parallel versus Sequential Gibbon}

\begin{figure*}
\begin{subfigure}[t]{0.5\linewidth}
\vspace{0pt}
  \input{gibbon_table}
  \caption{
    %
    $T_s$ is the run time of Sequential Gibbon.
    %
    $T_1$ and $T_{48}$ are the run times of Parallel Gibbon
    on 1 thread and on 48 threads respectively.
    $O$ is the single-thread percentage overhead: $O = (T_1 - T_s) / T_s * 100$.
    $S$ is the 48-thread speedup: $S = T_s / T_{48}$.
  }
  \label{fig:gibbon-table}
\end{subfigure}
\hspace{0.03\linewidth}
\begin{subfigure}[t]{0.42\linewidth}
  \vspace{-2pt}
  \includegraphics[scale=0.7]{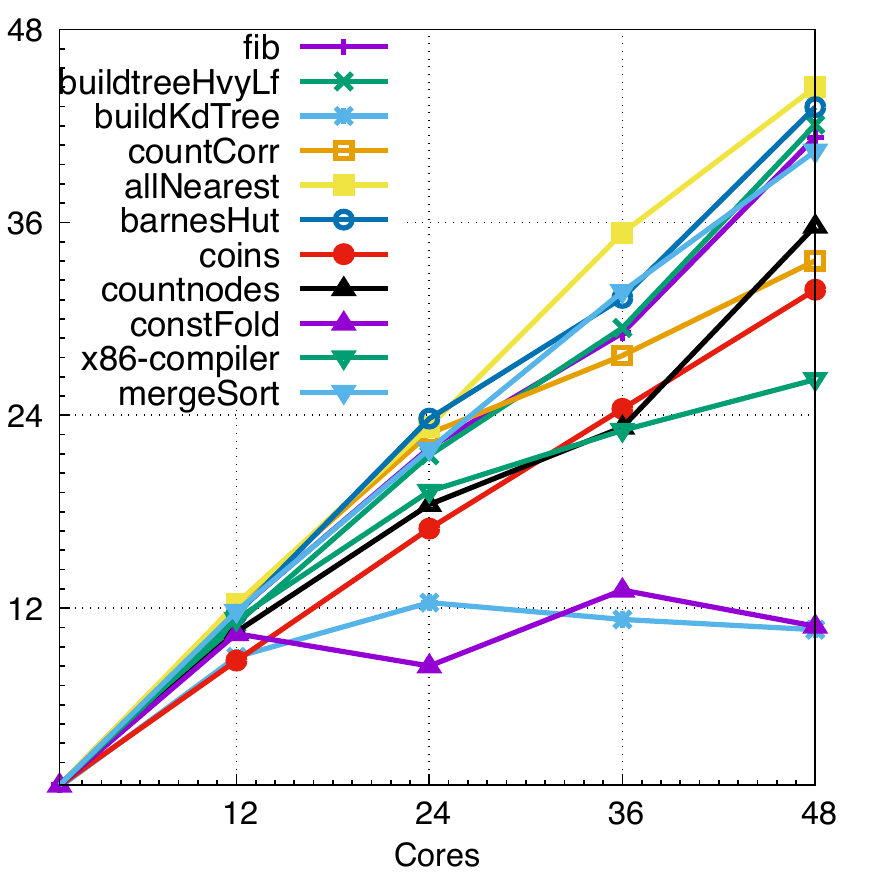}
  \vspace{9pt}
  \caption{
    Speedups relative to Sequential Gibbon.
  }
\label{fig:gibbon-speedup-plot}
\end{subfigure}
\caption{}
\label{fig:gibbon-results}
\end{figure*}

Figures~\ref{fig:gibbon-table} and \ref{fig:gibbon-speedup-plot} show the results
of comparing performance of benchmarks compiled using our
%
parallel
implementation, labeled ``Ours'', relative to Sequential Gibbon.
The quantities in the table can be interpreted as follows.
Column $T_s$ shows the run time of a sequential program.
which serves the purpose of a sequential baseline.
$T_1$ is the run time of a parallel program on a single thread, and
$O$ the percentage overhead relative to $T_s$, calculated as
$((T_1 - T_s) / T_s) * 100$.
$T_{48}$ is the run time of a parallel program on 48 threads and
$S$ is the speedup relative to $T_s$, calculated as $T_s / T_{48}$.
$R$ is the number of additional regions created to enable
parallel allocations, calculated as $R_{48} - R_{s}$.
For a majority of benchmarks, the overhead is under 3\%,
and the speedups range between 31.7$\times$ and 43.5$\times$.
These speedups match, or in cases such as \il{barnesHut} and \il{allNearest},
exceed those of optimized implementations that have been analyzed on
similar machines~\cite{pbbs,MPL,Heartbeat_17}.

With respect to the difference in speedups between different benchmarks,
we see the expected relationship among them which reflects their memory access patterns.
Compute-bound benchmarks such as \il{fib} scale very well,
whereas benchmarks such as \il{constFold} and \il{buildKdTree}
can become memory bound, and do not scale over a certain number of cores\footnote{For example, a simple parallel dot-product computation (in Cilk) has a similar linear access pattern and low arithmetic intensity to \il{constFold}, and it achieves only a 6X speedup on this same machine.}.
With respect to \il{buildKdTree}, a significant portion of its total running time is
spent in sorting the points at each node.
We observed that our \il{mergeSort} doesn't scale well on small inputs,
and since \il{buildKdTree} performs a series of smaller and smaller sorts,
it eventually runs into this, leading to lower scalability.
But its high overhead (14.6\%) and low speedup (10.6$\times$)
are in the same ballpark as an optimized C implementation
which \citet{adve_kdtree} analysed on a 32-core machine.
Figures~\ref{fig:ours_versus_others}
\iftoggle{EXTND}{
  and \ref{fig:others-benchmark-results} (given in the Appendix)
}{
  and 15 (given in the Appendix of the extended version~\cite{ParallelLoCal-tr})
}
show that \MPL{}, GHC and OCaml also scale similarly.
Overall, these results show that our technique is able to handle
parallelism in a mostly-serialized data representation effectively.

\subsubsection{Fragmentation:}
\label{subsec:fragmentation-results}

\new{
Traversals on a serialized heap are efficient because serialization minimizes
pointer-chasing and maximizes data locality.
But the heap produced by running intra-region allocations in parallel is
fragmented, which can affect the performance of subsequent traversals that
consume this heap, due to additional pointer dereferences and worse locality.
To measure this downstream effect,
we compare the run time of a single-threaded traversal operating
on a sequentially allocated value to that traversal
operating on a value allocated in parallel, which will be fragmented.
Thus, the thing whose run time is being compared---the traversal---stays the same
but it is given inputs that have different levels of fragmentation.
We also measure the amount of fragmentation introduced for parallelism by counting
the number of regions created solely to enable parallel allocations.
\Figref{fig:fragmentation-results} shows the results.

We use a subset of the benchmarks from \Secref{subsec:benchmarks}
whose output is a serialized value (the other benchmarks do not measure
the construction of new values),
and measure the time required to {\em traverse} the output.
For example, the benchmark \il{trav(constFold)}
constructs an input expression, runs constant folding over it,
and then sequentially traverses the resulting expression
(counts the number of leaves in it),
but the number reported is only the run time of the traversal
and it does not include the time taken to run \il{constFold} itself.
Each benchmark sample is the median of 9 runs,
such that each run allocates a value and then traverses it $N$ times,
where $N$ is set high enough to get the total run time over one second,
and the run time of a {\em single} traversal is reported.
$T_s$ is the time required to sequentially traverse a sequentially allocated
value, which is the baseline.

We compare against this baseline the performance of traversing
a value allocated in parallel with two levels of fragmentation:
\textit{optimum} and \textit{maximum}.
In the \textit{optimum} setting, the allocators are identical to
those used for measurements reported in \Figref{fig:gibbon-table}.
That is, they control fragmentation by
controlling the granularity of parallelism
(using the thresholds given in \Secref{subsec:benchmarks})
and are compiled using the {\em region-upon-steal} allocation
policy~(\Secref{subsec:parallel-allocations}).
%
$R_{\mathit{opt}}$ is the number of {\em additional} regions created to allocate a value in parallel using 48 threads.
For example, the sequential \il{constFold} uses a single region
and its parallel version requires 132 additional regions (133 regions total).
$T_{\mathit{opt}}$ is the time required to sequentially traverse the allocated value,
and $O_{\mathit{opt}}$ is the percentage overhead relative to $T_s$, calculated as
$O_{\mathit{opt}} = (T_{\mathit{opt}} - T_s) / T_s * 100$.
%
In the \textit{maximum} setting,
the allocators do not control fragmentation at all (no granularity control),
and they are compiled
using the {\em region-upon-spawn} allocation policy, which creates a fresh
region for every intra-region allocation task that is spawned.
This setting thus represents the {\em upper bound} on the amount of fragmentation that can be
introduced due to parallelism, where the heap essentially degenerates
to a full pointer-based representation.
$R_{\mathit{max}}$, $T_{\mathit{max}}$ and $O_{\mathit{max}}$ are the corresponding
numbers for this fragmentation setting when using 48 threads.
$O_{\mathit{max}}$ is the percentage overhead relative to $T_s$, calculated as
$O_{\mathit{max}} = (T_{\mathit{max}} - T_s) / T_s * 100$.
}

{

\begin{figure}

  \begin{tabular}{l c r ccc r ccc}
    \toprule
    & Seq Alloc. & \phantom{} & \multicolumn{3}{c}{Optimum Fragmentation} & \phantom{} & \multicolumn{3}{c}{Maximum Fragmentation} \\
    \cmidrule{2-2}
    \cmidrule{4-6}
    \cmidrule{8-10}
    Benchmark & $T_s$ && $\;\;\;$ $R_{\mathit{opt}}$ & $T_{\mathit{opt}}$ & $O_{\mathit{opt}}$ && $\;\;\;$ $R_{\mathit{max}}$ & $T_{\mathit{max}}$ & $O_{\mathit{max}}$ \\

    \midrule

    trav(\buildfib{}) & 0.99ms && $\;\;\;$ 341 & 1.02ms & 3.03 && $\;\;\;$ 262K & 11.45ms & 1056.6 \\

    trav(buildKdTree) & 6.22ms && $\;\;\;$ 31 & 6.64ms & 6.75 && $\;\;\;$ 262K & 55ms & 784.2 \\

    trav(coins) & 0.35s && $\;\;\;$ 9K & 0.37s & 5.71 && $\;\;\;$ 75M & 5.21s & 1388.6 \\

    trav(constFold) & 0.30s && $\;\;\;$ 132 & 0.32s & 6.67 && $\;\;\;$ 67M & 2.70s & 800 \\

    trav(x86-compiler) & 366.2$\mu s$ && $\;\;\;$ 1K & 377.5$\mu s$ & 3.09 && $\;\;\;$ 14K & 464.3$\mu s$ & 26.8 \\

    \midrule
    geomean & - && $\;\;\;$ - & - & 4.74 && $\;\;\;$ - & - & 476.9 \\
    \bottomrule
  \end{tabular}
  \caption{
    $T_s$ is the time required to sequentially traverse a sequentially allocated value.
    $R_{\mathit{opt}}$ is the number of {\em additional} regions created to allocate a value in parallel using 48 threads,
    $T_{\mathit{opt}}$ is the time required to sequentially traverse it,
    and $O_{\mathit{opt}}$ is the percentage overhead of the traversal: $O_{\mathit{opt}} = (T_{\mathit{opt}} - T_s) / T_s * 100$.
    $R_{\mathit{max}}$, $T_{\mathit{max}}$ and $O_{\mathit{max}}$ are the corresponding
    numbers for a value allocated in parallel using 48 threads with maximum fragmentation.
    $O_{\mathit{max}}$ is the percentage overhead relative to $T_s$ calculated as
    $O_{\mathit{max}} = (T_{\mathit{max}} - T_s) / T_s * 100$.
  }
\label{fig:fragmentation-results}
\end{figure}
}

\new{
Comparing $R_{\mathit{opt}}$ and $R_{\mathit{max}}$,
we see that in the optimum setting most of the allocated heap is still serialized
and uses only a small number of additional regions---less than 0.13\% of the maximum--- in most cases.
For \il{x86-compiler}, this percentage is higher (7.14\%) compared to the other benchmarks
because even in the optimum setting this benchmark
does not control the granularity of parallelism,
which is controlled only by the structure of the input
as we discuss in \Secref{subsec:compiler-case-study}.
In this case, the {\em region-upon-steal} allocation policy is responsible
for trimming the number of regions from 14K to 1K.
With respect to the run time of the traversal,
the overhead with optimum fragmentation is between 3.03\% to 6.75\%,
with a geomean of 4.74\%.
In addition to fragmentation,
the NUMA memory policy~\footnote{As described in the experimental setup, the shared memory on this machine is divided into two NUMA nodes and we use 48 threads evenly distributed across both nodes with the default {\em local} memory allocation policy.} has a significant impact on the overall overhead
because in the parallel version potentially
50\%
of memory accesses are at a non-local NUMA memory node.
If we run the experiment using a single NUMA node (\il{numactl $\;--$membind=0 $\;$ $\;--$physcpubind=1-24,49-71}),
the geomean overhead drops to 1.44\%,
with a significant reduction in the overheads for \il{buildKdTree} (0.96\%) and
\il{constFold} (2.95\%).
In the presence of maximum fragmentation, we see the expected result:
since heap degenerates to a full pointer-based representation,
the traversals are several times slower
and the benefits of using a serialized representation are lost.
This slowdown in traversing a pointer-based representation compared to a serialized one
is consistent with the results given in previous work~\cite{LoCal,Gibbon}.
Overall, these results show that using the granularity of parallelism to
guide the data representation works well in practice,
and gives us an efficient, {\em mostly-serialized} representation.
}

\subsection{Results: Gibbon versus Other Compilers}
\label{subsec:gibbon-versus-others}

\Figref{fig:ours_versus_others} shows the results of comparing performance of
our implementation to \MPL, OCaml, and GHC.
%
For each compiler, Column $T_s$ is the run time of a sequential program,
Column $T_{48}$ is the run time of a parallel program on 48 threads,
and an adjacent column to each shows the corresponding speedup (or slowdown) of our implementation
relative to this compiler.
For example, on 48 threads, \il{allNearest} is $3.95\times$ faster with our
implementation compared to OCaml.
\Figref{fig:scaling-subset} shows how a subset of benchmarks scale on 48 threads.
The scaling results for the remaining benchmarks
\iftoggle{EXTND}{
  are available in Appendix~\ref{sec:appendix-evaluation}.
}{
  are available in the Appendix of the extended version~\cite{ParallelLoCal-tr}.
}
With respect to self-relative comparisons,
on average, we scale similarly to \MPL{}, and {\em better} than OCaml or GHC.

Across all benchmarks,
on a single thread our \pargibbon{} offers a $1.93\times$, $2.53\times$, and $2.14\times$ geomean speedup compared to \MPL{}, OCaml, and GHC, respectively.
When utilizing 48 cores, our geomean speedup is $1.92\times$, $3.73\times$ and $4.01\times$.
Overall,
these results show that we start with a faster baseline in the sequential world,
and we're able to preserve the speedups in the parallel as well,
meaning that the use of dense representations
to improve sequential processing performance {\em coexists with scalable parallelism}.
The only benchmark for which our implementation is slower compared to others is \il{coins}.
This benchmark makes heavy use of linked-list operations
such as \il{cons}, \il{head}, and \il{tail},
and our implementation uses \il{malloc} to allocate memory
for every \il{cons}, which is inefficient.
Also, our dense representation currently offers no benefit when building
linked-lists by cons'ing onto existing lists.
All others, \MPL{}, OCaml and GHC, use a copying garbage collector~\cite{sivaramakrishnan2020retrofitting,MPL,marlow08_gc} allowing them to use a bump-allocator,
making them more efficient than our implementation.
%
%
\iftoggle{EXTND}{
  \Figref{fig:others-benchmark-results} in Appendix
}{
  Figure~15 in the Appendix of the extended version~\cite{ParallelLoCal-tr}
}
gives the {\em self-relative} performance
results for \MPL{}, OCaml and GHC.

\begin{figure}
\input{ours_versus_others}
\caption{
  Comparison of Ours with \MPL{}, OCaml, and GHC ---
  execution time in seconds, 
  and ratios to Ours. 
  $T_s$ is the run time of a sequential program, and
  $T_{48}$ is the run time of a parallel program on 48 threads.
  %
}
\label{fig:ours_versus_others}
\end{figure}

\begin{figure}
%
\includegraphics[scale=0.75]{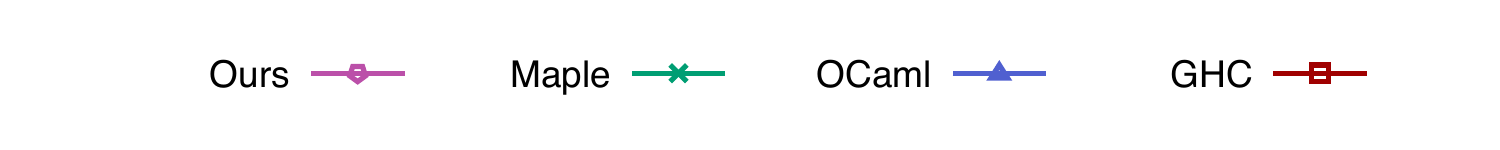}
\\
\hspace{0.02\linewidth}
\begin{subfigure}[t]{0.3\linewidth}
\includegraphics[scale=0.55]{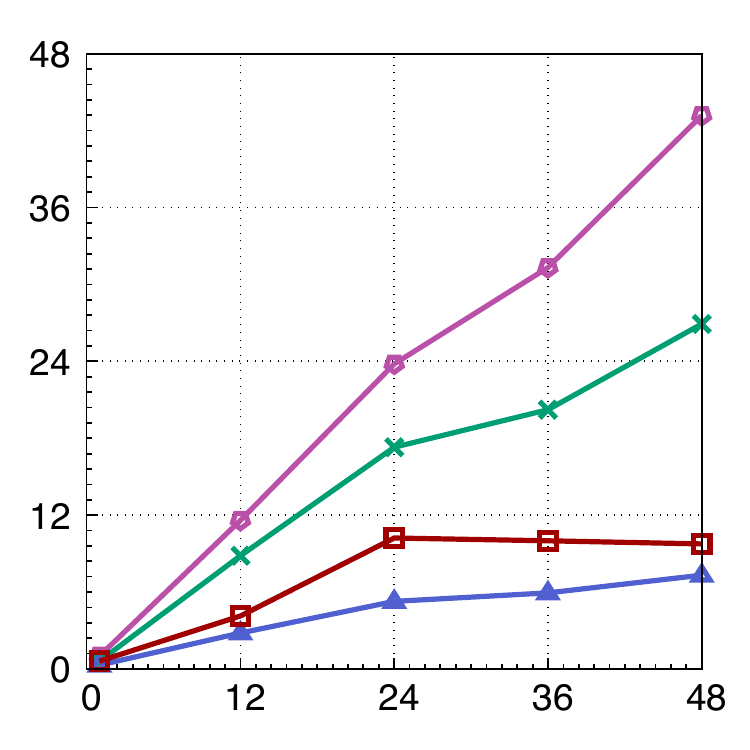}
\caption{barnesHut}
\end{subfigure}
\hspace{0.025\linewidth}
\begin{subfigure}[t]{0.3\linewidth}
\includegraphics[scale=0.55]{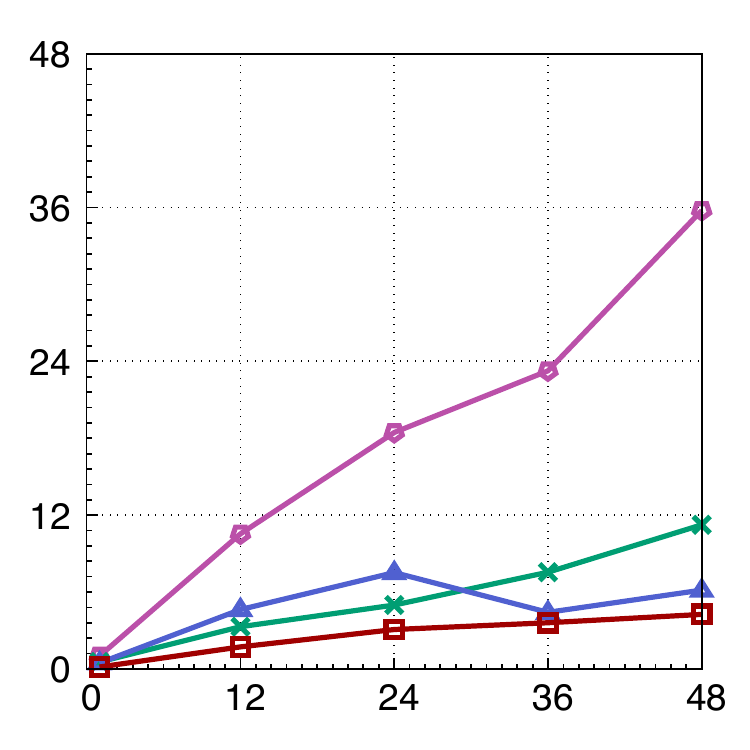}
\caption{countnodes}
\end{subfigure}
\hspace{0.025\linewidth}
\begin{subfigure}[t]{0.3\linewidth}
\includegraphics[scale=0.55]{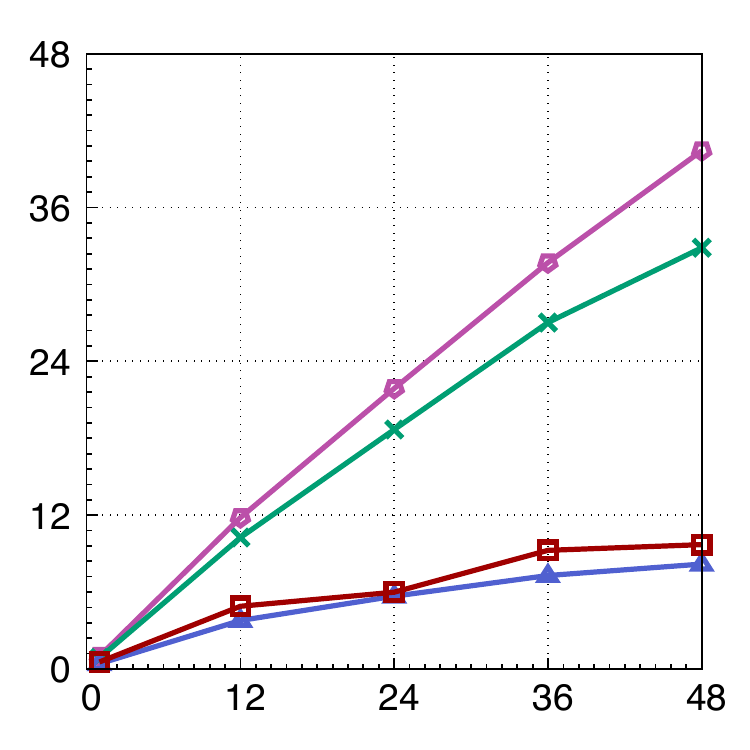}
\caption{mergeSort}
\end{subfigure}
\caption{
  Speedups on 1-48 threads relative to the fastest sequential baseline,
  which is Sequential Gibbon for these three benchmarks.
}
\label{fig:scaling-subset}
\end{figure}


\subsection{Results: x86-Compiler Case Study}
\label{subsec:compiler-case-study}

As an example of a complex benchmark which performs multiple traversals over
different datatypes, we implement a subset of
a compiler drawn from a university course~\cite{iu_compilers_book}.
Our version compiles to x86, from a source language that supports
integers and arithmetic and comparison operations on them,
booleans and operations such as \il{and} and \il{or},
and a conditional expression, \il{if}.
%
%
To compile this high level language, we first translate it to an
intermediate language similar to C,
in which the order of evaluation is explicit in its syntax.
The compiler is written in a nanopass style~\cite{sarkar_nanopass},
and is made up of five passes:
\begin{enumerate*}[label*=(\arabic*)]
\item \il{typecheck} type checks the source program,
\item \il{uniqify} freshens all bound variables to handle shadowing,
\item \il{explicateControl} translates to an
intermediate language similar to C,
\item \il{selectInstructions} generates x86 code which has variables in it,
\item and \il{assignHomes} maps each variable to a location on the stack.
\end{enumerate*}
The input to this benchmark is a synthetically generated, balanced syntax-tree
with conditional expressions at the top, followed by a sequence of let bindings.
The structure of the input program is used to control the granularity of parallelism.
%
The first three passes process the branches of conditionals in parallel,
and subsequent ones process every {\em block} of instructions in parallel.

\new{
On this compiler benchmark, Parallel Gibbon offers a
$1.24\times$, $1.11\times$, and $2.16\times$ speedup on a single thread, and
$1.02\times$, $2.20\times$ and $10.7\times$ speedup when utilizing 48 threads,
compared to \MPL{}, OCaml, and GHC, respectively.
Note that most of the run time and
also the self-relative parallel speedup of this benchmark is due to \il{assignHomes}.
%
The first four passes of the compiler have a linear memory access pattern
and low arithmetic intensity, much like \il{constFold}.
The passes inspect the input expression,
perform inserts or lookups on shallow environments
(which only contain entries for variables that are in scope at a point),
and then allocate the output.
But \il{assignHomes} works in a different way.
Since it needs to assign a unique stack location to every variable occurring in an expression,
it constructs an environment containing {\em all} variables in
the expression (as opposed to just those in scope at a point),
and then performs repeated lookups on it to
rewrite variable occurrences with stack location references.
This environment is significantly larger than those used by other passes,
making \il{assignHomes} much more work-intensive and suitable for parallel execution.

Here we have only taken the first step towards developing an efficient parallel compiler,
and there is ample opportunity for further investigation in this area.
In the future, we plan to expand the compiler's source language to include
constructs such as top-level function definitions and modules which
are a source of parallelism in many real compilers.
}


\section{Related Work}
\label{sec:related-work}

The most closely related work to this paper is, of course, Vollmer et al.'s LoCal~\cite{LoCal}, which was summarized in Section~\ref{subsec:local-primer}. As discussed there,
Vollmer et al.'s treatment only provided sequential semantics, while this paper
extends those semantics to incorporate parallelism.

This work, and LoCal, are related to several HPC approaches to serializing recursive trees into flat buffers for efficient traversal~\cite{goldfarb13sc, meyerovich11hotpar, makino90}. Notably, these approaches {\em must} maintain the ability to access the serialized trees in parallel, despite the elimination of pointers internal to the data structure, or risk sacrificing their performance goals. The key distinction that makes enabling parallelism in the HPC setting ``easier'' than in our setting is that these approaches are application-specific. The serialized layouts are tuned for trees whose structure and size are known prior to serialization, and the applications that consume these trees are specially-written to deal with the application-specific serialization strategies. Hence, offsets are either manually included in the necessary locations, or are not necessary as tree sizes can be inferred from application-specific information.

Work on more general approaches for packing recursive structures into buffers includes Cap'N Proto~\cite{capnproto}, which attempts to unify on-disk and in-memory representations of data structures, and Compact Normal Form (CNF)~\cite{cnf-icfp15}. Neither of these approaches have the same design goals as \seqcalc{} and \rparcalc{}:
both Cap'N Proto and CNF preserve internal pointers in their representations, eliding the problem of parallel access by invariably paying the cost (in memory consumption and lost spatial locality) of maintaining those pointers. We note that Vollmer et al. showed that \seqcalc{}'s representations enable faster sequential traversal than either of those two approaches~\cite{LoCal}, and Section~\ref{sec:evaluation} shows that our approach is comparable in {\em sequential} performance to \seqcalc{} despite also supporting parallelism.

There is a long line of work on flattening and nested data parallelism, where parallel computations over irregular structures are \textit{flattened} to operate over dense structures~\cite{nesl,keller98europar,bergstrom13ppopp}. However, these works do not have the same goals as ours. They focus on array
data, generating parallel code, and data layouts that promote data parallel access to the elements of the structure, rather than selectively trading off between parallel access to structures and efficient sequential access.

Efficient automatic memory management is a longstanding challenge for parallel functional
languages.
Recent work has addressed scalable garbage collection by structuring
the heap in a hierarchy of heaps, enabling task-private collections
~\cite{hierarchical-heaps-mutation}. There is work proposing a
split-heap collector that can handle a parallel lazy
language~\cite{marlow2009runtime} and a strict
one~\cite{multicore-ocaml}, and there is work on a
scalable, concurrent collector~\cite{ueno2016fully}.
A promising new line of work explores hierarchical heaps for parallel
functional programs~\cite{icfp-hierarchical-heaps}.
All of these designs focus on a conventional object model for algebraic datatypes that, unlike \rparcalc{},
assume a uniform, boxed representation.
In the future, we plan to investigate how results in these collectors relate to our model,
where objects may be laid out in a variety of ways.

\section{Conclusions and Future Work}
\label{sec:conclusions}

We have shown how a practical form of task parallelism can be reconciled with
dense data representations.  We demonstrated this result inside a compiler
designed to implicitly transform programs to operate on such dense
representations.  For a set of tree-manipulating programs we considered in
\Secref{sec:evaluation}, this experimental system yielded better performance
than existing best-in-class compilers.

To build on what we have presented in this paper, we plan to explore automatic
granularity control~\cite{Heartbeat_17,OracleGuided_19}; this would remove the
last major source of manual tuning in Gibbon programs,
which already automate data layout optimizations.
Parallel Gibbon with
automatic granularity control would represent the dream of implicitly parallel
functional programming with good absolute wall-clock performance.
While our current approach supports limited examples of data parallelism--friendly data structures beyond trees, such as dense arrays (\Secref{subsec:parallel-arrays}), we plan to further generalize our system by adding additional data structures that capture mutable sparse and dense multi-dimensional data.
%
%
We plan to support limited in-place mutation of densely-encoded algebraic data, by adding primitives based on linear types, which we expect to mesh well with the implicitly parallel functional paradigm.
While \pargibbon{} already out-performs competing parallel, functional approaches, we expect these additional features will both improve programmability (by relieving the programmer of the burden of granularity control) and performance (by supporting more efficient parallel structures and strategies).

\begin{acks}
  This work was supported in part by \grantsponsor{nsf}{National Science Foundation}{}
  awards \grantnum{spx-milind}{CCF-1725672}{}, \grantnum{spx-ryan}{CCF-1725679}
  and \grantnum{spx-milind-two}{CCF-1919197}{}, as well as
  by the \grantsponsor{EPSRC}{Engineering and Physical Sciences Research Council}{}
  award \grantnum{epsrc-granule}{EP/T013516/1}{}.
  %
  We would like to thank our shepherd, Cyrus Omar, as well as the anonymous
  reviewers for their suggestions and comments.
\end{acks}

\iftoggle{ARXIV}{
\input{ms.bbl}

}{
\bibliography{refs.bib}
}

\iftoggle{EXTND}{
\clearpage
\input{appendix}
}{}

\end{document}

%% file: defs.tex
\newcommand{\pargibbon}[0]{Parallel Gibbon}

\newcommand{\ourcalc}{\rparcalc{}}
\newcommand{\rparcalc}{LoCal$^{\text{par}}$}

\newcommand{\seqcalc}{LoCal}

\newcommand{\gramdef}{\; ::= \;}
\newcommand{\gramor}{\; | \;}
\newcommand{\keywd}[1]{\mathit{#1}}
\newcommand{\gramwd}[1]{\texttt{#1}}

\newcommand{\skeywd}[1]{\mathit{#1}}

\newcommand{\PROG}{\keywd{top}}
\newcommand{\DD}{\keywd{dd}}

\newcommand{\FD}{\keywd{fd}}
\newcommand{\DC}{\keywd{K}}

\newcommand{\EXPR}{\keywd{e}}
\newcommand{\sEXPR}{\skeywd{e}}
\newcommand{\DATA}{\gramwd{data}}
\newcommand{\TYP}{\keywd{\tau}}
\newcommand{\TYPC}{\keywd{\tau_c}}

\newcommand{\hTYP}{\keywd{\hat{\tau}}}
\newcommand{\sTYP}{\skeywd{\tau}}
\newcommand{\var}{\svar}
\newcommand{\svar}{x}
\newcommand{\fvar}{\sfvar}
\newcommand{\sfvar}{f}
\newcommand{\yvar}{y}

\newcommand{\ARROW}{\rightarrow}

\newcommand{\loc}{\skeywd{l}}

\newcommand{\concreteloc}[3]{\ensuremath{\langle #1, #2 \rangle ^{#3}}}
\newcommand{\reg}{\skeywd{r}}
\newcommand{\sreg}{\skeywd{r}}

\newcommand{\LE}{\keywd{le}}
\newcommand{\letpack}[3]{\gramwd{let}\;#1=#2\;\gramwd{in}\;#3}
\newcommand{\letloc}[3]{\gramwd{letloc}\;#1 = #2\;\gramwd{in}\;#3}
\newcommand{\letreg}[2]{\gramwd{letregion}\;#1\;\gramwd{in}\;#2}
\newcommand{\fapp}[2]{\fvar \;[#1]\; #2}
\newcommand{\TS}{\keywd{ts}}

\newcommand{\pat}{\keywd{pat}}

\newcommand{\ind}{\keywd{i}}
\newcommand{\indj}{\keywd{j}}
\newcommand{\concreteind}[1]{#1}

\newcommand{\indbef}[1]{#1\diamond}
\newcommand{\concretelocvar}{\keywd{cl}}

\newcommand{\indivar}[1]{\gramwd{ivar}\; #1}
\newcommand{\ivarid}{\var}

\newcommand{\indirection}[2]{\ensuremath{\text{\&}(#1,#2)}}

\newcommand{\VAL}{\keywd{v}}

\newcommand{\STOR}{\keywd{S}}
\newcommand{\stepsto}{\Rightarrow}

\newcommand{\rparstepsto}{\Longrightarrow_{rp}}

\newcommand{\caseclause}[2]{#1 \;\rightarrow \;#2}
\newcommand{\case}[2]{\gramwd{case}\; #1 \;\gramwd{of}\;#2}

\newcommand{\datacon}[3]{\ensuremath{#1 \;#2 \;#3}}

\newcommand{\litnum}{\keywd{n}}

\newcommand{\startr}[1]{\gramwd{start}\; #1}
\newcommand{\afterl}[1]{\gramwd{after}\; #1}
\newcommand{\tightoverset}[2]{%
  \mathop{#2}\limits^{\vbox to -.5ex{\kern-0.75ex\hbox{$#1$}\vss}}}
\newcommand\set[1]{\{ \, \ensuremath{#1} \,\}}

\newcommand{\heap}{\keywd{h}}
\newcommand{\heapval}{\keywd{hv}}

\newcommand{\partup}{\ensuremath{\parallel}}

\newcommand{\CENV}{\keywd{C}}
\newcommand{\TENV}{\keywd{\Gamma}}

\newcommand{\SENV}{\keywd{\Sigma}}
\newcommand{\MENV}{\keywd{M}}

\newcommand{\AENV}{\keywd{A}}
\newcommand{\NENV}{\keywd{N}}

\newcommand{\TASKVAR}{\keywd{T}}
\newcommand{\SEQSTATE}{\keywd{t}}

\newcommand{\TASKSET}{\mathbb{T}}
\newcommand{\CENVMAP}{\mathbb{C}}
\newcommand{\AENVMAP}{\mathbb{A}}
\newcommand{\NENVMAP}{\mathbb{N}}
\newcommand{\SENVMAP}{\mathbbl{\Sigma}}
\newcommand{\TENVMAP}{\mathbbl{\Gamma}}

\newcommand{\singlewriter}[3]{\ensuremath{HasSingleWriter(#1, #2, #3)}}
\newcommand{\getwriter}[3]{\ensuremath{GetSingleWriter(#1, #2, #3)}}
\newcommand{\isval}[1]{\ensuremath{IsVal(#1)}}
\newcommand{\taskcomplete}[1]{\ensuremath{TaskComplete(#1)}}
\newcommand{\xor}{\ensuremath{\oplus}}

\newcommand{\locreg}[2]{\ensuremath{{#1}^{#2}}}
\newcommand{\tyatlocreg}[3]{#1 \ensuremath{@} \locreg{#2}{#3}}
\newcommand{\styatlocreg}[3]{#1\ensuremath{@}\locreg{#2}{#3}}

\newcommand{\subst}[3]{\ensuremath{#1[#3/#2]}}

\newcommand{\typeofcon}{\keywd{TypeOfCon}}
\newcommand{\typeoffield}{\keywd{TypeOfField}}
\newcommand{\kargtys}{\keywd{ArgTysOfConstructor}}
\newcommand{\allocptr}[2]{\keywd{MaxIdx}(#1,#2)}

\newcommand{\followinm}{\keywd{Deref}}

\newcommand{\storelookupone}[3]{\hat{#1}(#2,#3)}
\newcommand{\maplookupone}[2]{\hat{#1}(#2)}

\newcommand{\mergeh}{\keywd{MergeH}}
\newcommand{\mergem}{\keywd{MergeM}}
\newcommand{\merges}{\keywd{MergeS}}
\newcommand{\tie}{\keywd{LinkFields}}
\newcommand{\linkfields}{\keywd{LinkFields}}

\newcommand{\ewitnesstext}{\ensuremath{\vdash_{\textit{ew}}}}
\newcommand{\ewitness}[4]{\ensuremath{#1;#2;#3 \ewitnesstext #4}}

\newcommand{\storewf}[7]{\ensuremath{#1;#2;#3;#4;#5 \vdash_{wf} #6;#7}}
\newcommand{\storewfcfa}[4]{\ensuremath{#1 \vdash_{wf_{cfc}} #3;#4}}
\newcommand{\storewfca}[5]{\ensuremath{#1;#2 \vdash_{wf_{ca}} #4;#5}}
\newcommand{\storewftasks}[1]{\ensuremath{\vdash_{wf_{tasks}} #1}}

\newcommand{\tcfun}{\ensuremath{\vdash_{fun}}}
\newcommand{\tcpat}{\ensuremath{\vdash_{pat}}}

\newcommand{\tctask}{\ensuremath{\vdash_{task}}}
\newcommand{\tctaskset}{\ensuremath{\vdash_{taskset}}}

\newcommand{\parmergemenv}[2]{\mergem(#1, #2)}
\newcommand{\parmergestor}[2]{\merges(#1, #2)}

\newcommand{\refwellformed}[2]{WF~\ref{#1};\ref{#2}}

\newcommand{\tfunctiondef}{T-Function-Definition}
\newcommand{\tdatacon}{T-DataConstructor}
\newcommand{\tdataconivars}{T-DataConstructor-Ivars}
\newcommand{\ddatacon}{D-DataConstructor}
\newcommand{\dletloctag}{D-LetLoc-Tag}
\newcommand{\dletlocafter}{D-LetLoc-After}

\newcommand{\dletlocstart}{D-LetLoc-Start}
\newcommand{\dapp}{D-App}
\newcommand{\dletregion}{D-LetRegion}
\newcommand{\tcase}{T-Case}
\newcommand{\dcase}{D-Case}
\newcommand{\tpat}{T-Pattern}
\newcommand{\tprogram}{T-Program}
\newcommand{\tllafter}{T-LetLoc-After}
\newcommand{\tlltag}{T-LetLoc-Tag}
\newcommand{\tllstart}{T-LetLoc-Start}
\newcommand{\tlregion}{T-LetRegion}
\newcommand{\tvar}{T-Var}
\newcommand{\ttaskset}{T-Taskset}
\newcommand{\ttasksetempty}{T-Taskset-Empty}
\newcommand{\ttask}{T-Task}
\newcommand{\tlet}{T-Let}
\newcommand{\dletexp}{D-Let-Expr}
\newcommand{\dletval}{D-Let-Val}
\newcommand{\dletexppar}{D-Par-Let}

\newcommand{\tapp}{T-App}
\newcommand{\tconcreteloc}{T-Concrete-Loc}
\newcommand{\dparsteptask}{D-Par-Step}
\newcommand{\emptytenv}{\emptyset}

\newcommand{\dregparsteptask}{D-Par-Step}
\newcommand{\dregparletexppar}{D-Par-Let-Fork}

\newcommand{\dregparletlocafternewreg}{D-LetLoc-After-NewReg}
\newcommand{\dregparcasejoin}{D-Par-Case-Join}

\newcommand{\dregpardataconjoin}{D-Par-DataConstructor-Join}

\newcommand{\overharpoon}[1]{\overrightharpoon{#1}}

\newcommand{\taskconfig}[3]{(#1, #2, #3)}

\newcommand{\MPL}{\text{MaPLe}}

\newcommand{\vsmaple}[1]{$\inferrule{\MPL{}}{\text{Ours}}{#1}$}
\newcommand{\vsocaml}[1]{$\inferrule{\text{OCaml}}{\text{Ours}}{#1}$}
\newcommand{\vsghc}[1]{$\inferrule{\text{GHC}}{\text{Ours}}{#1}$}
\newcommand{\buildfib}{buildtreeHvyLf}

\newcommand{\metafmerging}{
\[
\begin{array}{lcl}
  \merges(\STOR_1, \STOR_2) & = & \set{\reg \mapsto \mergeh(\heap_ 1, \heap_2) \mid (\reg \mapsto \heap_1) \in \STOR_1,  (\reg \mapsto \heap_2) \in \STOR_2} \\
                            & \cup & \set{\reg \mapsto \heap \mid (\reg \mapsto \heap) \in \STOR_1,  \reg \not\in dom(\STOR_2)} \\
                            & \cup & \set{\reg \mapsto \heap \mid (\reg \mapsto \heap) \in \STOR_2,  \reg \not\in dom(\STOR_1)} \\\\
  \mergeh(\heap_1, \heap_2) & = & \set{\concreteind{\ind_1} \mapsto \heapval \mid (\concreteind{\ind_1} \mapsto \heapval) \in \heap_1, (\concreteind{\ind_2} \mapsto \heapval) \in \heap_2, \concreteind{\ind_1} = \concreteind{\ind_2}} \\
                            & \cup & \set{\concreteind{\ind} \mapsto \heapval \mid (\concreteind{\ind} \mapsto \heapval) \in \heap_1, \concreteind{\ind} \not\in \set{\concreteind{\ind'} | \concreteind{\ind'} \in dom(\heap_2)}} \\
                            & \cup & \set{\concreteind{\ind} \mapsto \heapval \mid (\concreteind{\ind} \mapsto \heapval) \in \heap_2, \concreteind{\ind} \not\in \set{\concreteind{\ind'} | \concreteind{\ind'} \in dom(\heap_1)}} \\\\

  \mergem(\MENV_1, \MENV_2)
  & = & \set{\locreg{\loc}{\reg} \mapsto \concreteloc{\reg}{\concreteind{\ind_1}}{} \mid
  (\locreg{\loc}{\reg} \mapsto \concreteloc{\reg}{\concreteind{\ind_1}}{}) \in \MENV_1,
  (\locreg{\loc}{\reg} \mapsto \concreteloc{\reg}{\concreteind{\ind_2}}{}) \in \MENV_2,
  \ind_1 = \ind_2} \\
  & \cup & \set{\locreg{\loc}{\reg} \mapsto \concretelocvar \mid
  (\locreg{\loc}{\reg} \mapsto \concretelocvar) \in \MENV_1,
  \locreg{\loc}{\reg} \notin \MENV_2} \\
  & \cup & \set{\locreg{\loc}{\reg} \mapsto \concretelocvar \mid
  \locreg{\loc}{\reg} \notin \MENV_1,
  (\locreg{\loc}{\reg} \mapsto \concretelocvar) \in \MENV_2} \\
  & \cup & \set{\locreg{\loc}{\reg} \mapsto \concreteloc{\reg}{\indj}{} \mid
  (\locreg{\loc}{\reg} \mapsto \concreteloc{\reg}{\indivar{\ivarid}}{}) \in \MENV_1,
  (\locreg{\loc}{\reg} \mapsto \concreteloc{\reg}{\concreteind{\indj}}{}) \in \MENV_2} \\
  & \cup & \set{\locreg{\loc}{\reg} \mapsto \concreteloc{\reg}{\indj}{} \mid
  (\locreg{\loc}{\reg} \mapsto \concreteloc{\reg}{\concreteind{\indj}}{}) \in \MENV_1,
  (\locreg{\loc}{\reg} \mapsto \concreteloc{\reg}{\indivar{\ivarid}}{}) \in \MENV_2} \\
\end{array}
\]
}

%% file: haskell_style.tex
\usepackage{upquote}
\usepackage{listings}
\usepackage[]{xcolor}
\definecolor{MidnightBlue}{rgb}{0.1, 0.1, 0.44}
\definecolor{Bittersweet}{rgb}{1.0, 0.44, 0.37}
\definecolor{Cinnamon}{rgb}{0.82, 0.41, 0.12}
\definecolor{ForestGreen}{rgb}{0.0, 0.27, 0.13}
\definecolor{Gray}{rgb}{0.5, 0.5, 0.5}
\definecolor{RoyalPurple}{rgb}{0.47, 0.32, 0.66}
\definecolor{Maroon}{rgb}{0.5, 0.0, 0.0}
\definecolor{OtterBrown}{rgb}{0.4, 0.26, 0.13}
\definecolor{azure}{rgb}{0.0, 0.5, 1.0}
\definecolor{blush}{rgb}{0.87, 0.36, 0.51}

\definecolor{darkgreen}{rgb}{0,0.5,0}
\definecolor{darkred}{rgb}{0.5,0,0}
\definecolor{darkbrown}{rgb}{0.4, 0.26, 0.13}

\lstdefinestyle{haskell}{%
    language=Haskell,
    upquote=true,
    deletekeywords={case,class,data,default,deriving,do,in,instance,let,of,type,where,IO,ST,STM,if,then,else},
    morekeywords={[2]class,data,default,deriving,family,instance,type,where},
    morekeywords={[3]in,let,case,of,do,letpacked,letloc,letregion,start,after,if,then,else,spawn,sync,sizeof,cilk_spawn,cilk_sync,continuation_stolen, size_of,tie},
    morekeywords={[4]IORef,IO,ST,STM,STRef,TVar},
    literate=
        {Lambda}{{$\Lambda$}}1
        {\\\\}{{\char`\\\char`\\}}1
        {>->}{>->}3
        {>>=}{>>=}3
        {->}{{$\rightarrow$}}2
        {>=}{{$\geq$}}2
        {<-}{{$\leftarrow$}}2
        {<=}{{$\leq$}}2
        {=>}{{$\Rightarrow$}}2
        {|}{{$\mid$}}1
        {forall}{{$\forall$}}1
        {exists}{{$\exists$}}1
        {...}{{$\cdots$}}3
}

\lstdefinestyle{inline}{%
    style=haskell,
    basicstyle=\footnotesize\ttfamily,
    escapechar={\@},
    mathescape=true,
    literate=
        {\\}{{$\lambda$}}1
        {Lambda}{{$\Lambda$}}1        
        {\\\\}{{\char`\\\char`\\}}1
        {>->}{>->}3
        {>>=}{>>=}3
        {->}{{$\rightarrow$\space}}3    
        {>=}{{$\geq$}}2
        {<-}{{$\leftarrow$}}2
        {<=}{{$\leq$}}2
        {=>}{{$\Rightarrow$}}2
        {|}{{$\mid$}}1
        {forall}{{$\forall$}}1
        {exists}{{$\exists$}}1
        {...}{{$\cdots$}}3
}

\lstnewenvironment{code}
    {\lstset{style=haskell}%
      \csname lst@SetFirstLabel\endcsname}
    {\csname lst@SaveFirstLabel\endcsname}
    {}

\newcommand{\il}[1]{\lstinline[style=inline,mathescape=true];#1;}

\newcommand{\makeatchar}{\lstDeleteShortInline@}

%% file: formal_dynamics.tex
\newcommand{\mydots}{\ifmmode\mathinner{\ldotp\kern-0.2em\ldotp\kern-0.2em\ldotp}\else.\kern-0.13em.\kern-0.13em.\fi}

\newcommand{\rdcase}{
  \inferrule*[lab={\;\text{[\dcase]}}]
  {}{
    \STOR;\MENV;\case{\concreteloc{\reg}{\ind}{\locreg{\loc}{\reg}}}{[\mydots, \caseclause{\datacon{\DC}{}{(\overharpoon{\var : \tyatlocreg{\TYP}{\loc}{\reg}})}}{\EXPR}, \mydots]}
    \stepsto
    \\\\
    \STOR;\MENV';\subst{\EXPR}{\overharpoon{\var}}{\concreteloc{\reg}{\overharpoon{w}}{\overharpoon{\locreg{\loc}{\reg}}}}
    \\\\ {
         \begin{aligned}
           \text{where} \;
           & \; \MENV' = \MENV \cup \set{\overharpoon{\locreg{\loc}{\reg}_1} \mapsto \concreteloc{\reg}{\ind+1}{}, \mydots , \overharpoon{\locreg{\loc}{\reg}_{j+1}} \mapsto \concreteloc{\reg}{\overharpoon{w_{j+1}}}{}} \\[-2mm]
           & \; \ewitness{\overharpoon{\TYP_1}}{\concreteloc{\reg}{\ind+1}{}}{\STOR}{\concreteloc{\reg}{\overharpoon{w_1}}{}} \\[-1mm]
           & \; \ewitness{\overharpoon{\TYP_{j+1}}}{\concreteloc{\reg}{\overharpoon{w_j}}{}}{\STOR}{\concreteloc{\reg}{\overharpoon{w_{j+1}}}{}} \\[-1mm]
           & \; \DC = \storelookupone{\STOR}{\reg}{\ind}; \; \indj \in \set{1 , \mydots , \litnum - 1} ;
             \; \litnum = | \overharpoon{\var : \hTYP} | \\
         \end{aligned}
       }
  }
}

\newcommand{\rdregparsteptask}{
  \inferrule*[lab={\;\text{[\dregparsteptask]}}]
  {\STOR;\MENV;\EXPR \stepsto \STOR';\MENV';\EXPR'}
  {\TASKVAR_1, \mydots, \taskconfig{\hTYP}{\concretelocvar{}}{\STOR;\MENV;\EXPR}, \mydots \TASKVAR_n \rparstepsto
    \TASKVAR_1, \mydots,
    \taskconfig{\hTYP}{\concretelocvar{}}{\STOR';\MENV';\EXPR'}, \mydots \TASKVAR_n
  } 
}

\newcommand{\rdregparsteptaskapp}{
  \inferrule*[lab={\;\text{[\dregparsteptask]}}]
  {\STOR;\MENV;\EXPR \stepsto \STOR';\MENV';\EXPR'}
  {\TASKVAR_1, \mydots, \taskconfig{\hTYP}{\concretelocvar{}}{\STOR;\MENV;\EXPR}, \mydots \TASKVAR_n \rparstepsto \\\\
    \TASKVAR_1, \mydots,
    \taskconfig{\hTYP}{\concretelocvar{}}{\STOR';\MENV';\EXPR'}, \mydots \TASKVAR_n
  } 
}

\newcommand{\rdregparletlocafternewreg}{
  \inferrule*[lab={\;\text{[\dregparletlocafternewreg]}}]
  {}{
    \STOR;\MENV;\letloc{\locreg{\loc}{\reg}}{\afterl{\tyatlocreg{\TYP}{\loc_0}{\reg}}}{\EXPR} \stepsto \STOR';\MENV';\EXPR
    \\\\ {
          \begin{aligned}
            \\[-4mm]
            \text{where} \;
            & \; \concreteloc{\reg}{\indivar{\ivarid}}{\loc_0} = \maplookupone{\MENV}{\locreg{\loc_0}{\reg}} ; \reg' \; \text{fresh}  \\[-2mm]
            & \STOR' = \STOR \cup \set{\reg' \mapsto \emptyset} \\[-2mm]
            & \MENV' = \MENV \cup \set{\locreg{\loc}{\reg} \mapsto \concreteloc{\reg}{\indirection{\reg'}{0}}{}}
           \end{aligned}
         }
  }
}

\newcommand{\rdregparletlocafterseq}{
  \inferrule*[lab={\;\text{[\dletlocafter]}}]
  {}{
    \STOR;\MENV;\letloc{\locreg{\loc}{\reg}}{\afterl{\tyatlocreg{\TYP}{\loc_0}{\reg}}}{\EXPR} \stepsto
    \STOR;\MENV';\EXPR
    \\\\ {
          \begin{aligned}
            \\[-4mm]
            \text{where} \;
            & \; \concreteloc{\reg}{\concreteind{\ind}}{} = \maplookupone{\MENV}{\locreg{\loc_0}{\reg}} ;
            \; \ewitness{\TYP}{\concreteloc{\reg}{\concreteind{\ind}}{}}{\STOR}{\concreteloc{\reg}{\concreteind{\indj}}{}} \\[-2mm]
            & \; \MENV' = \MENV \cup \set{\locreg{\loc}{\reg} \mapsto \concreteloc{\reg}{\concreteind{\indj}}{}}
           \end{aligned}
         }
  }
}

\newcommand{\rdregpardataconjoin}{
  \inferrule*[lab={\;\text{[\dregpardataconjoin]}}]{}{
    \TASKVAR_1, \mydots, \taskconfig{\hTYP}{\concretelocvar{}}{\STOR;\MENV;\EXPR}, \mydots , \TASKVAR_n \rparstepsto
    \TASKVAR_1, \mydots, \TASKVAR', \mydots , \TASKVAR_n
    \\\\ {
          \begin{aligned}
            \\[-4mm]
            \text{where} \;
            & \; \EXPR = \datacon{\DC}{\locreg{\loc}{\reg}}{\overharpoon{\VAL}} ;
              \; \concreteloc{\reg}{\indivar{\ivarid}_{\indj}}{} = \overharpoon{\VAL_{\indj}} ;
              \; \TASKVAR_c \in \set{\TASKVAR_1, \mydots, \TASKVAR_n}
            \\[-1mm]
            & \; \TASKVAR_c = \taskconfig{\tyatlocreg{\TYP_c}{\loc_c}{\reg}}{\concreteloc{\reg}{\indivar{\ivarid_j}}{}}{\STOR_c;\MENV_c;\concreteloc{\reg}{\concreteind{\ind_c}}{\loc_c}} \\[-2mm]
            & \; \MENV' = \parmergemenv{\MENV_c}{\MENV} ;
              \; \STOR' = \parmergestor{\STOR_c}{\STOR} \\[-2mm]
            & \; \litnum = | \overharpoon{\VAL} | ;
              \; \overharpoon{\VAL'} = [\overharpoon{\VAL_1}, \mydots, \overharpoon{\VAL_{\indj-1}}, \concreteloc{\reg}{\ind_c}{\loc_c}, \overharpoon{\VAL_{\indj+1}}, \mydots, \overharpoon{\VAL_n}] \\[-2mm]
            & \; \TYP_{\indj} = \typeoffield(\DC,\indj) ; \\[-2mm]
            & \; \STOR'' = \tie(\STOR',\MENV, \TYP_{\indj}, \concreteloc{\reg}{\concreteind{\ind_c}}{\loc_c}) \;\; \text{if} \;j \neq n \;\; \text{else} \; \STOR' \\[-2mm]
            & \; \EXPR' = \datacon{\DC}{\locreg{\loc}{\reg}}{\overharpoon{\VAL'}} ;
            \; \TASKVAR' = \taskconfig{\hTYP}{\concretelocvar{}}{\STOR'';\MENV';\EXPR'} 
          \end{aligned}
    }
  }
}

\newcommand{\rdregparcasejoin}{
  \inferrule*[lab={\;\text{[\dregparcasejoin]}}]
    {}{\TASKVAR_1, \mydots, \TASKVAR_c, \mydots, \TASKVAR_n \rparstepsto
      \TASKVAR_1, \mydots, \TASKVAR_c', \mydots \TASKVAR_n{},
      \\\\{
        \begin{aligned}
          & \text{where} \\[-2mm]
          \hspace{4mm} & \; \TASKVAR_c = \taskconfig{\hTYP_c}{\concretelocvar{}_c}{\STOR_c;\MENV_c;\EXPR_c} \\[-2mm]
          & \; \EXPR_c = \case{\concreteloc{\reg}{\indivar{\ivarid_c}}{\locreg{\loc_c}{}}}{\overharpoon{\pat}} \\[-2mm]
          & \; \TASKVAR_p \in \set{\TASKVAR_1, \mydots, \TASKVAR_n} = \taskconfig{\tyatlocreg{\TYP_p}{\loc_p}{\reg}}{\concreteloc{\reg}{\indivar{\ivarid_c}}{}}{\STOR_p;\MENV_p;\concreteloc{\reg}{\concreteind{\ind_p}}{}} \\[-2mm]
          & \; \MENV_3 = \parmergemenv{\MENV_p}{\MENV_c} ;
            \; \STOR_3 = \parmergestor{\STOR_p}{\STOR_c} \\[-2mm]
          & \; \EXPR_c' = \subst{\case{\concreteloc{\reg}{\concreteind{\ind_p}}{{\locreg{\loc_p}{}}}}{\overharpoon{\pat}}}{\indivar{\ivarid_c}}{\concreteind{\ind_p}} \\[-2mm]
          & \; \TASKVAR_c' = \taskconfig{\hTYP_c}{\concretelocvar{}_c}{\STOR_3;\MENV_3;\EXPR_c'} 
        \end{aligned}
      }
    }
}

\newcommand{\rregpardcase}{
  \inferrule*[lab={\;\text{[\dcase]}}]
  {}{
    \STOR;\MENV;\case{\concreteloc{\reg}{\ind}{\locreg{\loc}{\reg}}}{[\mydots, \caseclause{\datacon{\DC}{}{(\overharpoon{\var : \tyatlocreg{\TYP}{\loc}{\reg}})}}{\EXPR}, \mydots]} \stepsto
    \STOR;\MENV';\EXPR'
    \\\\ {
         \begin{aligned}
           \text{where} \;
           & \; \EXPR' = \subst{\EXPR}{\overharpoon{\var}}{\concreteloc{\reg}{\overharpoon{w}}{\overharpoon{\locreg{\loc}{\reg}}}} \\[-2mm]
           & \; \MENV' = \MENV \cup \set{\overharpoon{\locreg{\loc}{\reg}_1} \mapsto \concreteloc{\reg}{\ind+1}{}, \mydots , \overharpoon{\locreg{\loc}{\reg}_{j+1}} \mapsto \concreteloc{\reg}{\overharpoon{w_{j+1}}}{}} \\[-2mm]
           & \; \ewitness{\overharpoon{\TYP_1}}{\concreteloc{\reg}{\concreteind{\ind}+1}{}}{\STOR}{\concreteloc{\reg}{\overharpoon{\concreteind{w_1}}}{}} \\[-1mm]
           & \; \ewitness{\overharpoon{\TYP_{j+1}}}{\concreteloc{\reg}{\overharpoon{\concreteind{w_j}}}{}}{\STOR}{\concreteloc{\reg}{\overharpoon{\concreteind{w_{j+1}}}}{}} \\[-1mm]
           & \; \DC = \STOR(\reg)(\concreteind{\ind}); \; \indj \in \set{1 , \mydots , \litnum - 1} ;
             \; \litnum = | \overharpoon{\var : \hTYP} | \\
         \end{aligned}
       }
  }
}

\newcommand{\rdregparletexppar}{
  \inferrule*[lab={\;\text{[\dregparletexppar]}}]
  {}{\TASKVAR_1, \mydots, \taskconfig{\hTYP}{\concretelocvar{}}{\STOR;\MENV;\EXPR}, \mydots \TASKVAR_n \rparstepsto
    \TASKVAR_1, \mydots,
    \taskconfig{\hTYP_1}{\concretelocvar{}_1'}{\STOR;\MENV;\EXPR_1}, \mydots \TASKVAR_n, \taskconfig{\hTYP}{\concretelocvar{}}{\STOR;\MENV_2;\EXPR_2'}
      \\\\{
        \begin{aligned}
          \\[-4mm]
          \text{where} \;
          & \; \EXPR = (\letpack{\var : \hTYP_1}{\EXPR_1}{\EXPR_2}) ;
            \; \hTYP_1 = \tyatlocreg{\TYP_1}{\loc_1}{\reg_1} \\[-2mm]
          & \; \var_1 \; \text{fresh} ;
            \; \concretelocvar{}_1' = \concreteloc{\reg_1}{\indivar{\ivarid_1}}{} ;
            \; \EXPR_2' = \subst{\EXPR_2}{\var}{\concretelocvar{}_1'} \\[-2mm]
          & \; \MENV = \set{\locreg{\loc_1}{\reg_1} \mapsto \concretelocvar{}_1} \cup \MENV' \\[-2mm]
          & \; \MENV_2 = \set{\locreg{\loc_1}{\reg_1} \mapsto \concretelocvar{}_1'} \cup \MENV'
        \end{aligned}
      }
  } 
}

\newcommand{\rdregparletexpparapp}{
  \inferrule*[lab={\;\text{[\dregparletexppar]}}]
  {}{\TASKVAR_1, \mydots, \taskconfig{\hTYP}{\concretelocvar{}}{\STOR;\MENV;\EXPR}, \mydots \TASKVAR_n \rparstepsto \\\\
    \TASKVAR_1, \mydots,
    \taskconfig{\hTYP_1}{\concretelocvar{}_1'}{\STOR;\MENV;\EXPR_1}, \mydots \TASKVAR_n, \taskconfig{\hTYP}{\concretelocvar{}}{\STOR;\MENV_2;\EXPR_2'}
      \\\\{
        \begin{aligned}
          \\[-4mm]
          \text{where} \;
          & \; \EXPR = (\letpack{\var : \hTYP_1}{\EXPR_1}{\EXPR_2}) ;
            \; \hTYP_1 = \tyatlocreg{\TYP_1}{\loc_1}{\reg_1} \\[-2mm]
          & \; \var_1 \; \text{fresh} ;
            \; \concretelocvar{}_1' = \concreteloc{\reg_1}{\indivar{\ivarid_1}}{} ;
            \; \EXPR_2' = \subst{\EXPR_2}{\var}{\concretelocvar{}_1'} \\[-2mm]
          & \; \MENV = \set{\locreg{\loc_1}{\reg_1} \mapsto \concretelocvar{}_1} \cup \MENV' \\[-2mm]
          & \; \MENV_2 = \set{\locreg{\loc_1}{\reg_1} \mapsto \concretelocvar{}_1'} \cup \MENV'
        \end{aligned}
      }
  } 
}

\newcommand{\rdregionpardatacon}{
  \inferrule*[lab={\;\text{[\ddatacon]}}]{}{
    \STOR;\MENV; \datacon{\DC}{\locreg{\loc}{\reg}}{\overharpoon{\VAL}} \stepsto
    \STOR';\MENV;\concreteloc{\reg'}{\ind'}{}
    \\\\ {
          \begin{aligned}
            \\[-4mm]
            \text{where} \;
            & \; \STOR' = \STOR \cup \set{\reg' \mapsto (\ind' \mapsto \DC)} ;
              \; \concreteloc{\reg'}{\concreteind{\ind'}}{} = \hat{\MENV}(\locreg{\loc}{\reg})
          \end{aligned}
    }
    }}

\newcommand{\rdregionparletlocstart}{
  \inferrule*[lab={\;\text{[\dletlocstart]}}]
  {}{\STOR;\MENV;\letloc{\locreg{\loc}{\reg}}{\startr{\reg}}{\EXPR} \stepsto \STOR;\MENV';\EXPR
    \\\\{
         \begin{aligned}
           \\[-4mm]
           \text{where} \;
           & \; \MENV' = \MENV \cup \set{\locreg{\loc}{\reg} \mapsto \concreteloc{\reg}{0}{}}
         \end{aligned}
        }
  }
}

\newcommand{\rdregionparletloctag}{
  \inferrule*[lab={\;\text{[\dletloctag]}}]
  {}{\STOR ; \MENV ;  \letloc{\locreg{\loc}{\reg}}{\locreg{\loc'}{\reg} + 1}{\EXPR} \stepsto
     \STOR;\MENV';\EXPR    
    \\\\ {
          \begin{aligned}
            \\[-4mm]
            \text{where} \;
            & \; \MENV' = \MENV \cup \set{\locreg{\loc}{\reg} \mapsto \concreteloc{\reg}{\concreteind{\ind} + 1}{}} ;
              \; (\locreg{\loc'}{\reg} \mapsto \concreteloc{\reg}{\concreteind{\ind}}{}) \in \MENV
           \end{aligned}
         }
}}

\newcommand{\rdregparletexp}{
  \inferrule*[lab={\;\text{[\dletexp]}}]
              {\STOR;\MENV;\EXPR_1 \stepsto \STOR';\MENV';\EXPR'_1 \\ \EXPR'_1 \neq \VAL}
              {\STOR;\MENV;\letpack{\var : \hTYP}{\EXPR_1}{\EXPR_2} \stepsto
               \STOR';\MENV';\letpack{\var : \hTYP}{\EXPR'_1}{\EXPR_2}}
}

\newcommand{\rdregparletval}{
  \inferrule*[lab={\;\text{[\dletval]}}]
    {}{\STOR;\MENV;\letpack{\var : \hTYP}{\VAL_1}{\EXPR_2} \stepsto \STOR;\MENV;\subst{\EXPR_2}{\var}{\VAL_1}} 
}

\newcommand{\rdregparapp}{
  \inferrule*[lab={\;\text{[\dapp]}}]
    {}{\STOR;
      \MENV;
      \fapp{\overharpoon{\locreg{\loc}{\reg}}}{\overharpoon{\VAL}}
      \stepsto \STOR;
      \MENV;
      \subst{\EXPR}{\overharpoon{\var}}{\overharpoon{\VAL}} \subst{}{\overharpoon{\locreg{\loc'}{\reg'}}}{\overharpoon{\locreg{\loc}{\reg}}}
      \\\\{
        \begin{aligned}
          \\[-4mm]
          \text{where} \;
          & \; \FD = Function(f) \\[-2mm]
          & \; \fvar : \forall _{\overharpoon{\locreg{\loc'}{\reg'}}}.
          \overharpoon{\hTYP_{\fvar}} \ARROW \hTYP_{\fvar} ; (\fvar \overharpoon{\var} = \EXPR) = Freshen(\FD)
        \end{aligned}
      }
    }
}

\newcommand{\rdregparletregion}{
  \inferrule*[lab={\;\text{[\dletregion]}}]
    {}{\STOR;\MENV;\letreg{\reg}{\EXPR} \stepsto \STOR \cup \set{\reg \mapsto \emptyset};\MENV;\EXPR{}
    }
}


%% file: formal_typings1.tex
\newcommand{\rtvar}{
  \inferrule*[lab={\;\text{[\tvar]}}]{\TENV(\var) = \tyatlocreg{\TYP}{\loc}{\reg} \\ \SENV(\locreg{\loc}{\reg})=\TYP}{\TENV;\SENV;\CENV;\AENV;\NENV \vdash \AENV; \NENV; \var : \tyatlocreg{\TYP}{\loc}{\reg}}
}

\newcommand{\rtconcreteloc}{
  \inferrule*[lab={\;\text{[\tconcreteloc]}}]{\SENV(\locreg{\loc}{\reg})=\TYP}{\TENV;\SENV;\CENV;\AENV;\NENV \vdash \AENV; \NENV; \concreteloc{r}{i}{l} : \tyatlocreg{\TYP}{\loc}{\reg}}
}

\newcommand{\rtlet}{
  \inferrule*[lab={\;\text{[\tlet]}}]{\TENV;\SENV;\CENV;\AENV;\NENV \vdash \AENV';\NENV';\EXPR_1 : \tyatlocreg{\TYP_1}{\loc_1}{\reg_1} \quad \locreg{\loc_1}{\reg_1} \in \NENV \quad \locreg{\loc_1}{\reg_1} \not\in \NENV'\\\\
    \TENV';\SENV';\CENV;\AENV';\NENV' \vdash \AENV'';\NENV'';\EXPR_2 : \tyatlocreg{\TYP_2}{\loc_2}{\reg_2} \quad \locreg{\loc_2}{\reg_2} \in \NENV}
                   {\TENV;\SENV;\CENV;\AENV;\NENV \vdash \AENV''; \NENV''; \letpack{\var : \tyatlocreg{\TYP_1}{\loc_1}{\reg_1}}{\EXPR_1}{\EXPR_2} : \tyatlocreg{\TYP_2}{\loc_2}{\reg_2} 
    \\\\{ \begin{aligned}
          \\[-4mm] \text{where} \;
          & \; \TENV' = \TENV \cup \set{\var \mapsto \tyatlocreg{\TYP_1}{\loc_1}{\reg_1}} ;
          \;   \SENV' = \SENV \cup \set{\locreg{\loc_1}{\reg_1} \mapsto \TYP_1}
          \end{aligned}
        }
   }
}

\newcommand{\rtlregion}{
  \inferrule*[lab={\;\text{[\tlregion]}}]{\TENV;\SENV;\CENV;\AENV';\NENV \vdash \AENV''; \NENV'; \EXPR : \tyatlocreg{\TYP}{\loc'}{\reg'} \quad \locreg{\loc'}{\reg'} \in \NENV}{\TENV;\SENV;\CENV;\AENV;\NENV \vdash \AENV''; \NENV'; \letreg{\sreg}{\sEXPR} : \tyatlocreg{\TYP}{\loc'}{\reg'} 
    \\\\{ \begin{aligned}
          \\[-4mm] \text{where} \;
          \AENV'= \AENV \cup \set{\reg \mapsto \emptyset}
          \end{aligned}
        }
    }
}

\newcommand{\rtllstart}{
  \inferrule*[lab={\;\text{[\tllstart]}}]{\AENV(\reg)=\emptyset \\ \locreg{\loc}{\reg} \not \in \NENV'' \\ \locreg{\loc'}{\reg'} \neq \locreg{\loc}{\reg} \\ \locreg{\loc'}{\reg'} \in \NENV \\ \TENV;\SENV;\CENV';\AENV';\NENV' \vdash \AENV'';\NENV'';\EXPR : \tyatlocreg{\TYP'}{\loc'}{\reg'}}{\TENV;\SENV;\CENV;\AENV;\NENV \vdash \AENV''; \NENV''; \letloc{\locreg{\loc}{\reg}}{\startr{\reg}}{\EXPR} : \tyatlocreg{\TYP'}{\loc'}{\reg'}
    \\\\ {\begin{aligned}
            \\[-4mm]
            \text{where} \; 
            \; \CENV' = \CENV \cup \set{\locreg{\loc}{\reg} \mapsto \startr{r}} ;
            \; \AENV' = \AENV \cup \set{\reg \mapsto \locreg{\loc}{\reg}} ;
            \; \NENV' = \NENV \cup \set{\locreg{\loc}{\reg}}
          \end{aligned}}
    }
}

\newcommand{\rtlltag}{
  \inferrule*[lab={\;\text{[\tlltag]}}]{\AENV(\reg)=\locreg{\loc'}{\reg} \\ \locreg{\loc'}{\reg}, \locreg{\loc''}{\reg''} \in \NENV \\ \locreg{\loc}{\reg} \not \in \NENV'' \\ \locreg{\loc}{\reg} \neq \locreg{\loc''}{\reg''} \\ \TENV;\SENV;\CENV';\AENV';\NENV' \vdash \AENV'';\NENV'';\EXPR : \tyatlocreg{\TYP''}{\loc''}{\reg''}}{\TENV;\SENV;\CENV;\AENV;\NENV \vdash \AENV''; \NENV''; \letloc{\locreg{\loc}{\reg}}{(\locreg{\loc'}{r} + 1)}{\EXPR} : \tyatlocreg{\TYP''}{\loc''}{\reg''}
  \\\\{\begin{aligned}
         \\[-4mm]
         \text{where} \;
         \; \CENV' = \CENV \cup \set{\locreg{\loc}{\reg} \mapsto (\locreg{\loc'}{\reg} + 1)} ;
         \; \AENV' = \AENV \cup \set{\reg \mapsto \locreg{\loc}{\reg}} ;
         \; \NENV' = \NENV \cup \set{\locreg{\loc}{\reg}}
       \end{aligned}
      }
  }
}

\newcommand{\rtllafter}{
  \inferrule*[lab={\;\text{[\tllafter]}}]{\AENV(\reg)=\locreg{\loc_1}{\reg} \\ \SENV(\locreg{\loc_1}{\reg}) = \TYP' \\ \locreg{\loc_1}{\reg} \not \in \NENV \\ \locreg{\loc}{\reg} \not \in \NENV'' \\ \locreg{\loc}{\reg} \neq \locreg{\loc'}{\reg'} \\ \TENV;\SENV;\CENV';\AENV';\NENV' \vdash \AENV'';\NENV'';\EXPR : \tyatlocreg{\TYP'}{\loc'}{\reg'}  \\ \locreg{\loc'}{\reg'} \in \NENV}{\TENV;\SENV;\CENV;\AENV;\NENV \vdash \AENV''; \NENV''; \letloc{\locreg{\loc}{\reg}}{\afterl{\tyatlocreg{\TYP'}{\loc_{1}}{\reg}}}{\EXPR} : \tyatlocreg{\TYP'}{\loc'}{\reg'}
  \\\\{\begin{aligned}
         \\[-4mm]
         \text{where} \;
         \; \CENV' = \CENV \cup \set{\locreg{\loc}{\reg} \mapsto \afterl{\tyatlocreg{\TYP'}{\loc_{1}}{\reg}}} ;
         \; \AENV' = \AENV \cup \set{\reg \mapsto \locreg{\loc}{\reg}} ;
         \; \NENV' = \NENV \cup \set{\locreg{\loc}{\reg}}
       \end{aligned}
      }
  }
}

\newcommand{\rtdatacon}{
  \inferrule*[lab={\;\text{[\tdatacon]}}]{\typeofcon(\DC)=\TYP \\ \typeoffield(\DC,\ind)=\overharpoon{\TYP_{\ind}'} \\\\ \locreg{\loc}{\reg} \in \NENV \\
 \AENV(\reg) = \overharpoon{\locreg{\loc_n}{\reg}} \quad \text{if}\;n \neq 0 \quad \text{else} \; \locreg{\loc}{\reg} \\\\
 \CENV(\overharpoon{\locreg{\loc_1}{\reg}}) = \locreg{\loc}{\reg} + 1 \\ \CENV(\overharpoon{\locreg{\loc_{\indj + 1}}{\reg}}) = \afterl{(\overharpoon{\TYP'_{\indj}} \ensuremath{@} \overharpoon{\locreg{\loc'_{\indj}}{\reg}})} \\\\
 \TENV;\SENV;\CENV;\AENV;\NENV \vdash \AENV;\NENV;\overharpoon{\VAL_{\ind}} : \overharpoon{\tyatlocreg{\TYP_{\ind}'}{\loc_{\ind}}{\reg}}
 }{\TENV;\SENV;\CENV;\AENV;\NENV \vdash \AENV';\NENV'; \datacon{\DC}{\locreg{\loc}{\reg}}{\overharpoon{\VAL}} : \tyatlocreg{\TYP}{\loc}{\reg}
 \\\\{\begin{aligned}
        \\[-4mm]
        \text{where} \;
        & \; \AENV' = \AENV \cup \set{\reg \mapsto \locreg{\loc}{\reg}} ;
          \; \NENV' = \NENV - \set{\locreg{\loc}{\reg}} \\[-2mm]
        & \; \litnum = |\overharpoon{\VAL}| ;
          \; \ind \in \keywd{I} = \set{1, \; \ldots \; , \litnum} ;
          \; \indj \in \keywd{I} - \set{\litnum}
       \end{aligned}
     }
   }
}

\newcommand{\rtdataconivars}{
  \inferrule*[lab={\;\text{[\tdataconivars]}}]{\typeofcon(\DC)=\TYP \\ \typeoffield(\DC,\ind)=\overharpoon{\TYP_{\ind}'} \\\\ \locreg{\loc}{\reg} \in \NENV \\
 \AENV(\reg) = \overharpoon{\locreg{\loc_n}{\reg}} \quad \text{if}\;n \neq 0 \quad \text{else} \; \locreg{\loc}{\reg} \\\\
 \CENV(\overharpoon{\locreg{\loc_1}{\reg}}) = \locreg{\loc}{\reg} + 1 \\ \CENV(\overharpoon{\locreg{\loc_{\indj + 1}}{\reg}}) = \afterl{(\overharpoon{\TYP'_{\indj}} \ensuremath{@} \overharpoon{\locreg{\loc'_{\indj}}{\reg}})} \\\\
 \exists_{\keywd{k} \in \keywd{I}}. \concreteloc{\reg}{\indivar{\ivarid}_{k}}{} = \overharpoon{\VAL_{k}} \\
 \TENV;\SENV;\CENV;\AENV;\NENV \vdash \AENV;\NENV;\overharpoon{\VAL_{\ind}} : \overharpoon{\tyatlocreg{\TYP_{\ind}'}{\loc_{\ind}}{\reg}}  
 }{\TENV;\SENV;\CENV;\AENV;\NENV \vdash \AENV';\NENV'; \datacon{\DC}{\locreg{\loc}{\reg}}{\overharpoon{\VAL}} : \tyatlocreg{\TYP}{\loc}{\reg}
 \\\\{\begin{aligned}
        \\[-4mm]
        \text{where} \;
        & \; \litnum = |\overharpoon{\VAL}| ;
          \; \ind \in \keywd{I} = \set{1, \; \ldots \; , \litnum} ;
          \; \indj \in \keywd{I} - \set{\litnum}
       \end{aligned}
     }
   }
}

\newcommand{\rtfunctiondef}{
  \inferrule*
   [lab={\;\text{[\tfunctiondef]}}]
   {\TENV;\SENV;\CENV;\AENV;\NENV \vdash \AENV;\NENV';
    \EXPR : \tyatlocreg{\TYP}{\loc}{\reg} \qquad \locreg{\loc}{\reg} \not \in \NENV'\\\\
    \forall_{i \in \set{1, \ldots, n}} . \exists_j . \overharpoon{\locreg{\loc_i}{\reg_i}} = \overharpoon{\locreg{\loc'_j}{\reg'_j}} \\ \exists_j . \locreg{\loc}{\reg} = \overharpoon{\locreg{\loc'_j}{\reg'_j}}
   }
   {\tcfun 
    \fvar : \forall _{\overharpoon{\locreg{l'}{r'}}}.
            \overharpoon{\tyatlocreg{\TYP}{\loc}{\reg}}
            \ARROW \tyatlocreg{\TYP}{\loc}{\reg};
    \fvar \overharpoon{\var} = \EXPR
    \\\\{\begin{aligned}
            \\[-4mm]
            \text{where} \;
            & \; \TENV = \set{\overharpoon{\var_1} \mapsto \overharpoon{{\tyatlocreg{\TYP_1}{\loc_1}{\reg_1}}}, \; \ldots \; , \overharpoon{\var_n} \mapsto \overharpoon{{\tyatlocreg{\TYP_n}{\loc_n}{\reg_n}}}}\\[-2mm]
            & \; \SENV = \set{\overharpoon{\locreg{\loc_1}{\reg_1}} \mapsto \overharpoon{\TYP_1}, \; \ldots \; , \overharpoon{\locreg{\loc_n}{\reg_n}} \mapsto \overharpoon{\TYP_n}} \\[-2mm]
            & \; \CENV = \emptyset; \; \AENV = \set{\reg \mapsto \locreg{\loc}{\reg}}; \; \NENV =  \set{\locreg{\loc}{\reg}}; \\[-2mm]
            & \; n = |\overharpoon{\var}| = |\overharpoon{\tyatlocreg{\TYP}{\loc}{\reg}}|\\[-2mm]
         \end{aligned}
        }
   }
}

\newcommand{\rtprogram}{
  \inferrule*
   [lab={\;\text{[\tprogram]}}]
   { \tcfun \overharpoon{\FD} \\
    \TENV;\SENV;\CENV;\AENV;\NENV \vdash \AENV';\NENV';\EXPR:\tyatlocreg{\TYP}{\loc}{\reg}}
   {\vdash_{prog} \AENV';\NENV';\overharpoon{\DD} \;; \overharpoon{\FD} \;; \EXPR : \tyatlocreg{\TYP}{\loc}{\reg} \\\\{\begin{aligned}
          \\[-4mm]
            \text{where} \;
            & \; \TENV = \emptyset ;
              \; \SENV = \emptyset \\[-2mm]
            & \; \CENV = \set{\locreg{\loc}{\reg} \mapsto \startr{\reg}} ;
              \; \AENV = \set{\reg \mapsto \locreg{\loc}{\reg}} ;
              \; \NENV = \set{\locreg{\loc}{\reg}}
          \end{aligned}
         }
   }
}

\newcommand{\rtapp}{
  \inferrule*[lab={\;\text{[\tapp]}}]{
    |\overharpoon{\locreg{\loc'}{\reg'}}| = |\overharpoon{\locreg{\loc''}{\reg''}}| \quad |\overharpoon{\VAL}| = |\overharpoon{\var}|\\\\
    \TENV;\SENV;\CENV;\AENV;\NENV \vdash \AENV;\NENV;\overharpoon{\VAL_i} : \overharpoon{\tyatlocreg{\TYP_i}{\loc_i}{\reg_i}} \\
    \locreg{\loc}{\reg} \in \NENV \\ \AENV(\reg) = \locreg{\loc}{\reg}\\\\
    \forall_i . \exists_j . \overharpoon{\locreg{\loc'''_i}{\reg'''_i}} = \overharpoon{\locreg{\loc''_j}{\reg''_j}} \wedge \overharpoon{\locreg{\loc_i}{\reg_i}} = \overharpoon{\locreg{\loc'_j}{\reg'_j}}  \quad \exists_j . \locreg{\loc'''}{\reg'''} = \overharpoon{\locreg{\loc''_j}{\reg''_j}} \wedge \locreg{\loc}{\reg} = \overharpoon{\locreg{\loc'_j}{\reg'_j}}
    }{\TENV;\SENV;\CENV;\AENV;\NENV \vdash \AENV;\NENV';
      \fapp{\overharpoon {\locreg{\loc'}{\reg'}}}{\overharpoon{\VAL} : \tyatlocreg{\TYP}{\loc}{\reg}}
     \\\\{\begin{aligned}
            \\[-4mm]
            \text{where} \;
            & \; \fvar : \forall _{\overharpoon{\locreg{\loc''}{r''}}}.
            \overharpoon{\tyatlocreg{\TYP_i}{\loc'''_i}{\reg'''_i}} \ARROW \tyatlocreg{\TYP}{\loc'''}{\reg'''} ; (\fvar \overharpoon{\var} = \EXPR) = Function(f)\\[-2mm]
             & \; \NENV' = \NENV - \set{\locreg{\loc}{\reg}}; 
              \; n = |{\overharpoon{\VAL}}| ;
              \; \ind \in \set{1, \; \ldots \;, n} 
          \end{aligned}
         }
     }
}

\newcommand{\rtcase}{
  \inferrule*[lab={\;\text{[\tcase]}}]{
     \TENV;\SENV;\CENV;\AENV;\NENV \vdash \AENV;\NENV; \VAL : \tyatlocreg{\TYP'}{\loc'}{\reg'} \\ \locreg{\loc}{\reg} \in \NENV \\\\
     \TYP';\TENV;\SENV;\CENV;\AENV;\NENV \vdash_{\pat} \AENV';\NENV'; \overharpoon{pat_{\ind}} : \tyatlocreg{\TYP}{\loc}{\reg}}
     {\TENV;\SENV;\CENV;\AENV;\NENV \vdash \AENV';\NENV';
      \case{\VAL}{\overharpoon{\pat} : \tyatlocreg{\TYP}{\loc}{\reg}}
      \\\\{\begin{aligned}
             \\[-4mm]
             \text{where} \;
               \; \litnum = |\overharpoon{\pat}| ;
               \; \ind \in \set{1, \; \ldots \; , \litnum}
             \end{aligned}
           }
      }
}

\newcommand{\rtpat}{
  \inferrule*[lab={\;\text{[\tpat]}}]{
     \typeofcon(\DC) = \TYP'' \\ \kargtys(\DC) = \overharpoon{\TYP'} \\ \SENV(\locreg{\loc}{\reg}) = \TYP \\\\
     \locreg{\loc}{\reg} \neq \overharpoon{\locreg{\loc'_i}{\reg'}} \\ \TENV';\SENV';\CENV;\AENV;\NENV \vdash \AENV';\NENV';\EXPR : \tyatlocreg{\TYP}{\loc}{\reg}}
     {\TYP'';\TENV;\SENV;\CENV;\AENV;\NENV \tcpat \AENV';\NENV';
     \caseclause{\datacon{\DC}{}{(\overharpoon{\var : \tyatlocreg{\TYP'}{\loc'}{\reg'}})}}{\EXPR} : \tyatlocreg{\TYP}{\loc}{\reg}
     \\\\{\begin{aligned}
            \\[-4mm]
            \text{where} \;
            & \; \TENV' = \TENV \cup \set{\overharpoon{\var_1} \mapsto \overharpoon{\TYP'_1} \ensuremath{@} \overharpoon{{\loc_1'}^{\reg'}}, \; \ldots \; , \overharpoon{\var_n} \mapsto \overharpoon{\TYP'_n} \ensuremath{@} \overharpoon{{\loc_n'}^{\reg'}}}\\[-2mm]
            & \; \SENV' = \SENV \cup \set{\overharpoon{\locreg{\loc_1'}{\reg'}} \mapsto \overharpoon{\TYP'_1}, \; \ldots \; , \overharpoon{\locreg{\loc_n'}{\reg'}} \mapsto \overharpoon{\TYP'_n}} \\[-2mm]
            & \; \ind \in \set{1, \ldots , n} ; 
              \; \litnum = |\overharpoon{\TYP'}| = |\overharpoon{\var : \tyatlocreg{\TYP'}{\loc'}{\reg}}|
         \end{aligned}
        }
     }
}

\newcommand{\rttasksetempty}{  
  \inferrule*[lab={\;\text{[\ttasksetempty]}}]{}{\TENVMAP;\SENVMAP;\CENVMAP;\AENVMAP;\NENVMAP \tctaskset \AENVMAP ; \NENVMAP ; \emptyset}
}

\newcommand{\rttaskset}{
  \inferrule*[lab={\;\text{[\ttaskset]}}]{
    \taskconfig{\hTYP}{\concretelocvar{}}{\STOR;\MENV;\EXPR} = \TASKVAR_i \quad
    \TENV = \TENVMAP(\concretelocvar{}) \quad
    \SENV = \SENVMAP(\concretelocvar{}) \quad
    \CENV = \CENVMAP(\concretelocvar{}) \quad
    \AENV = \AENVMAP(\concretelocvar{}) \quad
    \NENV = \NENVMAP(\concretelocvar{})
    \\\\
    \TENV;\SENV;\CENV;\AENV;\NENV \tctask \AENV';\NENV';\TASKVAR_i : \hTYP \quad
    \AENVMAP' = \AENVMAP \cup \set{\concretelocvar \mapsto \AENV'} \quad
    \NENVMAP' = \NENVMAP \cup \set{\concretelocvar \mapsto \NENV'} \\\\
    \TENVMAP;\SENVMAP;\CENVMAP;\AENVMAP';\NENVMAP' \tctaskset \AENVMAP'' ; \NENVMAP'' ; \set{\TASKVAR_1, \ldots, \TASKVAR_n}
  }
   {\TENVMAP;\SENVMAP;\CENVMAP;\AENVMAP;\NENVMAP \tctaskset  \AENVMAP'' ; \NENVMAP'' ; \set{\TASKVAR_1, \ldots, \TASKVAR_i, \ldots, \TASKVAR_n}
  }
}

\newcommand{\rttask}{
  \inferrule*[lab={\;\text{[\ttask]}}]{
    \TENV;\SENV;\CENV;\AENV;\NENV \vdash \AENV';\NENV';\EXPR : \hTYP
  }
             {\TENV;\SENV;\CENV;\AENV;\NENV \tctask \AENV'; \NENV' ; \taskconfig{\hTYP}{\concretelocvar{}}{\STOR;\MENV;\EXPR}}
}

%% file: fig1.tex
\begin{figure*}
\hspace{-1cm}
\begin{subfigure}[t]{0.43\linewidth}
\vspace{0pt}
\begin{tabular}{|c|c} 
\hline
\hspace{2mm}
{
\lstset{numbers=left,frame=l,numbersep=6pt,numberstyle=\color{black}}
\begin{code}
data Exp = Lit Int
         | Plus Exp Exp
         | Sub Exp Exp
         | Let Sym Exp Exp
         ...

constFold :: Exp -> Exp
constFold exp = case exp of
  Lit i -> Lit i
  Plus e1 e2 ->
    case (e1, e2) of
      (Lit i,Lit j) -> Lit (i+j)
      _ -> let (e3,e4) =
             ( constFold e1 @\partup@ @\label{line:parallel-tuple}@
               constFold e2 )
           in Plus e3 e4
  Sub e1 e2 ->
    ...
\end{code}
}
\\
\hline
\end{tabular}
\caption{Constant folding written using the front-end language for Gibbon (Haskell).}
\label{fig:example1-haskell}
\end{subfigure}
\hspace{0.04\linewidth}
\begin{subfigure}[t]{0.43\linewidth}
\vspace{0pt}
\begin{tabular}{|c|c}
\hline
\hspace{2mm}
{
\lstset{numbers=left,frame=l,numbersep=6pt,numberstyle=\color{black}}
\begin{code}
constFold: forall@\locreg{l_1}{r_1}@ @\locreg{l_2}{r_2}@. @\styatlocreg{Exp}{l_1}{r_1}@ -> @\styatlocreg{Exp}{l_2}{r_2}@
constFold [@\locreg{l_1}{r_1}@ @\locreg{l_2}{r_2}@] exp = case exp of
  Lit (i:@\styatlocreg{Int}{l_i}{r_1}@) -> (Lit @\locreg{l_2}{r_2}@ i) @\label{line:datacon}@
  Plus (e1: @\styatlocreg{Exp}{l_a}{r_1}@) (e2: @\styatlocreg{Exp}{l_b}{r_1}@) ->
    case (e1,e2) of
      (Lit(i:@\styatlocreg{Int}{l_c}{r_1}@), Lit(j:@\styatlocreg{Int}{l_d}{r_1}@))
        -> (Lit @\locreg{l_2}{r_2}@ (i+j))
      _
        ->
        letloc @\locreg{l_3}{r_2}@ = @\locreg{l_2}{r_2}@ + 1 in @\label{line:letloctag}@
        let e3 : @\styatlocreg{Exp}{l_3}{r_2}@ =
          constFold [@\locreg{l_a}{r_1}@ @\locreg{l_3}{r_2}@] e1 in
        letloc @\locreg{l_4}{r_2}@ = after(@\styatlocreg{Exp}{l_3}{r_2}@) in @\label{line:letlocafter}@
        let e4 : @\styatlocreg{Exp}{l_4}{r_2}@ =
          constFold [@\locreg{l_b}{r_1}@ @\locreg{l_4}{r_2}@] e2 in
        (Plus @\locreg{l_2}{r_2}@ e3 e4) @\label{line:joinpoint}@
  Sub (e1: @\styatlocreg{Exp}{l_a}{r_1}@) (e2: @\styatlocreg{Exp}{l_b}{r_1}@) ->
    ...
\end{code}
}
\hspace{1cm}
\\
\hline
\end{tabular}
\caption{\Figref{fig:example1-haskell} compiled into LoCal IR by Gibbon.}
\label{fig:example1}
\end{subfigure}
\vspace{-5mm}
\caption{\vspace{-3mm}}
\label{fig:example1-both}
\end{figure*}

%% file: formal_grammar.tex
\begin{displaymath}
  \begin{aligned}
  &\DC \in \; \textup{Data Constructors},\:\: \TYPC \in \; \textup{Type Constructors},\\
  &\var, \yvar,\fvar \in \; \textup{Variables},\:\:
  \loc, \locreg{l}{r} \in \; \textup{Symbolic Locations},\\
  &\reg \in \; \textup{Regions},\:\: \concreteind{\ind}, \concreteind{\indj} \in \; \textup{Concrete Region Indices},
  \end{aligned}
\end{displaymath}
\begin{displaymath}
  \begin{aligned}
    \textup{Top-Level Programs} && \PROG && \gramdef & \overharpoon{\DD} \;; \overharpoon{\FD} \;; \EXPR \\
    \textup{Datatype Declarations} && \DD && \gramdef & \DATA\;\TYPC = \overharpoon{\DC \; \overharpoon{\sTYP}\;} \\
    \textup{Function Declarations} && \FD && \gramdef & \fvar : \TS ; \fvar \overharpoon{\var} = \EXPR \\
    \textup{Located Types} && \hTYP && \gramdef & \tyatlocreg{\TYP}{\loc}{\reg} \\
    \textup{Types} && \TYP && \gramdef & \TYPC \\
    \textup{Type Scheme} && \TS && \gramdef &
      \forall _{\overharpoon{\locreg{l}{r}}}.
      \overharpoon{\hTYP} \ARROW \hTYP \\
      \textup{Extended Region Indices} && \indbef{\ind}, \indbef{\indj} && \gramdef & \concreteind{\ind} \gramor \indivar{\ivarid} \gramor \indirection{\reg}{\concreteind{\ind}} \\
    \textup{Concrete Locations} && \concretelocvar{} && \gramdef & \concreteloc{\reg}{\indbef{\ind}}{\loc}\\
    \textup{Values} && \VAL && \gramdef & \var \gramor \concretelocvar{}\\
    \textup{Expressions} && \EXPR && \gramdef & \VAL\\[-5pt]
    && && \gramor & \fapp{\overharpoon{\locreg{l}{r}}}{\overharpoon{\VAL}} \\
    && && \gramor & \datacon{\DC}{\keywd{\locreg{\loc}{\reg}}}{\overharpoon{\VAL}}\\
    && && \gramor & \letpack{\var:\hTYP}{\EXPR}{\EXPR} \\
    && && \gramor & \letloc{\locreg{\loc}{\reg}}{\LE}{\EXPR} \\
    && && \gramor & \letreg{\reg}{\EXPR} \\
    && && \gramor & \case{\VAL}{\overharpoon{\pat}} \\
    \textup{Pattern} && \pat && \gramdef &
    \caseclause{\datacon{\DC}{}{(\overharpoon{\var : \hTYP})}}{\EXPR} \\
    \textup{Location Expressions} && \LE && \gramdef
    & \startr{\reg} \\
    && && \gramor & \locreg{\loc}{r} + 1 \\
    && && \gramor & \afterl{\hTYP}
  \end{aligned}
\end{displaymath}

%% file: formal_dynamics_grammar.tex
\begin{displaymath}
  \begin{aligned}
    \textup{Store} && \STOR && \gramdef & \set{\reg_1 \mapsto \heap_1 , \; \ldots \; , \reg_{n} \mapsto \heap_{n}} \\
    \textup{Heap Values} && \heapval && \gramdef & \DC \gramor \indirection{\reg}{\concreteind{\ind}} \\
    \textup{Heap} && \heap && \gramdef & \set{\concreteind{\ind}_1 \mapsto \heapval_1 , \; \ldots \; , \concreteind{\ind}_{n} \mapsto \heapval_{n}}\\
    \textup{Location Map} && \MENV && \gramdef & \set{\locreg{\loc}{\reg_1}_1 \mapsto \concretelocvar{}_1 , \; \ldots \; , \locreg{\loc}{\reg_n}_n \mapsto \concretelocvar{}_n} \\
    \textup{Sequential States} && \SEQSTATE && \gramdef & \STOR ; \MENV ; \EXPR \\
    \textup{Parallel Tasks} && \TASKVAR && \gramdef & (\hTYP, \concretelocvar{}, \SEQSTATE) \\
    \textup{Task Set} && \TASKSET && \gramdef & \set { \TASKVAR_1, \ldots, \TASKVAR_n }
  \end{aligned}
\end{displaymath}

%% file: gibbon_table.tex
\setlength{\tabcolsep}{0.3em}
\renewcommand{\arraystretch}{1.15}
\begin{tabular}{@{}l cc cccc@{}}
    \toprule
    & Gibbon & \phantom{} & \multicolumn{4}{c}{Ours} \\
    \cmidrule{2-2}
    \cmidrule{4-7}
    Benchmark & $T_s$ && $T_1$ & O & $T_{48}$ & S \\

    \midrule
    

    fib & 12.8 && 12.8 & 0 & 0.31 & 41.3 \\

    \buildfib{} & 4.69 && 4.69 & 0 & 0.11 & 42.6 \\




    buildKdTree & 2.33 && 2.67 & 14.6 & 0.22 & 10.6 \\

    countCorr & 1.46  && 1.47 & 0.68 & 0.044 & 33.2 \\

    allNearest & 1.0 && 1.01 & 1 & 0.023 & 43.5 \\



    barnesHut & 3.21 && 3.21 & 0 & 0.074 & 43.4 \\



    coins & 3.04 && 3.13 & 3 & 0.096 & 31.7 \\

    countnodes & 0.21 && 0.21 & 0 & 0.006 & 35.0 \\

    constFold & 1.78 && 1.78 & 0 & 0.16 & 11.1 \\
    


    x86-compiler & 1.08 && 1.08 & 0 & 0.041 & 26.3 \\

    mergeSort & 1.58 && 1.60 & 1.27 & 0.039 & 40.5 \\

\iftoggle{isappendix}{
\midrule
average & - && - & 1.28 & - & 29.5 \\
}{
}
    \bottomrule
\end{tabular}

%% file: ours_versus_others.tex
\setlength{\tabcolsep}{0.02em}
\renewcommand{\arraystretch}{1.15}
\begin{tabular}{@{}l rrrr cccc r cccc r cccc@{}}
    \toprule
    & &&&& \multicolumn{4}{c}{\MPL{}} & \phantom{} & \multicolumn{4}{c}{OCaml} & \phantom{} & \multicolumn{4}{c}{GHC} \\
    \cmidrule{6-9}
    \cmidrule{11-14}
    \cmidrule{16-19}
    Benchmark &&&&& $T_s$ & \vsmaple{s} & $T_{48}$ & \vsmaple{48} && $T_s$ & \vsocaml{s} & $T_{48}$ & \vsocaml{48} && $T_s$ & \vsghc{s} & $T_{48}$ & \vsghc{48} \\
    \midrule


   fib &&&&& 37.4 & 2.92 & 1.06 & 2.93 && 21.1 & 1.65 & 0.50 & 1.61 && 31.9  & 2.5 & 0.76  & 2.45 \\

   \buildfib{} &&&&& 14.5 & 3.09 & 0.35 & 3.18 && 8.60 & 1.83 & 0.25 & 2.27  && 12.4 & 2.64 & 0.34  & 3.09 \\

    buildKdTree &&&&& 7.26 & 3.11 & 0.41 & 1.86 && 10.9 & 4.68 & 1.84 & 8.36 && 13.4 & 5.75 & 2.21 & 10.0 \\

    countCorr &&&&& 10.5 & 7.19 & 0.27 & 6.14 && 13.9 & 9.52 & 0.37 & 8.41 && 3.54 & 2.42 & 0.15 & 3.41 \\

    allNearest &&&&& 2.38 & 2.38 & 0.06 & 2.60 && 3.01 & 3.01 & 0.091 & 3.95 && 2.07 & 2.07 & 0.068 & 2.96 \\

    barnesHut &&&&& 5.05  & 1.57 & 0.12 & 1.62 && 10.9 & 3.40 & 0.44 & 5.94 && 4.97 & 1.55 & 0.33 & 4.46 \\


    coins &&&&& 1.71 & 0.56 & 0.05 & 0.52 && 1.05 & 0.34 & 0.036 & 0.37 && 0.82  & 0.27 & 0.085  & 0.88 \\

    countnodes &&&&& 0.37  & 1.76 & 0.019 & 3.16 && 0.46  & 2.19 & 0.034 & 5.67  && 1.45  & 6.90  & 0.049 & 8.16 \\

    constFold &&&&& 2.36 & 1.32 & 0.23 & 1.44 && 17.7 & 9.94 & 2.23 & 13.9 && 3.71 & 2.08 & 0.64 & 4.00 \\



    x86-compiler &&&&& 1.34 & 1.24 & 0.042  & 1.02 && 1.20 & 1.11 & 0.09 & 2.20 && 2.34 & 2.16 & 0.44 & 10.7 \\


    mergeSort &&&&& 1.74 & 1.10 & 0.047 & 1.20 && 3.83 & 2.42 & 0.19 & 4.87  && 2.74  & 1.73 & 0.16  & 4.10 \\

    \midrule
   geomean &&&&& - & $1.93\times$ & - & $1.92\times$  && - & $2.53\times$ & - & $3.73\times$ && - & $2.14\times$  & -  & $4.01\times$ \\

    \bottomrule
  \end{tabular}

%% file: ms.bbl

%% file: appendix.tex
\newpage
\onecolumn
\appendix

\toggletrue{isappendix}


\section{Appendix: Sample Gibbon Programs}
\label{sec:appendix-code-snippet}

\if{0}{
\begin{figure}[h!]
\begin{code}
type Point3d = (Float, Float, Float)
-- In Gibbon, the Kdtree type will be represented in a dense, mostly-serialized format.
data KdTree = KdLeaf { x :: Float, y :: Float, z :: Float }
            | KdNode { x :: Float, y :: Float, z :: Float
                     , total_elems   :: Int   -- Number of elements in this node
                     , split_axis    :: Int   -- (0: x, 1: y, 2: z)
                     , split_value   :: Float
                     , min :: Point3d, max :: Point3d
                     , left :: KdTree, right :: KdTree
                     }
            | KdEmpty

-- | Given an array of points, build a KdTree.
mkKdTree :: Array Point3d -> KdTree
mkKdTree axis pts = ...

-- | Return query's nearest-neighbor.
nearest :: KdTree -> Point3d -> Point3d
nearest tr query =
    case tr of
        KdEmpty -> (0.0,0.0,0.0)
        KdLeaf x y z -> (x,y,z)
        KdNode x y z _total_elems split_axis _split_value _min _max left right ->
            let pivot = (x,y,z)
                tst_query = get_coord split_xis query
                tst_pivot = get_coord split_xis pivot
            in if tst_query < tst_pivot
               then find_nearest pivot query tst_pivot tst_query left right
               else find_nearest pivot query tst_pivot tst_query right left

find_nearest :: Point3d -> Point3d -> Float -> Float -> KdTree -> KdTree -> Point3d
find_nearest pivot query tst_pivot tst_query good_side other_side =
  let best0 = nearest good_side query
      candidate1 = least_dist query best0 pivot
      -- Whether the difference between the splitting coordinate of the query point and current node
      -- is less than the distance (overall coordinates) from the search point to the current best.
      nearest_other_side = tst_query - tst_pivot
  in if (nearest_other_side * nearest_other_side) <= (dist query candidate1)
     then let candidate2 = nearest other_side query
          in least_dist query candidate1 candidate2
     else candidate1

-- | Return the point that is closest to a.
least_dist :: Point3d -> Point3d -> Point3d -> Point3d
least_dist a b c =
  let d1 = dist a b
      d2 = dist a c
  in if d1 < d2 then b else c

-- | Maps an array of points to an array of their nearest neighbor.
allNearest :: KdTree -> Array Point3d -> Array Point3d
allNearest tr ls = map (\p -> nearest tr p) ls
\end{code}
\caption{
A program to compute nearest-neighbors written in Gibbon's front-end language (Haskell).
}
\label{fig:gibbon-nearest-neighbors}
\end{figure}
}
\fi

\begin{figure}[h!]
\begin{code}
-- | Compute the nth fibonacci number, in parallel.
parfib :: Int -> Int -> Int
parfib cutoff n = if n <= 1
                  then n
                  else if n <= cutoff
                       then seqfib n
                       else let (x,y) = (parfib cutoff (n-1)) @\partup@
                                        (parfib cutoff (n-2))
                            in x + y    

-- | Parallel 'buildTreeHvyLf'.
buildTreeHvyLf :: Int -> Int -> Tree
buildTreeHvyLf cutoff i = if i <= 0
                          then Leaf (seqfib 20)
                          else if i <= cutoff
                               then buildTreeHvyLf_seq i
                               else let (x,y) = (buildTreeHvyLf cutoff (i-1)) @\partup@
                                                (buildTreeHvyLf cutoff (i-1))
                                    in Node i x y
             
-- --------------------------------------

type Point3d = (Float, Float, Float)

-- In Gibbon, the Kdtree type will be represented in a dense, mostly-serialized format:
data KdTree = KdLeaf { x :: Float, y :: Float, z :: Float }
            | KdNode { x :: Float, y :: Float, z :: Float
                     , total_elems   :: Int   -- Number of elements in this node
                     , split_axis    :: Int   -- (0: x, 1: y, 2: z)
                     , split_value   :: Float
                     , min :: Point3d, max :: Point3d
                     , left :: KdTree, right :: KdTree
                     }
            | KdEmpty

-- | Maps an array of points to an array of their nearest neighbor.
allNearest :: KdTree -> Array Point3d -> Array Point3d
allNearest tr ls =
    -- parallel map with a chunk-size of 1024.
    par_map 1024 (\p -> nearest tr p) ls
  where
    nearest :: KdTree -> Point3d -> Point3d
    nearest = ...

\end{code}
\caption{
Programs for \il{fib}, \buildfib{} and
\il{allNearest} in Gibbon's front-end language (Haskell).
}
\label{fig:gibbon-examples}
\end{figure}

\section{Appendix: Full Evaluation Details}
\label{sec:appendix-evaluation}

{
\begin{figure}
\vspace{-1cm}
\hspace{-0.03\linewidth}
\begin{subfigure}[t]{0.2\linewidth}
\includegraphics[scale=0.9]{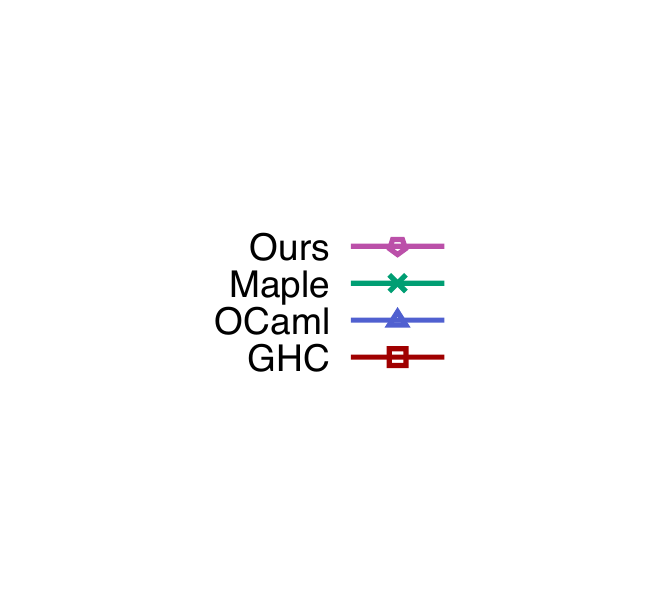}
\end{subfigure}
\hspace{0.19\linewidth}
\begin{subfigure}[t]{0.3\linewidth}
\includegraphics[scale=0.5]{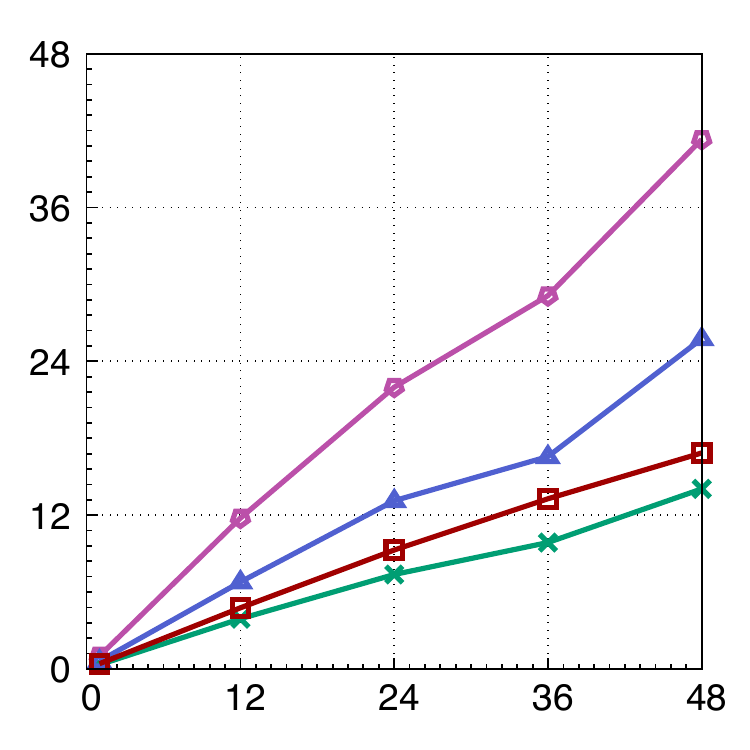}
\caption{fib}
\end{subfigure}
\hspace{0.01\linewidth}
\begin{subfigure}[t]{0.3\linewidth}
\includegraphics[scale=0.5]{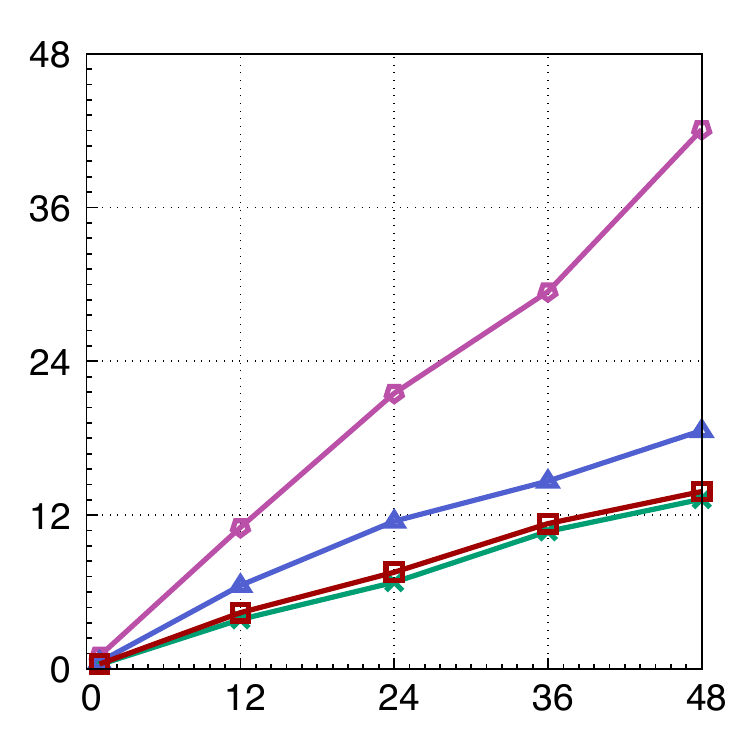}
\caption{buildtreeHvyLf}
\end{subfigure}
\\
\vspace{0.02\linewidth}
\hspace{0.04\linewidth}
\begin{subfigure}[t]{0.3\linewidth}
\includegraphics[scale=0.5]{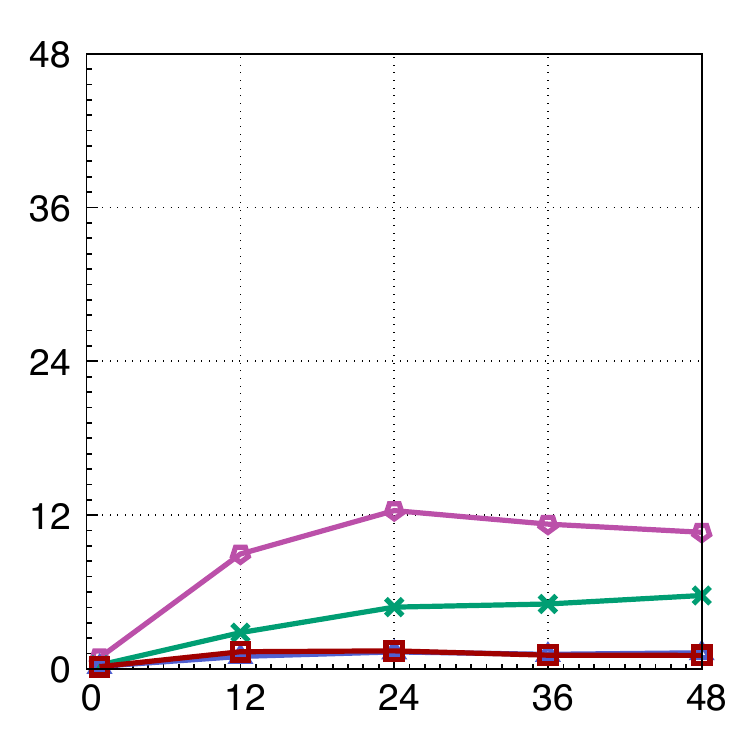}
\caption{buildKdTree}
\end{subfigure}
\hspace{0.005\linewidth}
\begin{subfigure}[t]{0.3\linewidth}
\includegraphics[scale=0.5]{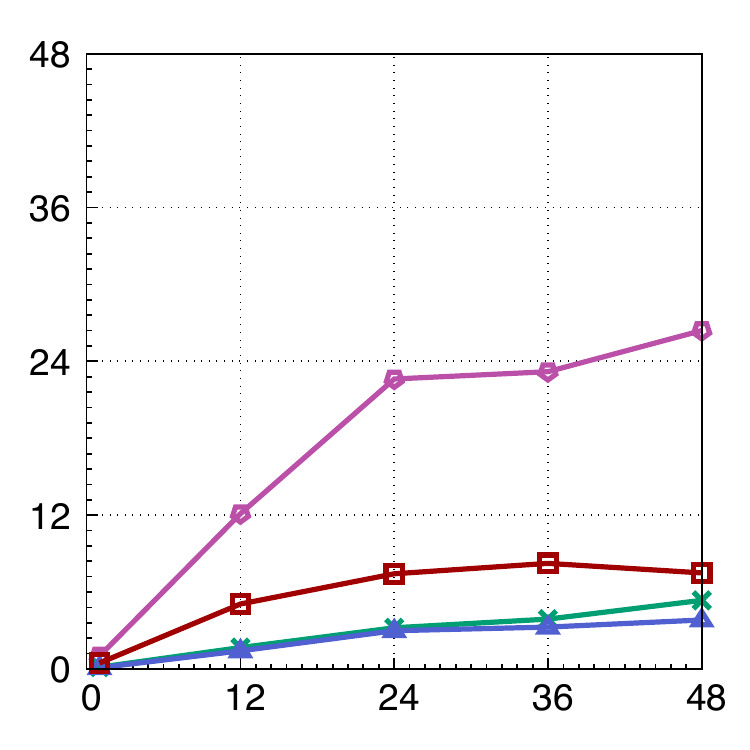}
\caption{countCorr}
\end{subfigure}
\hspace{0.02\linewidth}
\begin{subfigure}[t]{0.3\linewidth}
\includegraphics[scale=0.5]{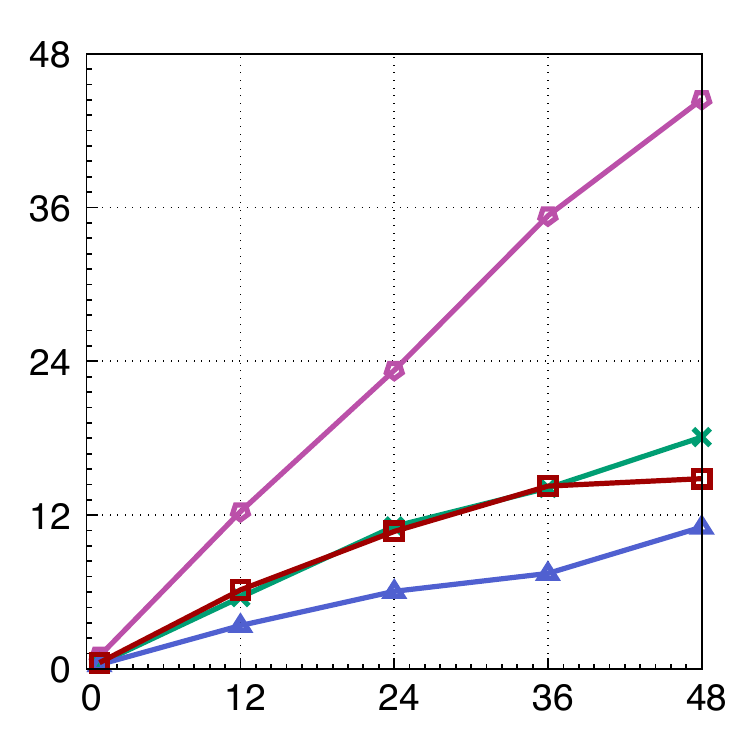}
\caption{allNearest}
\end{subfigure}
\\
\vspace{0.02\linewidth}
\hspace{0.03\linewidth}
\begin{subfigure}[t]{0.3\linewidth}
\includegraphics[scale=0.5]{figs/BarnesHut.pdf}
\caption{barnesHut}
\end{subfigure}
\hspace{0.022\linewidth}
\begin{subfigure}[t]{0.3\linewidth}
\includegraphics[scale=0.5]{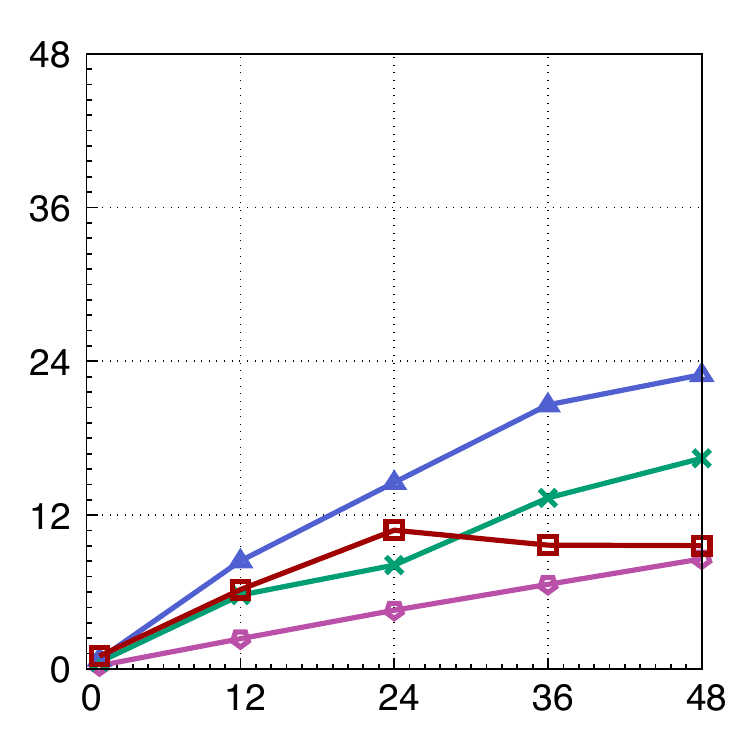}
\caption{coins}
\end{subfigure}
\hspace{0.005\linewidth}
\begin{subfigure}[t]{0.3\linewidth}
\includegraphics[scale=0.5]{figs/Countnodes.pdf}
\caption{countnodes}
\end{subfigure}
\\
\vspace{0.02\linewidth}
\hspace{0.05\linewidth}
\begin{subfigure}[t]{0.3\linewidth}
\includegraphics[scale=0.5]{figs/MergeSort.pdf}
\caption{mergeSort}
\end{subfigure}
\hspace{0.01\linewidth}
\begin{subfigure}[t]{0.3\linewidth}
\includegraphics[scale=0.5]{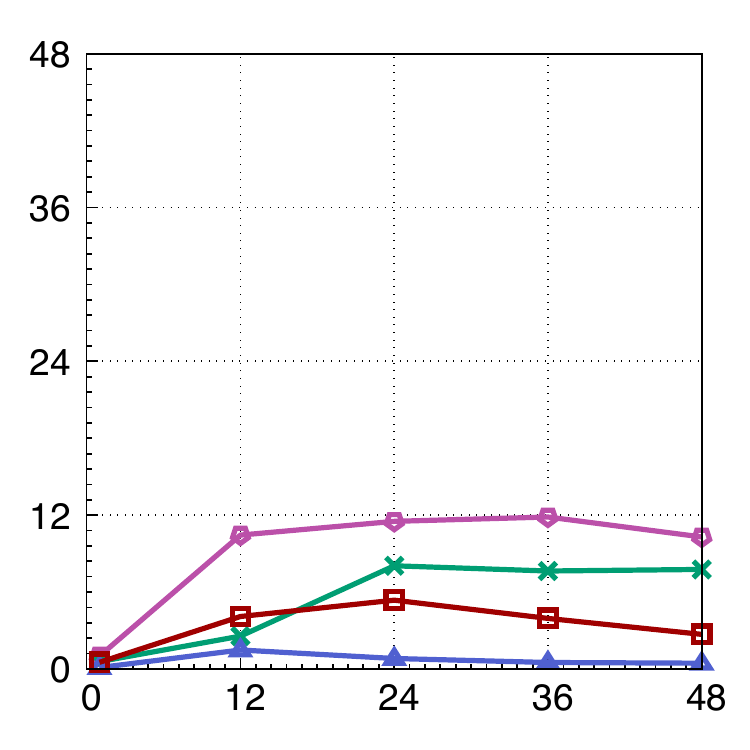}
\caption{constFold}
\end{subfigure}
\hspace{0.012\linewidth}
\begin{subfigure}[t]{0.3\linewidth}
\includegraphics[scale=0.5]{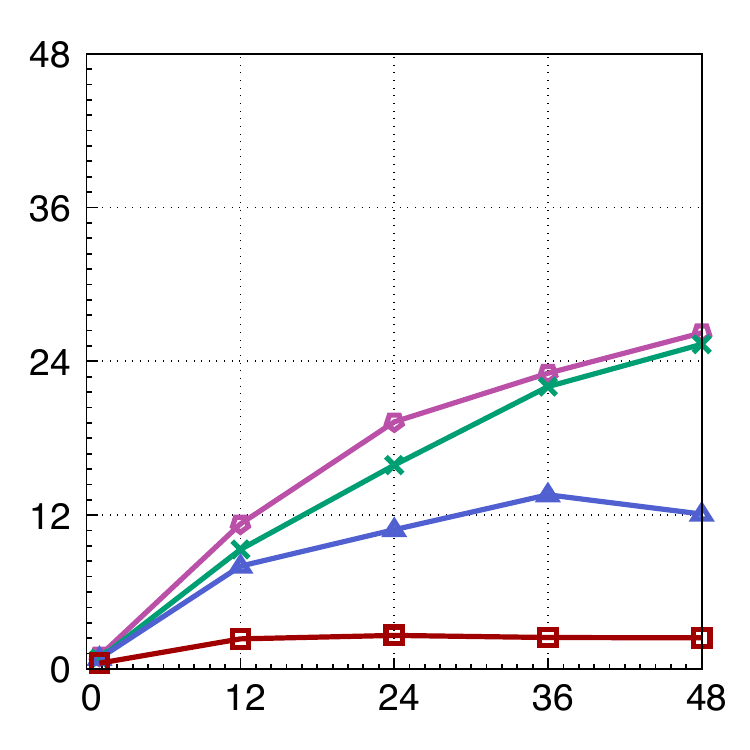}
\caption{x86-compiler}
\end{subfigure}
\caption{
Speedups relative to the fastest sequential baseline, which is Sequential Gibbon
for all benchmarks except \il{coins}, in which case sequential GHC is fastest.
}
\label{fig:scaling-all}
\end{figure}
}

In \Secref{sec:evaluation}, we presented the results for \MPL{}, OCaml, and GHC
by showing speedups/slowdowns of our implementation relative to them.
In this section, we give the full evaluation results.
To make this section somewhat self-contained, we also include a copy of
Figure~\ref{fig:gibbon-table}, which compares the performance of sequential
and parallel Gibbon.
\Figref{fig:others-benchmark-results} shows the self-relative comparisions
for \MPL{}, OCaml and GHC.
The quantities in the table can be interpreted as follows.
Column $T_s$ shows the run-time of a sequential program,
which serves the purpose of a sequential baseline.
$T_1$ is the run-time of a parallel program on a single thread, and
$O$ the percentage overhead relative to $T_s$,
calculated as $((T_1 - T_s) / T_s) * 100$.
$T_{48}$ is the run-time of a parallel program on 48 threads and
$S$ is the speedup relative to $T_s$, calculated as $T_s / T_{48}$.
These comparisions are {\em self-relative}, meaning that they compare sequential \MPL{} to
parallel \MPL{}, sequential OCaml to parallel OCaml, and sequential GHC to parallel GHC.
With respect to self-relative comparisons,
on average, we scale similarly to \MPL{}, and {\em better} than OCaml or GHC,
and the single-thread overhead across all four compilers is comparable to each other.
\Figref{fig:scaling-all} shows speedups on 1-48 threads
relative to the fastest sequential baseline, which is Sequential Gibbon
for all benchmarks except \il{coins}, for which GHC is fastest.

\begin{figure}[h!]
  \input{gibbon_table}
  \caption{
  A copy of \Figref{fig:gibbon-table} which compares performance of
  sequential and parallel Gibbon.
  }
\end{figure}

\begin{figure}[h!]
  \setlength{\tabcolsep}{0.4em}
  \input{others_appendix_table}  
  \caption{
    Execution time in seconds,
    self-relative overheads on a single thread (Columns 3, 8 and 13)
    and self-relative speedups on 48 threads (Columns 5, 10 and 15).
    $T_s$ is the run-time of a sequential program.
    %
    $T_1$ and $T_{48}$ are the run-times of a parallel program
    on 1 thread and on 48 threads respectively.
    $O$ is the single-thread percentage overhead: $O = (T_1 - T_s) / T_s * 100$.
    $S$ is the 48-thread speedup: $S = T_s / T_{48}$.
  }
  \label{fig:others-benchmark-results}
\end{figure}

\section{Appendix: Formalism}
\label{sec:appendix-formalism}

This section contains:
\begin{enumerate*}[label*=(\arabic*)]
\item typing rules for \ourcalc{} (\Secref{subsec:appendix-typing-rules}),
\item a complete version of the dynamic semantics (\Secref{subsec:appendix-dynamic-semantics}),
\item definitions of all metafunctions used in the formalism (\Secref{subsec:appendix-metafunctions}),
\item well-formedness judgments for various elements of the formal model (\Secref{subsec:well-formedness-judgments}),
\item and the complete proof of type-safety for
\rparcalc{} (\Secref{subsec:appendix-type-safety}).
\end{enumerate*}

\subsection{Typing rules for \ourcalc{}}
\label{subsec:appendix-typing-rules}

Our type system for \rparcalc{} requires some substantial extensions
to the original type system given by \citet{LoCal}.
These extensions address the need to handle multi-task configurations,
which require a number of new typing environments and rules.
Before we present these extensions, we recall the typing rule
for the configuration of a single task, which is mostly unchanged from
the original.
\[ \TENV;\SENV;\CENV;\AENV;\NENV \vdash \AENV'; \NENV'; \EXPR : \hTYP \]
The context for this judgement includes five different environments. 
First, $\TENV$ is a standard typing environment.
Second, $\SENV$ is a store-typing environment, mapping 
\emph{materialized} symbolic locations to their types. That is, every location in $\SENV$ {\em has been written} and contains a value of type $\SENV(l^r)$.
Third, $\CENV$ is a constraint environment, keeping
track of how symbolic locations relate to each other.
Fourth, $\AENV$ maps each region in scope to a location, and is used to symbolically
track the allocation and incremental construction of data structures;
Finally, $\NENV$ is a nursery of all symbolic locations that have been allocated,
but not yet written to.
Both $\AENV$ and $\NENV$ are threaded through the typing
rules, also occuring in the output of the judgement, to the right of the
turnstile.

{
\small
\begin{figure}[h!]
\begin{mathpar}
\rtdataconivars{}
\end{mathpar}
\caption{An additional typing rule for type checking an in-flight data constructor.}
\label{fig:tdatacon-ivars}
\end{figure}
}

Let us first consider the rule \textsc{\tdatacon} given in \Figref{fig:static-semantics}.
It starts by ensuring that the tag being written, and all the fields have the
correct type.
Along with that, the locations of all the fields of the constructor must
also match the expected constraints.
That is, the location of the first field
should be immediately after the constructor tag, and there should be
appropriate $\mathit{after}$ constraints for other fields in the
location constraint environment.
To indicate that the tag has been written, the allocation environment is extended,
and the location $\loc$ is removed from the nursery,
to prevent multiple writes to a location.
However, if any fields of the constructor have evaluated to an \il{ivar},
the tag {\em is not} written to the store (\dregpardataconjoin{}).
The rule \textsc{\tdataconivars{}} handles this case.
It does not extend the allocation environment, and also keeps the location $\loc$
in the nursery, so that a constructor tag may be written at this location
in the future, once this task has synchronized with those
allocating the fields.
This additional typing rule is necessary to satisfy a requirement of
the Top Level Preservation lemma (\ref{lemma:preservation}).

To generalize our typing rules to handle multi-task configurations, we
introduce new environments for variables $\TENVMAP$, store
typing $\SENVMAP$, allocation constraints $\CENVMAP$, allocation
pointers $\AENVMAP$, and nurseries $\NENVMAP$.
These environments extend their counterparts in the sequential
\seqcalc{} type system, and are needed to track state on a per-task
basis.
The precise typing rules to type check a parallel task $\TASKVAR$, and
a set of parallel tasks $\TASKSET$ are given in
the \Figref{fig:appendix-static-semantics-parallel}.
A parallel task $\TASKVAR$ is well-typed if its target expression
$\EXPR$ is well-typed, using the original \seqcalc{} typing rules, and
a task set $\TASKSET$ is well-typed if all tasks in in it are
well-typed.

{
\begin{figure}
\begin{displaymath}
  \begin{aligned}
    \textup{Typing Env. Map} && \TENVMAP && \gramdef & \; \set{\concretelocvar_1 \mapsto \TENV_1, \; \ldots \; , \concretelocvar_n \mapsto \TENV_n} \\
    \textup{Store Typing Map} && \SENVMAP && \gramdef & \; \set{\concretelocvar_1 \mapsto \SENV_1, \; \ldots \; , \concretelocvar_n \mapsto \SENV_n} \\
    \textup{Constraint Env. Map} && \CENVMAP && \gramdef & \; \set{\concretelocvar_1 \mapsto \CENV_1, \; \ldots \; , \concretelocvar_n \mapsto \CENV_n} \\
    \textup{Allocation Pointers Map} && \AENVMAP && \gramdef & \; \set{\concretelocvar_1 \mapsto \AENV_1, \; \ldots \; , \concretelocvar_n \mapsto \AENV_n} \\
    \textup{Nursery Map} && \NENVMAP && \gramdef & \; \set{\concretelocvar_1 \mapsto \NENV_1, \; \ldots \; , \concretelocvar_n \mapsto \NENV_n}
  \end{aligned}
\end{displaymath}
\caption{Extended grammar of \ourcalc{} for static semantics.}
\label{fig:appendix-static-semantics-parallel-grammar}
\end{figure}
}
%
{
\small      
\begin{figure}
\begin{mathpar}
\rttask{}
\quad
\rttasksetempty{} \\
\rttaskset{}
\end{mathpar}
\caption{A copy of the typing rules given in \Figref{fig:static-semantics-parallel}.}
\label{fig:appendix-static-semantics-parallel}
\end{figure}
}

{
\small
\begin{figure}[h!]
\begin{mathpar}
\mprset{flushleft}
\rtvar{}\quad\quad
\rtconcreteloc{}\\
\rtlet{}\quad\quad
\rtlregion{}\\
\rtllstart{}\\
\rtlltag{}\\
\rtllafter{}\\
\rtdatacon{}\quad\quad
\rtcase{}
\end{mathpar}
\caption{A copy of the typing rules for \seqcalc{} given in \cite{LoCal}}
\label{fig:static-semantics}
\end{figure}
}

{
\small
\begin{figure}[h!]
\begin{mathpar}
\mprset{flushleft}
\rtapp{}\\\\
\rtfunctiondef{}\\\\
\rtpat{}\\\\
\rtprogram{}
\end{mathpar}
\caption{A copy of remaining typing rules for \seqcalc{} given in \cite{LoCal}}
\label{fig:static-semantics2}
\end{figure}
}

\clearpage

\subsection{Dynamic semantics rules for \ourcalc{}}
\label{subsec:appendix-dynamic-semantics}

Figures~\ref{fig:dynamic-sequential-all} and~\ref{fig:dynamic-parallel-all} show
the complete dynamic semantics of \ourcalc{}.
These rules also appear in the main body of the paper,
but we include their copies here to make the appendix somewhat self-contained.
The driver which runs a \ourcalc{} program initially loads all data
types, functions, type checks them, and if successful, then seeds the
$Function$, $\typeofcon$, and $\typeoffield$ environments.
Let $\EXPR_0$ be the main expression.
If $\EXPR_0$ type checks with respect to the \tprogram{} rule, then
the main program is safe to run.
The initial configuration for the machine is a single task,
$\taskconfig{\hTYP}{\concreteloc{\reg}{0}{}}{\emptyset;\set{\loc \mapsto \concreteloc{\reg}{0}{}};\EXPR_0}$,
and the program can start taking evaluation steps from this initial configuration.
This configuration can be constructed automatically in a
straightforward way.

{
\small  
\begin{figure}[h!]
  \small
  \normalsize
  \begin{mathpar}
    \mprset{flushleft}
    \rdregparapp{}\quad\quad
    \rdregparletregion{}\\
    \rdregionparletloctag{}\\
    \rdregparletlocafterseq{}\quad\quad
    \rdregparletlocafternewreg{}\\
    \rdregionparletlocstart{}\quad
    \rdregionpardatacon{}\\
    \rdregparletexp{}\quad\quad
    \rdregparletval{}\\
    \rdcase{}
  \end{mathpar}
  \caption{A complete version of dynamic semantics rules (sequential transitions).}
  \label{fig:dynamic-sequential-all}
\end{figure}
}

{
\small  
\begin{figure}
  \small
  \normalsize
  \begin{mathpar}
    \mprset{flushleft}
    \rdregparsteptaskapp{}\\
    \rdregparletexpparapp{}\\
    \rdregparcasejoin\\
    \rdregpardataconjoin{}
  \end{mathpar}
  \caption{Copy of dynamic semantics rules (parallel transitions), given in \Figref{fig:dynamic-parallel}.}
  \label{fig:dynamic-parallel-all}
\end{figure}
}

\subsection{Definitions of metafunctions}
\label{subsec:appendix-metafunctions}

In the following, we define various metafunctions used in our formal model.

\subsubsection{Merging task memories}
\label{subsec:appendix-merging-memories-region-parallel}

We merge two stores by merging the heaps of all the regions that are
shared in common by the two stores, and then by combining with all
regions that are not shared.
We merge two heaps by taking the set of the all the heap values at
indices that are equal, and all the heap values at
indices in only the first and only the second heap.
The merging of location maps follows a similar pattern, but is slightly
complicated by its handling of locations that map to ivars.
In particular, for any location where one of the two location maps
holds an ivar and the other one holds a concrete index, we assign
to the resulting location map the concrete index, because the concrete
index contains the more recent information.

{
\begin{figure}[h!]
\metafmerging
\caption{Metafunctions for merging task memories.}
\label{fig:appendix-merging-task-memories}
\end{figure}
}

\subsubsection{End-Witness judgement}
\label{subsec:appendix-end-witness-judgement}

The end-witness provides a naive computational interpretation
of the process for finding the index one past the end of a
given concrete location, with its given type.
This rule is mostly the same as the one given for the original, sequential \seqcalc{},
but includes an additional case for handling indirection pointers.
%
To compute the end-witness of an indirection pointer,
the judgement first reads the address $\concreteloc{\reg'}{\ind'_s}{}$ which is written
at $\concreteloc{\reg}{\ind_s}{}$,
and returns the end-witness of the value allocated at $\concreteloc{\reg'}{\ind'_s}{}$.

{
\begin{figure}[h!]
\noindent\paragraph{Case (A)} $\ewitness{\TYPC}{\concreteloc{\reg}{\concreteind{\ind_{s}}}{}}{\STOR}{\concreteloc{\reg}{\concreteind{\ind_{e}}}{}}$:

\begin{enumerate}
\item \label{ewitness:impl1} $\STOR(\reg)(\ind_s) = \DC'$ \; \text{such that} \\
      $\; \DATA\;\TYPC = \overharpoon{\DC_1 \; \overharpoon{\sTYP}_1\;} \; | \; \ldots \; | \; \DC' \; \overharpoon{\sTYP}' \; | \; \ldots \; | \; \overharpoon{\DC_m \; \overharpoon{\sTYP}_m\;}$

\item \label{ewitness:impl4} $\overharpoon{\concreteind{w_0}} = \concreteind{\ind_s}+1$

\item \label{ewitness:impl2}
  $\ewitness{\overharpoon{\TYP'_1}}{\concreteloc{\reg}{\overharpoon{\concreteind{w_0}}}{}}{\STOR}{\concreteloc{\reg}{\overharpoon{\concreteind{w_1}}}{}} \wedge$ \\
  $\ewitness{\overharpoon{\TYP'_{j+1}}}{\concreteloc{\reg}{\overharpoon{\concreteind{w_j}}}{}}{\STOR}{\concreteloc{\reg}{\overharpoon{\concreteind{w_{j+1}}}}{}}$
  \\ where $\concreteind{\indj} \in \set{1,\ldots,n-1} ; n = | \overharpoon{\TYP'} |$

\item \label{ewitness:impl3}
  $\ind_e = \overharpoon{\concreteind{w_n}}$
\end{enumerate}


\noindent\paragraph{Case (B)} $\ewitness{\TYPC}{\concreteloc{\reg}{\concreteind{\ind_{s}}}{}}{\STOR}{\concreteloc{\reg'}{\concreteind{\ind'_{e}}}{}}$:

\begin{enumerate}

\item $\STOR(\reg)(\ind_s) = \indirection{\reg'}{\concreteind{\ind_s'}}$

\item $\ewitness{\TYPC}{\concreteloc{\reg'}{\concreteind{\ind_{s}'}}{}}{\STOR}{\concreteloc{\reg'}{\concreteind{\ind_{e}'}}{}}$
\end{enumerate}

\caption{The end-witness rule.}
\label{fig:end-witness}
\end{figure}
}

\subsubsection{Linking fields of a data constructor}
\label{subsec:appendix-linking-fields}


Given a store $\STOR$, a location map $\MENV$, and the addresses of the
$n^{th}$ and $(n+1)^{th}$ fields of a data constructor $\DC$,
this metafunction conditionally establishes a link between these fields.
If the location map $\MENV$ maps the symbolic location of the $(n+1)^{th}$ field,
$\locreg{\loc_2}{\reg_1}$, to an indirection pointer $\indirection{\reg_2}{\ind_2}$,
it means that these fields were computed in parallel with each other,
and thus would have been allocated into separate regions.
To link such fields, we use the end-witness judgement to compute an address
$\concreteloc{\reg_1}{\ind_e}{}$ which is one past the end of the $n^{th}$ field,
and we then write an indirection pointer pointing to the starting address of the $(n+1)^{th}$
field at $\concreteloc{\reg_1}{\ind_e}{}$.
This establishes the desired link.
If the fields have been allocated into the same region, we do not have to link them,
and the store $\STOR$ is returned unchanged in this case.
A precondition for using this metafunction is that the $n^{th}$ field must be fully
allocated.
%

{
\begin{figure}[h!]
\[
\begin{array}{lcl}
  \tie(\STOR,\MENV,\TYP_1,\concreteloc{\reg_1}{\concreteind{\ind_1}}{\loc_1}, \concretelocvar^{\loc_2}) = \STOR \cup \set{\reg_1 \mapsto (\concreteind{\ind_e} \mapsto \indirection{\reg_2}{\concreteind{\ind_2}})} \\  
  \text{where} \;\; (\locreg{\loc_2}{\reg_1} \mapsto \concreteloc{\reg_1}{\indirection{\reg_2}{\concreteind{\ind_2}}}{}) \in \MENV
  \; \text{and} \; \ewitness{\TYP_1}{\concreteloc{\reg_1}{\concreteind{\ind_1}}{}}{\STOR}{\concreteloc{\reg_1}{\concreteind{\ind_{e}}}{}}\\\\
  \tie(\STOR,\MENV,\TYP_1,\concreteloc{\reg_1}{\ind_1}{\loc_1}, \concretelocvar^{\loc_2}) = \STOR \\  
  \text{where} \;\; (\locreg{\loc_2}{\reg_1} \mapsto \concreteloc{\reg_1}{\indirection{\reg_2}{\concreteind{\ind_2}}}{}) \notin \MENV
\end{array}
\]
\caption{Metafunction for linking fields of a data constructor.}
\label{fig:appendix-linking-fields}
\end{figure}
}

\subsubsection{Derefrencing indirections in $\MENV$}
\label{subsec:maplookup}

If a location $\loc$ maps to an indirection pointer $\indirection{\reg'}{\ind}$ in $\MENV$,
it is derefrenced by returning the address $\concreteloc{\reg'}{\ind}{}$.
Otherwise, the mapping contained in $\MENV$ is returned unchanged.

{
\begin{figure}[h!]
\[
\begin{array}{l}
\maplookupone{\MENV}{\loc} = \concreteloc{\reg}{\indbef{\indj}}{} \;\; \text{where}
\begin{array}{l}
(\loc \mapsto \concretelocvar) \in \MENV(\loc) \\
\concreteloc{\reg}{\indbef{\indj}}{} = \followinm(\MENV, \concretelocvar)
\end{array} \\\\
\followinm(\MENV,\concreteloc{\reg}{\indirection{\reg'}{\concreteind{\ind}}}{}) = \concreteloc{\reg'}{\concreteind{\ind}}{} \\
\followinm(\MENV,\concreteloc{\reg}{\concreteind{\ind}}{}) = \concreteloc{\reg}{\concreteind{\ind}}{} \\
\followinm(\MENV,\concreteloc{\reg}{\indivar{\ivarid}}{}) = \concreteloc{\reg}{\indivar{\ivarid}}{}
\end{array}
\]
\caption{}
\label{fig:maplookup}
\end{figure}
}

\subsubsection{Other global environments and metafunctions}
\label{subsec:other-metafunctions}

\begin{itemize}
\item $Function(f)$: An environment that maps a function $f$ to its definition $\FD$.

\item $Freshen(\FD)$: A metafunction that freshens all bound variables in function definition
$\FD$ and returns the resulting function definition.

\item $TypeOfCon(\DC):$ An environment that maps a data constructor to its type.

\item $TypeOfField(\DC,i)$: A metafunction that returns the type of the \il{i}'th field
of data constructor $\DC$.

\item $ArgTysOfConstructor(\DC)$: An environment that maps a data constructor to its field types.

\item $\allocptr{\reg}{\STOR}$ = $\max(\set{-1} \cup \set{ \indj \; | \; {(\reg \mapsto (\indj \mapsto \DC)) \in \STOR}})$: A metafunction that returns the highest allocated address in the store,
or -1 if nothing has been allocated yet.

\item $\isval{\EXPR}$: A metafunction that checks if an expression is a value or not.

\item $Ivars(\EXPR)$: A metafunction that yields the set of ivars that occur in the term $\EXPR$.

\item $ \singlewriter{\TASKSET}{\hTYP}{\indivar{\ivarid}} =
|\set{\taskconfig{\hTYP}{\concreteloc{\reg}{\indivar{\ivarid}}{\loc}}{\STOR;\MENV;\EXPR} \; | \; \exists_{\STOR,\MENV,\EXPR} . \taskconfig{\hTYP}{\concreteloc{\reg}{\indivar{\ivarid}}{\loc}}{\STOR;\MENV;\EXPR} \in \TASKSET}| = 1
$ \\
A metafunction that checks if there is exactly one task $\TASKVAR \in \TASKSET$
which can supply a value of type $\hTYP$ for $\indivar{\ivarid}$.

\item \getwriter{\TASKSET}{\hTYP}{\indivar{\ivarid}}:
A metafunction that returns a unique task $\TASKVAR \in \TASKSET$
which can supply a value of type $\hTYP$ for $\indivar{\ivarid}$
if it exists, or returns -1 otherwise.



\item $\taskcomplete{\taskconfig{\hTYP}{\concretelocvar}{\STOR;\MENV;\EXPR}} = \isval{\EXPR}$:
A metafunction that checks whether a task has evaluated to a value.
%


\item $deepSupersetEqS(\SENVMAP_1, \SENVMAP_2) = \SENVMAP_1 \supseteq \SENVMAP_2 \vee
( (\concretelocvar \mapsto \SENV_2) \in \SENVMAP_2 \Rightarrow
  (\concretelocvar \mapsto \SENV_1) \in \SENVMAP_1 \wedge \SENV_1 \supseteq \SENV_2
)
$:
A metafunction that checks if the first argument is a {\em deep} superset of the second.
In other words, the first argument can contain a mapping $(\concretelocvar \mapsto \SENV)$
which is not present in the second one and thus be a super set at the outer level.
Or a specific mapping within the first argument can be a super set of the corresponding
mapping within the second, and be a super set at an inner level.
Note that it uses an inclusive-or, and both these conditions may be simultaneously true.

\item $deepSupersetEqC(\CENVMAP_1, \CENVMAP_2)$:
A metafunction identical to $deepSupersetEqS$, but for $\CENVMAP$.

\item $a \xor b = (a \wedge \neg b) \vee (\neg a \wedge b)$:
A metafunction for the exclusive-or logical operator.

\end{itemize}

\subsection{Well-formedness judgments}
\label{subsec:well-formedness-judgments}

In this section we present certain well-formedness criteria for
various elements of the formal model.
Some of these are similar to the corresponding judgments for sequential \seqcalc{} given
in \cite{LoCal},
but they are extended to handle ivars, and indirections in the location map and store.
We give an additional judgement to check the well-formedness of the set of
tasks executing in a parallel machine.
Because there are many requirements specified inside the various
well-formedness judgments, we introduce notation for referring
to requirements individually.
For example, the notation
\refwellformed{sec:well-formedness-allocation}{wf:impl-linear-alloc2}
refers to the judgement
$\storewfca{\AENV}{\NENV}{}{\MENV}{\STOR}$,
specified in Section~\ref{sec:well-formedness-allocation},
and in that judgement, rule number~\ref{wf:impl-linear-alloc2}.

\subsubsection{Well-formedness of a task set}
\label{sec:well-formedness-tasks}

This judgement specifies certain invariants that should hold
for all tasks executing in a parallel machine
for the task set to be well-formed.
Rule~\ref{wf:ivar-present} specifies that for all ivars referenced in an
expression being computed by a certain task,
there is a corresponding ivar in the location map,
and there is exactly one other task in the task set which supplies a
well-typed value for that ivar.
%
Rule~\ref{wf:task-store-wf} references a separate judgement for well-formedness of the
store with respect to the location map of a parallel task.

\paragraph{Judgement form}

$\SENVMAP;\CENVMAP;\AENVMAP;\NENVMAP \storewftasks{\TASKSET}$

\paragraph{Definition}

\paragraph{}

$\forall \taskconfig{\hTYP}{\concretelocvar{}}{\STOR;\MENV;\EXPR} \in \TASKSET.$

$
\SENV = \SENVMAP(\concretelocvar) ; \;
\CENV = \CENVMAP(\concretelocvar) ; \;
\AENV = \AENVMAP(\concretelocvar) ; \;
\NENV = \NENVMAP(\concretelocvar)
$

\begin{enumerate}
\item \label{wf:ivar-present} $
\concreteloc{\reg}{\indivar{\ivarid}}{\loc} \in Ivars(\EXPR) \wedge
 \emptyset;\SENV;\CENV;\AENV;\NENV \vdash \AENV' ; \NENV' ; \concreteloc{\reg}{\indivar{\ivarid}}{\loc} : \hTYP' \Rightarrow\\
 (\locreg{\loc}{\reg} \mapsto \concreteloc{\reg}{\indivar{\ivarid}}{\loc}) \in \MENV \wedge
 \singlewriter{\TASKSET}{\hTYP'}{\indivar{\ivarid}}
 $
 

\item \label{wf:task-store-wf} $\storewf{\SENV}{\CENV}{\AENV}{\NENV}{\TASKSET}{\MENV}{\STOR}$
\end{enumerate}

\subsubsection{Well-formedness of a store}
\label{sec:well-formedness}

\paragraph{Judgement form}

$\storewf{\SENV}{\CENV}{\AENV}{\NENV}{\TASKSET}{\MENV}{\STOR}$

The well-formedness judgement for a parallel task's
store specifies the valid layouts of the store by using the location
map and the various environments from the typing judgement.
%
Rule~\ref{wf:map-store-consistency} specifies that, for each location $\locreg{\loc}{\reg}$
in the store-typing environment,
there is a corresponding address in the location map.
There are two possible ways in which the value at location $\locreg{\loc}{\reg}$
may be allocated --- (1) sequentially, by the current task, or
(2) in parallel, by a different task.
The first disjunct holds when the value is allocated sequentially by the current task.
In this case, the address in the location map is a concrete index, 
and it satisfies a corresponding end-witness judgement.
Otherwise, the second disjunct holds.
In this case, the address is an ivar,
there is exactly one task which allocates a value at this ivar,
its location map contains a concrete index as location $\locreg{\loc}{\reg}$'s address,
and if it has evaluated to a value, the concrete index satisfies
a corresponding end-witness judgement.
Note that these disjuncts are connected with an {\em exclusive-or}, and only of them
can hold at a time.
%
%
Rules~\ref{wf:cfc} and~\ref{wf:ca} reference the judgments for well-formedness concerning
in-flight constructor applications (\Secref{sec:well-formedness-constructors}) and correct allocation in
regions (\Secref{sec:well-formedness-allocation}), respectively.
Finally, Rule~\ref{wf:impl1} specifies that the nursery and store-typing environments reference
no common locations, which is a way of reflecting that each location is either in the process
of being constructed and in the nursery, or allocated and in the store-typing environment, but
never both.

\newcommand{\mapstoreconsistency}{
\ensuremath{
(\locreg{\loc}{\reg} \mapsto \TYP) \in \SENV \Rightarrow \\
            (\concreteloc{\reg}{\ind_1}{} = \hat{\MENV}(\locreg{\loc}{\reg}) \wedge
            \ewitness{\TYP}{\concreteloc{\reg}{\ind_1}{}}{\STOR}{\concreteloc{\reg}{\ind_2}{}}) \xor \\
            (\concreteloc{\reg}{\indivar{\ivarid}}{} = \hat{\MENV}(\locreg{\loc}{\reg}) \wedge
             \exists_{\STOR',\MENV',\EXPR'}. \taskconfig{\TYP}{\concreteloc{\reg}{\indivar{\ivarid}}{}}{\STOR';\MENV';\EXPR'} = \getwriter{\TASKSET}{\hTYP}{\indivar{\ivarid}} \wedge
             \concreteloc{\reg}{\ind_1}{} = \hat{\MENV'}(\locreg{\loc}{\reg}) \wedge \\
             (\isval{\EXPR'} \Rightarrow \ewitness{\TYP}{\concreteloc{\reg}{\ind_1}{}}{\STOR'}{\concreteloc{\reg}{\ind_2}{}})
             )
}}

\paragraph{Definition}

\begin{enumerate}

    \item \label{wf:map-store-consistency} $\mapstoreconsistency{}$ 

    \item \label{wf:cfc} $\storewfcfa{\CENV}{}{\MENV}{\STOR}$

    \item \label{wf:ca} $\storewfca{\AENV}{\NENV}{}{\MENV}{\STOR}$

    \item \label{wf:impl1} $dom(\SENV) \cap \NENV = \emptyset $

\end{enumerate}

\subsubsection{Well-formedness of constructor application}
\label{sec:well-formedness-constructors}

\paragraph{Judgement form}

$\storewfcfa{\CENV}{}{\MENV}{\STOR}$

The well-formedness judgement for constructor application specifies the various constraints
that are necessary for ensuring correct formation of constructors, dealing with constructor
application being an incremental process that spans multiple \ourcalc{} instructions.
Rule~\ref{wfconstr:constraint-start} specifies that, if a location corresponding to the
first address in a region is in the constraint environment, then there is a corresponding
entry for that location in the location map.
Rule~\ref{wfconstr:constraint-tag} specifies that, if a location corresponding to the address one past a constructor
tag is in the constraint environment, then there are corresponding locations for the address
of the tag and the address after it in the location map.
Rule~\ref{wfconstr:constraint-after} specifies that, if a location corresponding to the address
one past after a fully allocated constructor application is in the constraint environment,
then there are corresponding locations
for the address of the start of that constructor application, and
for the address one past the constructor application in the location map.
There are two possible ways in which the values at locations $\locreg{\loc'}{\reg}$
and $\locreg{\loc}{\reg}$ may be allocated ---
(1) sequentially, by a single task, and thus in a single region, or
(2) in parallel with each other, by separate tasks, and thus in separate regions.
The first disjunct holds when the values are allocated sequentially by a single task.
In this case, the starting address of the constructor application is a concrete index,
and an end-witness for it also exists.
The address one past the constructor application is this end-witness.
The second disjunct holds when the values are allocated in parallel by separate tasks.
In this case, the location map $\MENV$ and store $\STOR$ belong to the task
that allocates the value at location $\locreg{\loc}{\reg}$.
The starting address of the constructor application is an ivar,
and the address one past the constructor application is an indirection pointer
pointing to a separate region which has been allocated in the store.
The third disjunct holds when the values are allocated in parallel by separate tasks,
but after those tasks have synchronized with each other, and their memories merged
(\ref{subsec:appendix-merging-memories-region-parallel}).
The merge resets the starting address of the constructor application back to
a concrete index, and an end-witness for it now exists.
The store contains an indirection pointer at this end-witness which points
to the start of the region that contains the value at location $\locreg{\loc}{\reg}$.
In other words, a link between the values at locations $\locreg{\loc'}{\reg}$ and
$\locreg{\loc}{\reg}$ exists.
The address one past the constructor application is the corresponding indirection pointer.
Note that these disjuncts are connected with an {\em exclusive-or}, and only of them
can hold at a time.

\paragraph{Definition}

\newcommand{\wfconstrstart}{
\ensuremath{
(\locreg{\loc}{\reg} \mapsto \startr{\reg}) \in \CENV \Rightarrow
(\locreg{\loc}{\reg} \mapsto \concreteloc{\reg}{0}{}) \in \MENV 
}}

\newcommand{\wfconstrtag}{
\ensuremath{
\locreg{\loc}{\reg} \mapsto (\locreg{\loc'}{\reg} + 1)) \in \CENV \Rightarrow
\concreteloc{\reg}{\ind_l}{} = \hat{\MENV}(\locreg{\loc'}{\reg}) \wedge
\concreteloc{\reg}{\ind_l + 1}{} = \hat{\MENV}(\locreg{\loc}{\reg})
}}

\newcommand{\wfconstrafter}{
\ensuremath{
(\locreg{\loc}{\reg} \mapsto \afterl{\tyatlocreg{\TYP}{\loc'}{\reg}}) \in \CENV \Rightarrow \\            
            (\concreteloc{\reg}{\ind_1}{} = \hat{\MENV}(\locreg{\loc'}{\reg}) \wedge
            \ewitness{\TYP}{\concreteloc{\reg}{\ind_1}{}}{\STOR}{\concreteloc{\reg}{\ind_2}{}} \wedge
            \concreteloc{\reg}{\ind_2}{} = \hat{\MENV}(\locreg{\loc}{\reg}))
            \xor \\
            ( \concreteloc{\reg}{\indivar{\ivarid_1}}{} = \hat{\MENV}(\locreg{\loc'}{\reg}) \wedge
                  (\locreg{\loc}{\reg} \mapsto \concreteloc{\reg}{\indirection{\reg_2}{0}}{}) \in \MENV \wedge
                  \set{\reg_2 \mapsto \heap} \in \STOR)            
             \xor \\            
            (\concreteloc{\reg}{\ind_1}{} = \hat{\MENV}(\locreg{\loc'}{\reg}) \wedge
            \ewitness{\TYP}{\concreteloc{\reg}{\ind_1}{}}{\STOR}{\concreteloc{\reg}{\ind_2}{}} \wedge
              (\locreg{\loc}{\reg} \mapsto \concreteloc{\reg}{\indirection{\reg_2}{0}}{}) \in \MENV \wedge
                 \STOR(\reg)(\ind_2) = \indirection{\reg_2}{0}
                 )
}}

\begin{enumerate}
    \item \label{wfconstr:constraint-start} $\wfconstrstart{}$

    \item \label{wfconstr:constraint-tag} $\wfconstrtag{}$

    \item \label{wfconstr:constraint-after} $\wfconstrafter{}$

\end{enumerate}

\subsubsection{Well-formedness concerning allocation}
\label{sec:well-formedness-allocation}

\paragraph{Judgement form}

$\storewfca{\AENV}{\NENV}{}{\MENV}{\STOR}$

The well-formedness judgement for safe allocation specifies the various properties
of the location map and store that enable continued safe allocation, avoiding in particular
overwriting cells, which could, if possible, invalidate overall type-safety.
Rule~\ref{wf:impl-linear-alloc} requires that, if a location $\locreg{\loc}{\reg}$ is in both the allocation
and nursery environments, i.e., that address represents an in-flight
constructor application, then there is a corresponding location in the location
map and the address of that location is the highest address in the store.
Alternatively, the value at location $\locreg{\loc}{\reg}$ is allocated in parallel by
a separate task, in a separate region,
and its address is the corresponding indirection pointer.
Rule~\ref{wf:impl-linear-alloc2} requires that, if there is an address in the allocation
environment and that address is fully allocated, then the address of that location is the
highest address in the store.
Rule~\ref{wf:impl-write-once} requires that, if there is an address in the nursery, then
there is a corresponding location in the location map, but nothing at the corresponding
address in the store.
Finally, Rule~\ref{wf:impl-empty-region} requires that, if there is a region that has been
created but for which nothing has yet been allocated, then there can be no addresses
for that region in the store.

\paragraph{Definition}

\begin{enumerate}
    \item \label{wf:impl-linear-alloc} $ ((\reg \mapsto \locreg{\loc}{\reg}) \in \AENV \wedge \locreg{\loc}{\reg} \in \NENV) \Rightarrow \\
          ((\locreg{\loc}{\reg} \mapsto \concreteloc{\reg}{\ind}{}) \in \MENV \wedge
          \ind > \allocptr{\reg}{\STOR})
          \xor
          (\locreg{\loc}{\reg} \mapsto \indirection{\reg_2}{\ind_2}) \in \MENV
          $

    \item \label{wf:impl-linear-alloc2} $ ((\reg \mapsto \locreg{\loc}{\reg}) \in \AENV \wedge 
    \, \concreteloc{\reg}{\ind_s}{} = \hat{\MENV}(\locreg{\loc}{\reg}) \wedge
    \locreg{\loc}{\reg} \not \in \NENV \wedge
    \, \ewitness{\TYP}{\concreteloc{\reg}{\ind_s}{}}{\STOR}{\concreteloc{\reg}{\ind_e}{}}) \Rightarrow
          \ind_e > \allocptr{\reg}{\STOR}
          $

    \item \label{wf:impl-write-once} $ \locreg{\loc}{\reg} \in \NENV \Rightarrow
          \concreteloc{\reg}{\ind}{} = \hat{\MENV}(\locreg{\loc}{\reg}) \wedge
          (\reg \mapsto (\ind \mapsto \DC)) \not \in \STOR
          $

    \item \label{wf:impl-empty-region} $(\reg \mapsto \emptyset) \in \AENV \Rightarrow
    (\reg \mapsto \emptyset) \in \STOR$
\end{enumerate}

\subsection{Type-Safety}
\label{subsec:appendix-type-safety}

We state the type-safety theorem as follows:
\begin{theorem}[Type Safety]
  \label{theorem:type-safety}
  \begin{displaymath}
  \begin{aligned}
  \text{If} \;\; & \emptyset;\SENVMAP;\CENVMAP;\AENVMAP;\NENVMAP \vdash_{taskset} \AENVMAP' ; \NENVMAP' ; \TASKSET \; \wedge \; \SENVMAP;\CENVMAP;\AENVMAP;\NENVMAP \storewftasks{\TASKSET} \\
  \text{and} \;\; & \TASKSET \rparstepsto^{n} \TASKSET' \\
  \text{then, either} \;\; & \forall \TASKVAR \in \TASKSET'. \; \taskcomplete{\TASKVAR} \\
  \text{or} \;\; & \exists \; \TASKSET''. \; \TASKSET' \rparstepsto \TASKSET''.
  \end{aligned}
  \end{displaymath}
\end{theorem}
This theorem states that, if a given task set $\TASKSET$ is well typed
and its overall store is well formed, and if $\TASKSET$
makes a transition to some task set $\TASKSET'$ in $n$ steps,
then either all tasks in
$\TASKSET'$ are fully evaluated or $\TASKSET'$ can take
a step to some task set $\TASKSET''$.
As usual, we prove this theorem by showing progress and preservation.

\begin{lemma}[Top Level Progress]
  \label{lemma:single-task-progress}
  \begin{displaymath}
  \begin{aligned}
  \text{If} \;\; & \emptyset;\SENVMAP;\CENVMAP;\AENVMAP;\NENVMAP \tctaskset \AENVMAP' ; \NENVMAP' ; \TASKSET \\
  \text{and} \;\; & \SENVMAP;\CENVMAP;\AENVMAP;\NENVMAP \storewftasks{\TASKSET} \\
  \text{then} \;\; & \forall \TASKVAR \in \TASKSET.\; \taskcomplete{\TASKVAR} \\
  \text{else} \;\; & \TASKSET \rparstepsto \TASKSET'.
  \end{aligned}
  \end{displaymath}
\end{lemma}

\begin{bproof}

If every $\TASKVAR \in \TASKSET$ has evaluated to a value, we have no further
proof obligations.
Otherwise, the obligation is to show that there is at least one task which
can take a sequential or a parallel step.
By inversion on \ttaskset{}, and the typing rule given in the premise of this lemma,
we know that all tasks in the task set $\TASKSET$ are well-typed.
%
%
Then, by inversion on \ttask{}, we know that the expression $\EXPR$
it is evaluating is also well-typed.
%
We show that there is at least one $\TASKVAR \in \TASKSET$ which
can take a sequential or a parallel step,
by performing induction on the typing derivation of the expression $\EXPR$
that a task is evaluating.

\begin{bcase}[\tlet, $\; \EXPR = \letpack{\var : \hTYP_1}{\EXPR_1}{\EXPR_2}$]

\begin{mathpar}
\rtlet{}
\end{mathpar}

Because $\EXPR$ is not a value, the proof obligation is to show that there is at least one task which
can take a step. That is, there is a rule in the dynamic semantics whose
left-hand side matches the machine configuration $\TASKSET$.
There are two rules that can match.

\begin{enumerate}

\item
\begin{mathpar}
\mprset{flushleft}
\rdregparletexppar{}
\end{mathpar}

By inversion on \dregparletexppar{}, the only obligation is to estalish that
$\concretelocvar_1 = \hat{\MENV}(\locreg{\loc_1}{\reg_1})$.
%
%
By applying the rule \refwellformed{sec:well-formedness-tasks}{wf:task-store-wf} for
the well-formedness of the task set given in the premise of this lemma,
we can obtain a well-formedness result for the store and location map used by this task:
$\storewf{\SENV}{\CENV}{\AENV}{\NENV}{\TASKSET}{\MENV}{\STOR}$.
%
%
We can then perform inversion on the well-formedness of the store,
and apply the rule \refwellformed{sec:well-formedness}{wf:ca} to obtain
$\storewfca{\AENV}{\NENV}{}{\MENV}{\STOR}$.
Finally, we apply \refwellformed{sec:well-formedness-allocation}{wf:impl-write-once}
to obtain $\concreteloc{\reg}{\ind}{} = \hat{\MENV}(\locreg{\loc_1}{\reg_1})$.
Thus we have discharged the proof obligation,
and this task can take a parallel step using \dregparletexppar{}.

\item
\begin{mathpar}
\mprset{flushleft}
\rdregparsteptask{}
\end{mathpar}

To obtain this result, we use Lemma~\ref{lemma:single-thread-progress}
for single thread progress.
There are two preconditions in order to use this lemma:
$\emptyset;\SENV;\CENV;\AENV;\NENV \vdash \AENV';\NENV';\EXPR : \hTYP$, and
$\storewf{\SENV}{\CENV}{\AENV}{\NENV}{\MENV}{\STOR}$.
The first precondition is already established at the start of the proof,
by performing inversion on \ttaskset{}, and then on \ttask{}.
The second precondition can be discharged
by performing inversion on the
on the well-formedness of the task set given in the premise
of this lemma, and then applying the rule
\refwellformed{sec:well-formedness-tasks}{wf:task-store-wf}.

\end{enumerate}

\end{bcase}

\begin{bcase}[\tdataconivars, $\;\EXPR = \datacon{\DC}{\keywd{\locreg{\loc}{\reg}}}{\overharpoon{\VAL}}$]

\begin{mathpar}
\rtdataconivars{}
\end{mathpar}

Because $\EXPR$ is not a value, the proof obligation is to show that there is a task which
can take a step. That is, there is a rule in the dynamic semantics whose
left-hand side matches the machine configuration $\TASKSET$.
There are a single rule that can match.


\begin{mathpar}
\rdregpardataconjoin{}
\end{mathpar}

If any of the fields of the data constructor have evaluated to an ivar,
the only rule that can match is \dregpardataconjoin{}.
By inversion on \dregpardataconjoin{}, the only obligation then is to find a task $\TASKVAR_c$
which supplies a value of that ivar.
To obtain this result we perform inversion on the well-formedness of the task set given in the premise
of this lemma, and apply the rule \refwellformed{sec:well-formedness-tasks}{wf:ivar-present}.
Thus, we know that exactly one such task $\TASKVAR_c$ exists.
There are two possible states in which the task $\TASKVAR_c$ may be in ---
it may have evaluated its expression to a value, or not.
If the former is true, then the rule \dregpardataconjoin{} matches, and this task
can take a step.
Otherwise, the $\TASKVAR_c$, or one of its child tasks can take a step,
and thus this case.


%



\end{bcase}

\begin{bcase}[\tcase, $\;\EXPR = \case{\VAL}{\overharpoon{\pat} : \tyatlocreg{\TYP}{\loc}{\reg}}$]

Because $\EXPR$ is not a value, the proof obligation is to show that there is a task which
can take a step. That is, there is a rule in the dynamic semantics whose
left-hand side matches the machine configuration $\TASKSET$.
There are two rules that can match: \dregparcasejoin{}, or \dparsteptask{}.
With respect to \dregparcasejoin{}, it can can be discharged by using similar reasoning
as the \tdataconivars{} case.
As for \dparsteptask{}, we use Lemma~\ref{lemma:single-thread-progress}
for single thread progress like earlier.

\end{bcase}

\begin{bcase}

The cases for \tdatacon{}, \tlregion{}, \tlltag{}, \tllstart{}, \tllafter{}, \tvar{}, and \tconcreteloc{}
use the Lemma~\ref{lemma:single-thread-progress}
for single thread progress, and can be discharged by using similar reasoning to
the previous cases.

\end{bcase}

$\blacksquare$

\end{bproof}

\begin{lemma}[Single Thread Progress]
\label{lemma:single-thread-progress}
\begin{displaymath}
  \begin{aligned}
  \text{If} \;\; & \emptyset;\SENV;\CENV;\AENV;\NENV \vdash \AENV';\NENV';\EXPR : \hTYP \\
  \text{and} \;\; & \storewf{\SENV}{\CENV}{\AENV}{\NENV}{\MENV}{\STOR} \\
  \text{then} \;\; & \EXPR \; \mathit{value} \\
  \text{else} \;\; & \STOR;\MENV;\EXPR \stepsto \STOR';\MENV';\EXPR'.
  \end{aligned}
\end{displaymath}
\end{lemma}

\begin{bproof}

The proof is by induction on the typing derivation of $\EXPR$.

\begin{bcase}[\tllafter{}, $\; \EXPR = \letloc{\locreg{\loc}{\reg}}{\afterl{\tyatlocreg{\TYP'}{\loc_{1}}{\reg}}}{\EXPR'}$
]

\begin{mathpar}
\rtllafter{}
\end{mathpar}

Because $\EXPR$ is not a value, the proof obligation is to show that there is a task which
can take a step. That is, there is a rule in the dynamic semantics whose
left-hand side matches the machine configuration $\STOR;\MENV;\EXPR$.
There are two rules that can match: \dletlocafter{}, or
\dregparletlocafternewreg{}.
They both have a similar precondition: $\concretelocvar{} = \hat{\MENV}(\locreg{\loc_1}{\reg})$.
To obtain this result, we need to use
rule \refwellformed{sec:well-formedness}{wf:map-store-consistency}
of the well-formedness of the store given in the premise of this lemma.
This rule requires that $\SENV(\locreg{\loc_1}{\reg}) = \TYP'$,
which can be obtained by inversion on \tllafter{}.
By inversion on
\refwellformed{sec:well-formedness}{wf:map-store-consistency},
we can establish that either
$\concreteloc{\reg}{\ind}{} = \hat{\MENV}(\locreg{\loc_1}{\reg})$, or
$\concreteloc{\reg}{\indivar{\ivarid}}{} = \hat{\MENV}(\locreg{\loc_1}{\reg})$
holds.
The individual requirements for each case are handled by the following case analysis.

\begin{enumerate}

\item $\concreteloc{\reg}{\ind}{} = \hat{\MENV}(\locreg{\loc_1}{\reg})$
\begin{mathpar}
\rdregparletlocafterseq{}
\end{mathpar}

If $\concreteloc{\reg}{\ind}{} = \hat{\MENV}(\locreg{\loc_1}{\reg})$,
the only rule that can match is \dletlocafter{}.
The only remaining obligation is to show that an end-witness $\indj$ for the value
allocated at the address $\concreteloc{\reg}{\ind}{}$ exists:
$\ewitness{\TYP'}{\concreteloc{\reg}{\ind}{}}{\STOR}{\concreteloc{\reg}{\indj}{}}$.
This can be discharged
by applying the rule \refwellformed{sec:well-formedness}{wf:map-store-consistency}.
If $\concreteloc{\reg}{\ind}{} = \hat{\MENV}(\locreg{\loc_1}{\reg})$,
the first disjunct of \refwellformed{sec:well-formedness}{wf:map-store-consistency} holds,
and thus discharges our required obligation.

\item $\concreteloc{\reg}{\indivar{\ivarid}}{} = \hat{\MENV}(\locreg{\loc_1}{\reg})$
\begin{mathpar}
\rdregparletlocafternewreg{}
\end{mathpar}

This rule has no remaining obligations, and matches directly, and thus this case.

\end{enumerate}

\end{bcase}

\begin{bcase}

The cases for \tdatacon{}, \tlet, \tlregion{}, \tlltag{}, \tllstart{}, \tvar{}, \tconcreteloc{},
\tapp are similar to the proof of progress for sequential \seqcalc{} given in \cite{LoCal}.

\end{bcase}

$\blacksquare$

\end{bproof}

\begin{lemma}[Top Level Preservation]
  \label{lemma:preservation}
  \begin{displaymath}
  \begin{aligned}
  \text{If} \;\; & \emptyset;\SENVMAP;\CENVMAP;\AENVMAP;\NENVMAP \vdash_{taskset} \AENVMAP' ; \NENVMAP' ; \TASKSET \\
  \text{and} \;\; & \SENVMAP;\CENVMAP;\AENVMAP;\NENVMAP \storewftasks{\TASKSET} \\
  \text{and} \;\; & \TASKSET \rparstepsto \TASKSET' \\
  \text {then for some} \;\;& deepSuperSetEqS(\SENVMAP', \SENVMAP), \;
                              deepSuperSetEqC(\CENVMAP', \CENVMAP), \;
                               \AENVMAP'' \supseteq \AENVMAP', \; \NENVMAP'' \supseteq \NENVMAP'.\\
  &   \emptyset;\SENVMAP';\CENVMAP';\AENVMAP'';\NENVMAP'' \vdash_{taskset} \AENVMAP''' ; \NENVMAP''' ; \TASKSET' \\
  \text{and} \;\; & \SENVMAP';\CENVMAP';\AENVMAP'';\NENVMAP'' \storewftasks{\TASKSET'}.
  \end{aligned}
  \end{displaymath}
\end{lemma}

\begin{bproof}

The proof is by induction on the given derivation of the dynamic semantics.

\begin{bcase}[\dregparletexppar{}]

\begin{mathpar}
\rdregparletexppar{}
\end{mathpar}

Let $\SENV = \SENVMAP(\concretelocvar)$,
$\CENV = \CENVMAP(\concretelocvar)$,
$\AENV = \AENVMAP(\concretelocvar)$, and
$\NENV = \AENVMAP(\concretelocvar)$
be the environments corresponding to the task
$\taskconfig{\hTYP}{\concretelocvar{}}{\STOR;\MENV;\EXPR}$.
We instantiate the new environment maps as:
\[
\begin{array}{l}
\SENV' = \SENV \cup \set{\locreg{\loc_1}{\reg_1} \mapsto \hTYP_1} \\
\SENVMAP' = \SENVMAP \cup \set{\concretelocvar'_1 \mapsto \SENV, \concretelocvar \mapsto \SENV'} \\
\CENVMAP' = \CENVMAP \cup \set{\concretelocvar'_1 \mapsto \CENV} \\
\AENVMAP'' = \AENVMAP' \cup \set{\concretelocvar'_1 \mapsto \AENV} \\
\NENVMAP'' = \NENVMAP' \cup \set{\concretelocvar'_1 \mapsto \NENV}
\end{array}
\]
The task executing the body of the let expression has the target location
$\concretelocvar$.
In $\SENVMAP'$, we include an updated entry,
$(\concretelocvar \mapsto \SENV')$,
to establish the allocation of the bound expression.
And we extend the environments $\SENVMAP'$, $\CENVMAP'$, $\AENVMAP'$, and $\NENVMAP'$
to contain a entry for $\concretelocvar'_1$,
which is the target location of the task executing the bound expression,
as the incoming environments won't an entry for $\concretelocvar'_1$,
which is a fresh concrete location generated by \dregparletexppar{}.

\begin{enumerate}

\item

The first obligation is to show that the result $\TASKSET'$ of the evaluation step
is well-typed with respect to the environments $\SENVMAP'$, $\CENVMAP'$, $\AENVMAP''$,
and $\NENVMAP''$.
By inversion on \ttaskset{}, the obligation then is to show that all tasks in
$\TASKSET'$ are well-typed.
By the typing rule given in the premise of this lemma
and by inversion on \ttaskset{},
we can directly establish the well-typedness of the tasks in $\TASKSET$.
Thus there are only two remaining obligations, namely to show that the two new tasks
spawned by \dregparletexppar{} are well-typed.

\begin{enumerate}


\item $ \text{Let} \;
        \SENV = \SENVMAP(\concretelocvar'_1), \;
        \CENV = \CENVMAP(\concretelocvar'_1), \;
        \AENV = \AENVMAP(\concretelocvar'_1), \;
        \NENV = \AENVMAP(\concretelocvar'_1). \; \text{Obl:} \;
\emptyset;\SENV;\CENV;\AENV;\NENV \vdash_{task} \AENV'; \NENV';\taskconfig{\hTYP_1}{\concretelocvar{}_1'}{\STOR;\MENV;\EXPR_1}$

By inversion on \ttask{}, we see that in order to prove that this task is well-typed,
we must show that the expression $\EXPR_1$
is well-typed with respect to the environments $\SENV$, $\CENV$, $\AENV$, and $\NENV$.
Concretely, the obligation is to show that
$\emptyset;\SENV;\CENV;\AENV;\NENV \vdash \AENV'; \NENV'; \EXPR_1 : \hTYP_1$
holds.
We can discharge this obligation as follows.
By the typing rule given in the premise of this lemma, and by inversion on \ttaskset{},
we can establish that the task $\taskconfig{\hTYP}{\concretelocvar{}}{\STOR;\MENV;\EXPR}$ is
well-typed.
By inversion on \ttask{}, we establish that the expression $\EXPR$ is well-typed:
$\TENV;\SENV;\CENV;\AENV;\NENV \vdash \AENV'; \NENV'; \EXPR : \hTYP$.
Then by inversion on \tlet{}, we can obtain the desired result.

\item $ \text{Let} \;
        \SENV' = \SENVMAP(\concretelocvar), \;
        \CENV = \CENVMAP(\concretelocvar), \;
        \AENV = \AENVMAP(\concretelocvar), \;
        \NENV = \AENVMAP(\concretelocvar). \; \text{Obl:} \;
\emptyset;\SENV';\CENV;\AENV;\NENV \vdash_{task} \AENV'; \NENV';\taskconfig{\hTYP}{\concretelocvar{}}{\STOR;\MENV_2;\EXPR_2'}$

This obligation can be discharged by using similar reasoning as above.
The only difference is that in order for the expression $\EXPR'_2$ to typecheck,
\tlet{} requires the location of bound expression, $\locreg{\loc_1}{\reg_1}$,
to be in the store typing environment.
The environment $\SENV'$ that is instantiated for this task fulfils this requirement,
and thus this obligation is discharged.

\end{enumerate}

\item

The second obligation is to show that the result of the evaluation step
is well-formed.
The individual requirements, labeled 
\refwellformed{sec:well-formedness-tasks}{wf:ivar-present} -
\refwellformed{sec:well-formedness-tasks}{wf:task-store-wf},
are handled by the following case analysis.

\begin{itemize}
\item Case (\refwellformed{sec:well-formedness-tasks}{wf:ivar-present}):
      for each ivar $\concreteloc{\reg}{\indivar{\ivarid}}{}$ in the result $\TASKSET'$,
      there exists exactly one task in $\TASKSET'$ which supplies a well-typed value for it.

By the well-formedness of the task set given in the premise of this lemma,
this already holds for all ivars in the task set $\TASKSET$.
The only remaining obligation then is to show that a corresponding unique task exists
for the only new ivar introduced by \dregparletexppar{},
namely \concreteloc{\reg_1}{\indivar{\ivarid_1}}{}.
%
This obligation discharges straightforwardly by inversion on \dregparletexppar{},
and the typing rule given in the premise of this lemma.



\item Case(\refwellformed{sec:well-formedness-tasks}{wf:task-store-wf}):
      the stores of all tasks in the result $\TASKSET'$ are well-formed.

By the well-formedness of the task set given in the premise of this lemma,
we can directly establish the well formedness of stores of all tasks in the task set
$\TASKSET$.
Thus there are only two remaining obligations, to show that the stores of the two new tasks
spawned by \dregparletexppar{} are well-formed.
\begin{enumerate}

\item
Case $\taskconfig{\hTYP_1}{\concretelocvar{}_1'}{\STOR;\MENV;\EXPR_1}$;
$\storewf{\SENV}{\CENV}{\AENV'}{\NENV'}{\TASKSET'}{\STOR}{\MENV}$:

Since all of the environments remain unchanged in this task,
this case follows immediately by inversion on the well-formedness of the task set
given in the premise of this lemma.

\item
Case $\taskconfig{\hTYP}{\concretelocvar{}}{\STOR;\MENV_2;\EXPR_2'}$;
and $\storewf{\SENV'}{\CENV}{\AENV'}{\NENV'}{\TASKSET'}{\STOR}{\MENV_2}$:

This task uses the updated environment $\SENV' = \SENV \cup \set{\locreg{\loc_1}{\reg_1} \mapsto \hTYP_1}$, so we must show the store $\STOR$, and location map $\MENV_2$ are well-formed.
The individual requirements, labeled 
\refwellformed{sec:well-formedness}{wf:map-store-consistency} -
\refwellformed{sec:well-formedness}{wf:impl1},
are handled by the following case analysis.

\begin{itemize}
\item Case (\refwellformed{sec:well-formedness}{wf:map-store-consistency}):

$ (\locreg{\loc}{\reg} \mapsto \TYP) \in \SENV \Rightarrow \\
            (\concreteloc{\reg}{\ind_1}{} = \hat{\MENV}(\locreg{\loc}{\reg}) \wedge
            \ewitness{\TYP}{\concreteloc{\reg}{\ind_1}{}}{\STOR}{\concreteloc{\reg}{\ind_2}{}}) \xor \\
            (\concreteloc{\reg}{\indivar{\ivarid}}{} = \hat{\MENV}(\locreg{\loc}{\reg}) \wedge
             \exists_{\STOR',\MENV',\EXPR'}. \taskconfig{\TYP}{\concreteloc{\reg}{\indivar{\ivarid}}{}}{\STOR';\MENV';\EXPR'} = \getwriter{\TASKSET}{\hTYP}{\indivar{\ivarid}} \wedge
             \concreteloc{\reg}{\ind_1}{} = \hat{\MENV'}(\locreg{\loc}{\reg}) \wedge \\
             (\isval{\EXPR'} \Rightarrow \ewitness{\TYP}{\concreteloc{\reg}{\ind_1}{}}{\STOR'}{\concreteloc{\reg}{\ind_2}{}})
             )
          $

By the well formedness of the task set given in the premise of this lemma,
we establish the well formedness of stores of all tasks in the task set
$\TASKSET$, and we obtain the result $\storewf{\SENV}{\CENV}{\AENV}{\NENV}{\TASKSET}{\STOR}{\MENV}$.
Thus, for all locations in $\SENV$, the above already holds.
Then the only obligation is to show that it holds for the new location added to $\SENV'$,
$\locreg{\loc_1}{\reg_1}$.
For $\locreg{\loc_1}{\reg_1}$,
the second disjunct follows straightforwardly by inversion on \dregparletexppar{}.

\item Cases (\refwellformed{sec:well-formedness}{wf:cfc}), (\refwellformed{sec:well-formedness}{wf:ca}), and (\refwellformed{sec:well-formedness}{wf:impl1}):

The remaining three cases also discharge straightforwardly by inversion on \dregparletexppar{},
and on the well-formedness of the task set given in the premise of this lemma.

\end{itemize}
\end{enumerate}

\end{itemize}
\end{enumerate}

\end{bcase}

\begin{bcase}[\dregpardataconjoin{}]

\begin{mathpar}
\rdregpardataconjoin{}
\end{mathpar}

Let
$\SENV = \SENVMAP(\concretelocvar)$,
$\CENV = \CENVMAP(\concretelocvar)$,
$\AENV = \AENVMAP(\concretelocvar)$, and
$\NENV = \AENVMAP(\concretelocvar)$
be the environments corresponding to the task
$\taskconfig{\hTYP}{\concretelocvar{}}{\STOR;\MENV;\EXPR}$,
and let
$\SENV_c = \SENVMAP(\concreteloc{\reg}{\indivar{\ivarid}}{})$,
$\CENV_c = \CENVMAP(\concreteloc{\reg}{\indivar{\ivarid}}{})$,
$\AENV_c = \AENVMAP(\concreteloc{\reg}{\indivar{\ivarid}}{})$, and
$\NENV_c = \AENVMAP(\concreteloc{\reg}{\indivar{\ivarid}}{})$
be the environments corresponding to the task
$\taskconfig{\tyatlocreg{\TYP_c}{\loc_c}{\reg_c}}{\concretelocvar{}}{\STOR;\MENV;\EXPR}$.

We instantiate the new environment maps as:
\[
\begin{array}{l}
\SENVMAP' = \SENVMAP \cup \set{\concretelocvar \mapsto (\SENV \cup \SENV_c)} \\
\CENVMAP' = \CENVMAP \cup \set{\concretelocvar \mapsto (\CENV \cup \CENV_c)}
\end{array}
\]

\begin{enumerate}

\item

The first obligation is to show that the result $\TASKSET'$ of the evaluation step
is well-typed with respect to the environments $\SENVMAP'$, $\CENVMAP'$, $\AENVMAP'$,
and $\NENVMAP'$.
By inversion on \ttaskset{}, the obligation then is to show that all tasks in
$\TASKSET'$ are well-typed.
By the typing rule given in the premise of this lemma
and by inversion on \ttaskset{},
we can directly establish the well-typedness of the tasks in $\TASKSET$.
Thus the only remaining obligation, is to show that the task that is
updated by the rule \dregpardataconjoin{} is well-typed.
By inversion on \ttask{}, we need to show that the expression it is
evaluating is well-typed.
Concretely, the proof obligation is:
\begin{displaymath}
\emptyset;\SENV';\CENV';\AENV';\NENV' \vdash
 \AENV'';\NENV'';
 \datacon{\DC}{\locreg{\loc}{\reg}}{\overharpoon{\VAL'}} : \hTYP
\end{displaymath}
where $\overharpoon{\VAL'} = [\overharpoon{\VAL_1}, \ldots, \overharpoon{\VAL_{\indj-1}}, \concreteloc{\reg}{\ind_c}{}, \overharpoon{\VAL_{\indj+1}}, \ldots, \overharpoon{\VAL_n}]$.
We discharge this case by showing that $\overharpoon{\VAL'}$ and
$\overharpoon{\VAL}$ have the same type.
In order to establish this result,
we first perform inversion on the typing judgement given in the premise of this lemma,
and then on \ttaskset{} and \ttask{},
an obtain a result that $\datacon{\DC}{\locreg{\loc}{\reg}}{\overharpoon{\VAL}}$
is well-typed.
Since $\overharpoon{\VAL'}$ is obtained by replacing the $j^{\text{th}}$ value in
$\overharpoon{\VAL}$,
namely $\concreteloc{\reg}{\indivar{\ivarid_j}}{}$,
with the value $\concreteloc{\reg}{\ind_c}{}$,
if the values
$\concreteloc{\reg}{\indivar{\ivarid_j}}{}$ and $\concreteloc{\reg}{\ind_c}{}$,
are of the same type,
then $\overharpoon{\VAL'}$ and $\overharpoon{\VAL}$ should also have the same type.
By performing inversion on the well-formedness of the task set given in the premise
of this lemma, and then applying the rule
\refwellformed{sec:well-formedness-tasks}{wf:ivar-present},
we establish that there is exactly one task in the task set which supplies
a well-typed value for $\indivar{\ivarid_j}$.
Thus, $\concreteloc{\reg}{\ind_c}{}$ must have the same type as
$\concreteloc{\reg}{\indivar{\ivarid_j}}{}$, and thus this case.

\item
The second obligation is to show that the result of the evaluation step
is well-formed.
The individual requirements, labeled 
\refwellformed{sec:well-formedness-tasks}{wf:ivar-present} -
\refwellformed{sec:well-formedness-tasks}{wf:task-store-wf},
are handled by the following case analysis.
\begin{itemize}

\item Case (\refwellformed{sec:well-formedness-tasks}{wf:ivar-present}):
      for each ivar $\concreteloc{\reg}{\indivar{\ivarid}}{}$ in the result $\TASKSET'$,
      there exists exactly one task in $\TASKSET'$ which supplies a well-typed value for it

By inversion on the well-formedness of the task set given in premise of this lemma,
this already holds for all ivar in $\TASKSET$.
Since \dregparletlocafternewreg{} doesn't introduce any new ivars,
this case discharges straightforwardly.

\item Case (\refwellformed{sec:well-formedness-tasks}{wf:task-store-wf}):
      $\storewf{\SENV}{\CENV}{\AENV'}{\NENV'}{\TASKSET'}{\STOR''}{\MENV'}$

By inversion on the well-formedness of the task set given in the premise of this lemma,
we establish that all tasks in the task set $\TASKSET$ are well formed.
By applying \refwellformed{sec:well-formedness-tasks}{wf:task-store-wf},
we obtain the following:
\begin{align}
& \label{prf:wf-datacon-join-eqn2} \storewf{\SENV}{\CENV}{\AENV}{\NENV}{\TASKSET}{\MENV}{\STOR} \\
& \label{prf:wf-datacon-join-eqn3} \storewf{\SENV_c}{\CENV_c}{\AENV_c}{\NENV_c}{\TASKSET}{\MENV_c}{\STOR_c}
\end{align}
The proof obligation is to show that the store $\STOR''$
is well-formed with respect to the location map $\MENV'$,
and the environments $\SENV'$, $\CENV'$, $\AENV'$, and $\NENV'$,
where $\MENV' = \parmergemenv{\MENV}{\MENV_c}$.
To discharge this case,
we perform a case analysis on the definition of $\mergem$.

\begin{itemize}

\item Case 1:

   $
  {\set{\locreg{\loc}{\reg} \mapsto \concreteloc{\reg}{\ind_1}{} \mid
  (\locreg{\loc}{\reg} \mapsto \concreteloc{\reg}{\ind_1}{}) \in \MENV,
  (\locreg{\loc}{\reg} \mapsto \concreteloc{\reg}{\ind_2}{}) \in \MENV_c,
  \ind_1 = \ind_2}}
  $

   Identitical entries in $\MENV$ and $\MENV_c$ remain unchanged in the merged location map $\MENV'$.
   For such entries, this case holds straightforwardly by using Results~(\ref{prf:wf-datacon-join-eqn2}) and (\ref{prf:wf-datacon-join-eqn3}).

\item Case 2:

  ${\set{\locreg{\loc}{\reg} \mapsto \concretelocvar \mid
  (\locreg{\loc}{\reg} \mapsto \concretelocvar) \in \MENV,
  \locreg{\loc}{\reg} \notin \MENV_c}}$ and
  ${\set{\locreg{\loc}{\reg} \mapsto \concretelocvar \mid
  \locreg{\loc}{\reg} \notin \MENV,
  (\locreg{\loc}{\reg} \mapsto \concretelocvar) \in \MENV_c
  }}$

   Entries that are not shared in common by $\MENV$ and $\MENV_c$ remain unchanged in the merged location map $\MENV'$.
   For such entries, this case holds straightforwardly by using Results~(\ref{prf:wf-datacon-join-eqn2}) and (\ref{prf:wf-datacon-join-eqn3}).

\item Case 3:
   ${\set{\locreg{\loc}{\reg} \mapsto \concreteloc{\reg}{\indj}{} \mid
  (\locreg{\loc}{\reg} \mapsto \concreteloc{\reg}{\indivar{\ivarid}}{}) \in \MENV,
  (\locreg{\loc}{\reg} \mapsto \concreteloc{\reg}{\indj}{}) \in \MENV_c
  }}$

  When a location maps to an ivar in $\MENV$, and to a concrete index in $\MENV_c$, 
  the merged location map $\MENV'$ keeps the concrete index and discards the ivar.
  To discharge this case, 
  we must show that even when the location map is thus updated, the store $\STOR''$
  is well-formed.
  The individual requirements, labeled 
  \refwellformed{sec:well-formedness}{wf:map-store-consistency} -
  \refwellformed{sec:well-formedness}{wf:impl1},
  are handled by the following case analysis.

\begin{itemize}

\item Case (\refwellformed{sec:well-formedness}{wf:map-store-consistency}):

$ (\locreg{\loc}{\reg} \mapsto \TYP) \in \SENV \Rightarrow \\
            (\concreteloc{\reg}{\ind_1}{} = \hat{\MENV}(\locreg{\loc}{\reg}) \wedge
            \ewitness{\TYP}{\concreteloc{\reg}{\ind_1}{}}{\STOR}{\concreteloc{\reg}{\ind_2}{}}) \xor \\
            (\concreteloc{\reg}{\indivar{\ivarid}}{} = \hat{\MENV}(\locreg{\loc}{\reg}) \wedge
             \exists_{\STOR_2,\MENV_2,\EXPR_2}. \taskconfig{\TYP}{\concreteloc{\reg}{\indivar{\ivarid}}{}}{\STOR_2;\MENV_2;\EXPR_2} = \getwriter{\TASKSET}{\hTYP}{\indivar{\ivarid}} \wedge
             \concreteloc{\reg}{\ind_1}{} = \hat{\MENV_2}(\locreg{\loc}{\reg}) \wedge \\
             (\isval{\EXPR_2} \Rightarrow \ewitness{\TYP}{\concreteloc{\reg}{\ind_1}{}}{\STOR_2}{\concreteloc{\reg}{\ind_2}{}})
             )$

By inversion on Result~(\ref{prf:wf-datacon-join-eqn2}), we know that
\refwellformed{sec:well-formedness}{wf:map-store-consistency} holds for $\STOR$.
And since $(\locreg{\loc}{\reg} \mapsto \concreteloc{\reg}{\indivar{\ivarid}}{}) \in \MENV$,
the second disjunct of \refwellformed{sec:well-formedness}{wf:map-store-consistency} must hold.
Since this $\locreg{\loc}{\reg}$ is being updated to map to a concrete index,
in order to discharge this case we must show that the first disjunct of
\refwellformed{sec:well-formedness}{wf:map-store-consistency} now holds for it.
There are two obligations: $\concreteloc{\reg}{\indj}{} = \hat{\MENV'}(\locreg{\loc}{\reg})$,
and $\ewitness{\TYP}{\concreteloc{\reg}{\indj}{}}{\STOR''}{\concreteloc{\reg}{\indj_2}{}}$.
Since the second disjunct of \refwellformed{sec:well-formedness}{wf:map-store-consistency} holds for the store $\STOR$ and location map $\MENV$, we obtain the following:

(i) $\concreteloc{\reg}{\indivar{\ivarid}}{} = \hat{\MENV}(\locreg{\loc}{\reg})$

(ii) $\exists_{\STOR_2,\MENV_2,\EXPR_2}. \taskconfig{\TYP}{\concreteloc{\reg}{\indivar{\ivarid}}{}}{\STOR_2;\MENV_2;\EXPR_2} = \getwriter{\TASKSET}{\hTYP}{\indivar{\ivarid}}$

(iii) $\concreteloc{\reg}{\ind_1}{} = \hat{\MENV_2}(\locreg{\loc}{\reg})$

(iv) $\isval{\EXPR_2} \Rightarrow \ewitness{\TYP}{\concreteloc{\reg}{\ind_1}{}}{\STOR_2}{\concreteloc{\reg}{\ind_2}{}}$

By applying \refwellformed{sec:well-formedness-tasks}{wf:ivar-present},
we can establish that there is exactly one task in $\TASKSET$ that can provide
a well-typed value for an ivar.
Thus, the task being merged by \dregpardataconjoin{} must be this unique task.
As a result, the location map $\MENV_2$, store $\STOR_2$, and expression $\EXPR_2$,
must be same as $\MENV_c$, $\STOR_c$, and $\concreteloc{\reg}{\ind_c}{}$.
Thus, we obtain:

(v) $\concreteloc{\reg}{\indj}{} = \hat{\MENV_c}(\locreg{\loc}{\reg})$

(vi) $\isval{\concreteloc{\reg}{\ind_c}{}} \Rightarrow \ewitness{\TYP}{\concreteloc{\reg}{\indj}{}}{\STOR_c}{\concreteloc{\reg}{\indj_2}{}}$

The first obligation, $\concreteloc{\reg}{\indj}{} = \hat{\MENV'}(\locreg{\loc}{\reg})$,
discharges by inspection on $\mergem{}$, which inserts
the mapping $(\locreg{\loc}{\reg} \mapsto \concreteloc{\reg}{\indj}{})$ in $\MENV'$.
The second obligation, $\ewitness{\TYP}{\concreteloc{\reg}{\indj}{}}{\STOR''}{\concreteloc{\reg}{\indj_2}{}}$,
discharges since $\isval{\concreteloc{\reg}{\ind_c}{}}$ is true,
and the store $\STOR''$ contains all the allocations performed in $\STOR_c$.
Thus, this case.

\item \refwellformed{sec:well-formedness}{wf:cfc} $\storewfcfa{\CENV}{\MENV'}{}{\STOR''}$

\begin{itemize}[label=$\circ$]
    \item Case (\refwellformed{sec:well-formedness-constructors}{wfconstr:constraint-start}):

    $\wfconstrstart{}$

    Since the address of a location at the start of a region cannot be an ivar,
    this case discharges trivially.

    \item Case (\refwellformed{sec:well-formedness-constructors}{wfconstr:constraint-tag}):

            $\wfconstrstart{}$

    Since the address of a location one past the start of a region cannot be an ivar,
    this case discharges trivially.

    \item Case (\refwellformed{sec:well-formedness-constructors}{wfconstr:constraint-after}):

    $ (\locreg{\loc}{\reg} \mapsto \afterl{\tyatlocreg{\TYP}{\loc'}{\reg}}) \in \CENV \Rightarrow \\            
            (\concreteloc{\reg}{\ind_1}{} = \hat{\MENV}(\locreg{\loc'}{\reg}) \wedge
            \ewitness{\TYP}{\concreteloc{\reg}{\ind_1}{}}{\STOR}{\concreteloc{\reg}{\ind_2}{}} \wedge
            \concreteloc{\reg}{\ind_2}{} = \hat{\MENV}(\locreg{\loc}{\reg}))
            \xor \\
            ( \concreteloc{\reg}{\indivar{\ivarid_1}}{} = \hat{\MENV}(\locreg{\loc'}{\reg}) \wedge
                  (\locreg{\loc}{\reg} \mapsto \concreteloc{\reg}{\indirection{\reg_2}{0}}{}) \in \MENV \wedge
                  \set{\reg_2 \mapsto \heap} \in \STOR)            
             \xor \\            
            (\concreteloc{\reg}{\ind_1}{} = \hat{\MENV}(\locreg{\loc'}{\reg}) \wedge
            \ewitness{\TYP}{\concreteloc{\reg}{\ind_1}{}}{\STOR}{\concreteloc{\reg}{\ind_2}{}} \wedge
              (\locreg{\loc}{\reg} \mapsto \concreteloc{\reg}{\indirection{\reg_2}{0}}{}) \in \MENV \wedge
                 \STOR(\reg)(\ind_2) = \indirection{\reg_2}{0}
                 )
            $

By inversion on Result~(\ref{prf:wf-datacon-join-eqn2}), we know that
\refwellformed{sec:well-formedness-constructors}{wfconstr:constraint-after} holds for $\STOR$.
And since $(\locreg{\loc}{\reg} \mapsto \concreteloc{\reg}{\indivar{\ivarid}}{}) \in \MENV$,
the second disjunct of \refwellformed{sec:well-formedness-constructors}{wfconstr:constraint-after}
must hold.
But if $\locreg{\loc}{\reg}$ is being updated to map to a concrete index instead of
an ivar by the merge,
and since this case looks at state of the machine after a merge,
we must show that the
third disjunct of
\refwellformed{sec:well-formedness-constructors}{wfconstr:constraint-after} now holds
in order to discharge this case.
There are three obligations:

$\bullet \; \concreteloc{\reg}{\indj}{} = \hat{\MENV'}(\locreg{\loc}{\reg})$

This obligation discharges straightforwardly by the premise of this case,
and by inspection of $\mergem{}$, which updates $\locreg{\loc}{\reg}$ to map to
a concrete index instead of an ivar.

$\bullet \; \ewitness{\TYP}{\concreteloc{\reg}{\indj}{}}{\STOR''}{\concreteloc{\reg}{\indj_2}{}}$

By Result~\ref{prf:wf-datacon-join-eqn3}, we get
$\storewf{\SENV}{\CENV}{\AENV}{\NENV}{\TASKSET}{\MENV_c}{\STOR_c}$.
And since $(\locreg{\loc}{\reg} \mapsto \concreteloc{\reg}{\indj}{})$, the
first disjunct of \refwellformed{sec:well-formedness}{wf:map-store-consistency} must hold.
The only precondition of \refwellformed{sec:well-formedness}{wf:map-store-consistency},
$(\locreg{\loc}{\reg} \mapsto \TYP) \in \SENV$,
discharges straightforwardly by inversion on \tllafter{}, which is the only rule that
adds a constraint $(\locreg{\loc_1}{\reg} \mapsto \afterl{\tyatlocreg{\TYP}{\loc}{\reg}})$ to $\CENV$.
Thus, we get
$\ewitness{\TYP}{\concreteloc{\reg}{\indj}{}}{\STOR_c}{\concreteloc{\reg}{\indj_2}{}}$.
Since the $\STOR''$ contains all the allocations performed in $\STOR_c$,
$\ewitness{\TYP}{\concreteloc{\reg}{\indj}{}}{\STOR''}{\concreteloc{\reg}{\indj_2}{}}$ must
hold as well.

$\bullet \; (\locreg{\loc_1}{\reg} \mapsto \concreteloc{\reg}{\indj_2}{}) \in \MENV' \vee
            (\STOR''(\reg)(\indj_2) = \indirection{\reg_2}{\indj_3} \wedge
            (\locreg{\loc_1}{\reg} \mapsto \concreteloc{\reg}{\indirection{\reg_2}{\indj_3}}{}) \in \MENV')
$

We can discharge this case by showing that the second disjunct holds.
There are two obligations: $\STOR''(\reg)(\indj_2) = \indirection{\reg_2}{\indj_3}$,
and $(\locreg{\loc_1}{\reg} \mapsto \concreteloc{\reg}{\indirection{\reg_2}{\indj_3}}{}) \in \MENV'$.
The first obligation discharges by inspecting the definition of $\tie$, which
writes the appropriate indirection to the store.
To discharge the second obligation, we perform inversion on Result~\ref{prf:wf-datacon-join-eqn2}.
By inversion, we know that \refwellformed{sec:well-formedness-constructors}{wfconstr:constraint-after}
must hold for $\MENV$.
And since $(\locreg{\loc}{\reg} \mapsto \concreteloc{\reg}{\indivar{\ivarid}}{}) \in \MENV$, the
second disjunct of \refwellformed{sec:well-formedness-constructors}{wfconstr:constraint-after}
must hold for $\MENV$.
Thus, we get
$(\locreg{\loc_1}{\reg} \mapsto \concreteloc{\reg}{\indirection{\reg_2}{\indj_3}}{}) \in \MENV$.
Since $\MENV'$ contains this mapping from $\MENV$,
$(\locreg{\loc_1}{\reg} \mapsto \concreteloc{\reg}{\indirection{\reg_2}{\indj_3}}{}) \in \MENV'$
holds as well.

\item Case  $\storewf{\SENV'}{\CENV'}{\AENV'}{\NENV'}{\TASKSET'}
  {\set{\locreg{\loc}{\reg} \mapsto \concreteloc{\reg}{\indj}{} \mid
  (\locreg{\loc}{\reg} \mapsto \concreteloc{\reg}{\indj}{}) \in \MENV,
  (\locreg{\loc}{\reg} \mapsto \concreteloc{\reg}{\indivar{\ivarid}}{}) \in \MENV_c}}
  {\STOR''}
  $

This case is identical to the previous one.

\end{itemize}

\item Case \refwellformed{sec:well-formedness}{wf:ca}

The individual requirements, labeled 
\refwellformed{sec:well-formedness-allocation}{wf:impl-linear-alloc} -
\refwellformed{sec:well-formedness-allocation}{wf:impl-empty-region},
are handled by the following case analysis.

\begin{itemize}[label=$\circ$]

\item Case (\refwellformed{sec:well-formedness-allocation}{wf:impl-linear-alloc}):

$ ((\reg \mapsto \locreg{\loc}{\reg}) \in \AENV' \wedge \locreg{\loc}{\reg} \in \NENV') \Rightarrow \\
          ((\locreg{\loc}{\reg} \mapsto \concreteloc{\reg}{\ind}{}) \in \MENV' \wedge
          \ind > \allocptr{\reg}{\STOR})
          \xor
          (\locreg{\loc}{\reg} \mapsto \indirection{\reg_2}{\ind_2}) \in \MENV'
          $

By the well-formedness of the store given in the premise of this lemma,
the above already holds for locations in the environments $\AENV$ and $\NENV$.
This case discharges straightforwardly by using Results~(\ref{prf:wf-datacon-join-eqn2}) and (\ref{prf:wf-datacon-join-eqn3}) since \dregpardataconjoin{} doesn't introduce any new
locations in $\AENV$ and $\NENV$.

\item Case (\refwellformed{sec:well-formedness-allocation}{wf:impl-linear-alloc2})

$ ((\reg \mapsto \locreg{\loc}{\reg}) \in \AENV \wedge 
    \, \concreteloc{\reg}{\ind_s}{} = \hat{\MENV}(\locreg{\loc}{\reg}) \wedge
    \locreg{\loc}{\reg} \not \in \NENV' \wedge
    \, \ewitness{\TYP}{\concreteloc{\reg}{\ind_s}{}}{\STOR}{\concreteloc{\reg}{\ind_e}{}}) \Rightarrow
          \ind_e > \allocptr{\reg}{\STOR}
          $

By inversion on \tdataconivars{}, $\locreg{\loc}{\reg} \in \NENV'$, and thus this
case discharges immediately.

\item Case (\refwellformed{sec:well-formedness-allocation}{wf:impl-write-once}):

$ \locreg{\loc}{\reg} \in \NENV' \Rightarrow
          \concreteloc{\reg}{\ind}{} = \hat{\MENV'}(\locreg{\loc}{\reg}) \wedge
          (\reg \mapsto (\ind \mapsto \DC)) \not \in \STOR'
          $

By the well-formedness of the store given in the premise of this lemma,
the above already holds for all locations in $\NENV$.
This case discharges straightforwardly by using Results~(\ref{prf:wf-datacon-join-eqn2}) and (\ref{prf:wf-datacon-join-eqn3}) since \dregpardataconjoin{} doesn't introduce any new
locations in $\NENV$.


\item \refwellformed{sec:well-formedness-allocation}{wf:impl-empty-region}:

$(\reg \mapsto \emptyset) \in \AENV \Rightarrow (\reg \mapsto \emptyset) \in \STOR$

This case discharges because, from the premise of the
lemma, this property holds for the original environment
$\AENV$ and store $\STOR$, and, by inversion on \tdataconivars{},
continues to hold for $\AENV'$ and $\STOR'$.

\end{itemize}

\item Case (\refwellformed{sec:well-formedness}{wf:impl1}):
$dom(\SENV) \cap \NENV = \emptyset $

This case discharges because, from the premise of the
lemma, this property holds for the original environments
$\NENV$ and $\SENV$, and, by inversion on \tdatacon{},
continues to hold for $\NENV$ and $\SENV$.

\end{itemize}

\end{itemize}

\end{itemize}

\end{enumerate}

\end{bcase}

\begin{bcase}[\dregparcasejoin{}]

\begin{mathpar}
\rdregparcasejoin{}
\end{mathpar}

This case can be proved using similar reasoning as \dregpardataconjoin{}.

\end{bcase}

\begin{bcase}[\dregparsteptask{}]

\begin{mathpar}
\rdregparsteptask{}
\end{mathpar}

Let $\SENV = \SENVMAP(\concretelocvar)$,
$\CENV = \CENVMAP(\concretelocvar)$,
$\AENV = \AENVMAP(\concretelocvar)$, and
$\NENV = \AENVMAP(\concretelocvar)$
be the environments corresponding to the task
$\taskconfig{\hTYP}{\concretelocvar{}}{\STOR;\MENV;\EXPR}$.
We instantiate the new environment maps as:
\[
\begin{array}{l}
\SENVMAP' = \SENVMAP ; \;
\CENVMAP' = \CENVMAP
\end{array}
\]
By inversion on \ref{sec:well-formedness-tasks},
the two obligations are two show that the resulting $\TASKSET'$ is well typed,
and that the stores of all tasks in it are well formed.
By the typing rule given in the premise of this lemma
and by inversion on \ttaskset{},
we can directly establish the well-typedness of the tasks in $\TASKSET$.
Using similar reasoning, we establish that the stores of all tasks in $\TASKSET$
are well-formed.
Thus, there are only two remaining obligations, namely
to show that the task resulting from the sequential step is well-typed,
and that its store is well-formed.
And by inversion on \ttask{}, in order to show that a task is well-typed,
we must show that the expression it is evaluating is well-typed.
Thus, there are two obligations that we must prove, namely
$\emptytenv;\SENV';\CENV';\AENV';\NENV' \vdash \AENV'';\NENV'';\EXPR' : \hTYP$, and
$\storewf{\SENV'}{\CENV'}{\AENV'}{\NENV'}{\MENV'}{\STOR'}$
We can discharge both of these by using
Lemma~\ref{lemma:single-thread-preservation}
for single thread preservation.
There are three preconditions in order to use this lemma, which are
handled by the following case analysis.

\begin{enumerate}

\item $\emptytenv;\SENV;\CENV;\AENV;\NENV \vdash \AENV';\NENV';\EXPR : \hTYP$

By inversion on \ttaskset{}, and the typing rule given in the premise of this lemma,
we know that all tasks in the task set $\TASKSET$ are well-typed.
Thus, we obtain the result
$\SENV;\CENV;\AENV;\NENV \vdash_{task} \AENV'; \NENV'; \taskconfig{\hTYP}{\concretelocvar{}}{\STOR;\MENV;\EXPR}$.
Then, by inversion on \ttask{}, we obtain the result that the expression
$\EXPR$ is also well-typed.
Thus, this obligation is discharged.

\item $\storewf{\SENV}{\CENV}{\AENV}{\NENV}{\MENV}{\STOR}$

By the well-formedness of the task set given in the premise of this lemma,
we establish the well-formedness of stores of all tasks in the task set $\TASKSET$,
and we obtain the result $\storewf{\SENV}{\CENV}{\AENV}{\NENV}{\TASKSET}{\STOR}{\MENV}$,
thus discharging this obligation.

\item $\STOR;\MENV;\EXPR \stepsto \STOR';\MENV';\EXPR$.

This obligation discharges straightforwardly by inversion on \dregparsteptask{}.

\end{enumerate}

\end{bcase}

$\blacksquare$

\end{bproof}

\begin{lemma}[Single Thread Preservation]
  \label{lemma:single-thread-preservation}
  \begin{displaymath}
    \begin{aligned}
      \text{If} \;\; & \emptytenv;\SENV;\CENV;\AENV;\NENV \vdash \AENV';\NENV';\EXPR : \hTYP \\
      \text{and} \;\; & \storewf{\SENV}{\CENV}{\AENV}{\NENV}{\MENV}{\STOR}\\
      \text{and} \;\; & \STOR;\MENV;\EXPR \stepsto \STOR';\MENV';\EXPR' \\
      \text{then for some} \;\; & \SENV' \supseteq \SENV, \CENV' \supseteq \CENV ,\\
      & \emptytenv;\SENV';\CENV';\AENV';\NENV' \vdash \AENV'';\NENV'';\EXPR' : \hTYP \\
      \text{and} \;\; & \storewf{\SENV'}{\CENV'}{\AENV'}{\NENV'}{\MENV'}{\STOR'}.
    \end{aligned}
  \end{displaymath}
\end{lemma}

\begin{bproof}

The proof is by induction on the given derivation of the dynamic semantics.

\begin{bcase}[\dregparletlocafternewreg{}]

\begin{mathpar}
\rdregparletlocafternewreg{}
\end{mathpar}

\begin{enumerate}

\item
The first obligation is to show that the result of the evaluation step
is well-typed, that is
$\emptytenv;\SENV;\CENV';\AENV';\NENV' \vdash \AENV''; \NENV''; \EXPR' : \hTYP$,
where $\hTYP = \tyatlocreg{\TYP}{\loc'}{\reg'}$.
This proof obligation follows straightforwardly by inversion on \tllafter{}.

\item
The second obligation is to show that the result of the evaluation step
is well-formed: $\storewf{\SENV}{\CENV'}{\AENV'}{\NENV'}{\TASKSET'}{\STOR'}{\MENV'}$
The individual requirements, labeled 
\refwellformed{sec:well-formedness}{wf:map-store-consistency} -
\refwellformed{sec:well-formedness}{wf:impl1},
are handled by the following case analysis.

\begin{itemize}

\item Case (\refwellformed{sec:well-formedness}{wf:map-store-consistency}):

$ (\locreg{\loc}{\reg} \mapsto \TYP) \in \SENV \Rightarrow \\
            (\concreteloc{\reg}{\ind_1}{} = \hat{\MENV}(\locreg{\loc}{\reg}) \wedge
            \ewitness{\TYP}{\concreteloc{\reg}{\ind_1}{}}{\STOR}{\concreteloc{\reg}{\ind_2}{}}) \xor \\
            (\concreteloc{\reg}{\indivar{\ivarid}}{} = \hat{\MENV}(\locreg{\loc}{\reg}) \wedge
             \exists_{\STOR',\MENV',\EXPR'}. \taskconfig{\TYP}{\concreteloc{\reg}{\indivar{\ivarid}}{}}{\STOR';\MENV';\EXPR'} = \getwriter{\TASKSET}{\hTYP}{\indivar{\ivarid}} \wedge
             \concreteloc{\reg}{\ind_1}{} = \hat{\MENV'}(\locreg{\loc}{\reg}) \wedge \\
             (\isval{\EXPR'} \Rightarrow \ewitness{\TYP}{\concreteloc{\reg}{\ind_1}{}}{\STOR'}{\concreteloc{\reg}{\ind_2}{}})
             )
          $ 

By the well-formedness of the store given in the premise of this lemma,
the above already holds for all locations in the location environment $\MENV$.
The obligation discharges by inspecting the only new location
in $\MENV'$, namely $\locreg{\loc}{\reg}$, which
is fresh and therefore cannot be in the domain of $\SENV$.

\item Case (\refwellformed{sec:well-formedness}{wf:cfc}):
$\storewfcfa{\CENV'}{\MENV'}{}{\STOR'}$

Of the requirements for this judgement, the only one that is
not satisfied immediately by the well-formedness of the store
given in the premise of the lemma is requirement
\refwellformed{sec:well-formedness-constructors}{wfconstr:constraint-after}.
The specific requirement is to establish that:

$ (\locreg{\loc}{\reg} \mapsto \afterl{\tyatlocreg{\TYP}{\loc'}{\reg}}) \in \CENV \Rightarrow \\            
            (\concreteloc{\reg}{\ind_1}{} = \hat{\MENV}(\locreg{\loc'}{\reg}) \wedge
            \ewitness{\TYP}{\concreteloc{\reg}{\ind_1}{}}{\STOR}{\concreteloc{\reg}{\ind_2}{}} \wedge
            \concreteloc{\reg}{\ind_2}{} = \hat{\MENV}(\locreg{\loc}{\reg}))
            \xor \\
            ( \concreteloc{\reg}{\indivar{\ivarid_1}}{} = \hat{\MENV}(\locreg{\loc'}{\reg}) \wedge
                  (\locreg{\loc}{\reg} \mapsto \concreteloc{\reg}{\indirection{\reg_2}{0}}{}) \in \MENV \wedge
                  \set{\reg_2 \mapsto \heap} \in \STOR)            
             \xor \\            
            (\concreteloc{\reg}{\ind_1}{} = \hat{\MENV}(\locreg{\loc'}{\reg}) \wedge
            \ewitness{\TYP}{\concreteloc{\reg}{\ind_1}{}}{\STOR}{\concreteloc{\reg}{\ind_2}{}} \wedge
              (\locreg{\loc}{\reg} \mapsto \concreteloc{\reg}{\indirection{\reg_2}{0}}{}) \in \MENV \wedge
                 \STOR(\reg)(\ind_2) = \indirection{\reg_2}{0}
                 )
            $

The second disjunct follows immediately by inversion on \dregparletlocafternewreg{}.

\item Case (\refwellformed{sec:well-formedness}{wf:ca}):
$\storewfca{\AENV'}{\NENV'}{}{\MENV'}{\STOR'}$

The individual requirements, labeled 
\refwellformed{sec:well-formedness-allocation}{wf:impl-linear-alloc} -
\refwellformed{sec:well-formedness-allocation}{wf:impl-empty-region},
are handled by the following case analysis.

\begin{itemize}

\item Case Case (\refwellformed{sec:well-formedness-allocation}{wf:impl-linear-alloc}):

$ ((\reg \mapsto \locreg{\loc}{\reg}) \in \AENV' \wedge \locreg{\loc}{\reg} \in \NENV') \Rightarrow \\
          ((\locreg{\loc}{\reg} \mapsto \concreteloc{\reg}{\ind}{}) \in \MENV' \wedge
          \ind > \allocptr{\reg}{\STOR})
          \xor
          (\locreg{\loc}{\reg} \mapsto \indirection{\reg_2}{\ind_2}) \in \MENV'
          $

By the well-formedness of the store given in the premise of this lemma,
the above already holds for locations in the environments $\AENV$ and $\NENV$.
The obligation discharges by inspecting the only new location
added to environments, namely $\locreg{\loc}{\reg}$.
The second disjunct, namely
$(\locreg{\loc}{\reg} \mapsto \indirection{\reg_2}{\ind_2}) \in \MENV'$,
follows immediately by inversion on \dregparletlocafternewreg{}.

\item Case (\refwellformed{sec:well-formedness-allocation}{wf:impl-linear-alloc2})

$ ((\reg \mapsto \locreg{\loc}{\reg}) \in \AENV \wedge 
    \, \concreteloc{\reg}{\ind_s}{} = \hat{\MENV}(\locreg{\loc}{\reg}) \wedge
    \locreg{\loc}{\reg} \not \in \NENV \wedge
    \, \ewitness{\TYP}{\concreteloc{\reg}{\ind_s}{}}{\STOR}{\concreteloc{\reg}{\ind_e}{}}) \Rightarrow
          \ind_e > \allocptr{\reg}{\STOR}
          $

By inversion on \tllafter{}, $\locreg{\loc}{\reg} \in \NENV'$, and thus this
case discharges immediately.

\item Case (\refwellformed{sec:well-formedness-allocation}{wf:impl-write-once}):

$ \locreg{\loc}{\reg} \in \NENV' \Rightarrow
          \concreteloc{\reg}{\ind}{} = \hat{\MENV'}(\locreg{\loc}{\reg}) \wedge
          (\reg \mapsto (\ind \mapsto \DC)) \not \in \STOR'
          $

Both conjuncts discharge immediately by inversion on \dregparletlocafternewreg{}.
Since $(\locreg{\loc}{\reg} \mapsto \indirection{\reg'}{0}) \in \MENV'$,
we obtain $\concreteloc{\reg'}{0}{} = \hat{\MENV'}$, thus satisfying the first obligation.
And since $\reg'$ is fresh, $(\reg' \mapsto (0 \mapsto \DC)) \not \in \STOR'$
holds as well, thus discharging this case.

\item Case (\refwellformed{sec:well-formedness-allocation}{wf:impl-empty-region}):
$(\reg \mapsto \emptyset) \in \AENV \Rightarrow (\reg \mapsto \emptyset) \in \STOR$

This case discharges because, from the premise of the
lemma, this property holds for the original environment
$\AENV$ and store $\STOR$, and, by inversion on \tllafter{},
continues to hold for $\AENV'$ and $\STOR'$.

\end{itemize}

\item Case (\refwellformed{sec:well-formedness}{wf:impl1}):
$dom(\SENV) \cap \NENV = \emptyset $

Because it is a bound location, $\loc \not \in dom(\SENV)$, and by inversion on \tllafter{}
$\loc \in \NENV'$, which discharges this obligation.

\end{itemize}


\end{enumerate}

\end{bcase}

\begin{bcase}


The cases for \dletexp{}, \dletval{}, \dletregion, \dletloctag{}, \dletlocstart{}, \dletlocafter, and \dapp{}
are similar to the proof of preservation for sequential \seqcalc{} given in \cite{LoCal}.


\end{bcase}

$\blacksquare$
\end{bproof}

%% file: others_appendix_table.tex
\setlength{\tabcolsep}{0.14em}
\renewcommand{\arraystretch}{1.2}
\begin{tabular}{@{}l ccccc r ccccc r ccccc@{}}
    \toprule
    & \multicolumn{5}{c}{\MPL{}} & \phantom{} & \multicolumn{5}{c}{OCaml} & \phantom{} & \multicolumn{5}{c}{GHC} \\
    \cmidrule{2-6}
    \cmidrule{8-12}
    \cmidrule{14-18}
    Benchmark & $T_s$ & $T_1$ & O & $T_{48}$ & S && $T_s$ & $T_1$ & O & $T_{48}$ & S && $T_s$ & $T_1$ & O & $T_{48}$ & S \\
    \midrule


   fib & 37.4 & 38.6 & 3.2 & 0.91 & 41.1 && 21.1  & 21.1  & 0 & 0.5  & 42.2 && 31.9  & 31.8  & -0.3 & 0.76 & 42  \\

    \buildfib{} & 14.5 & 14.6 & 0.69 & 0.35 & 41.4 && 8.6 & 8.6 & 0 & 0.25 & 34.4 && 12.4 & 12.4 & 0 & 0.34 & 36.5  \\




   buildKdTree & 7.26 & 7.65 & 5.37 & 0.41 & 17.7 && 10.9  & 11.7  & 7.34 & 1.84 & 5.92 && 13.4  & 13.6 & 1.5 & 2.21 & 6.06  \\

    countCorr & 10.5 & 10.5 & 0 & 0.27 & 38.9  && 13.9 & 15.0 & 7.9 & 0.37 & 37.6 && 3.54 & 3.57 & 0.84 & 0.15 & 23.6 \\

    allNearest & 2.38 & 2.4 & 0.84 & 0.06 & 39.6 && 3.01 & 3.09 & 2.65 & 0.091 & 33.1 && 2.07  & 2.04 & -1.4 & 0.068 & 30.4  \\


    barnesHut & 5.05  & 5.33 & 5.5 & 0.12 & 42 && 10.9 & 11 & 0.91 & 0.44 & 24.7 && 4.97 & 5.29 & 6.4 & 0.33 & 15.1  \\



    coins & 1.71 & 1.51  & -11\% & 0.05 & 34.2 && 1.05  & 1.07  & 1.9  & 0.036  & 29.1  && 0.82 & 0.82 & 0  & 0.085  & 9.6  \\

    countnodes & 0.37  & 0.38  & 2.7 & 0.019 & 19.5  && 0.46 & 0.45  & -2.17  & 0.034 & 13.5  && 1.45 & 1.46 & 0.68  & 0.049 & 29.6  \\

    constFold & 2.36  & 3.29  & 39.4 & 0.23 & 10.3 && 17.7 & 19.2 & 8.5 & 2.23 & 7.94 && 3.71 & 3.99 & 7.55 & 0.64 & 5.80 \\



    x86-compiler & 1.34 & 1.33 & -0.74 & 0.042 & 31.9 && 1.2  & 1.3 & 8.33 & 0.09 & 13.3  && 2.34 & 2.38  & 1.7  & 0.44 & 5.31  \\


    mergeSort & 1.74 & 1.84 & 5.75 & 0.048 & 36.3 && 3.83  & 3.94  & 2.87 & 0.19 & 20.16 && 2.74  & 2.95 & 7.66 & 0.16 & 17.1  \\

   \midrule
    average & - & - & 3.29 & - & 33.7  && - & - & 4.14 & - & 15.3 && - & - & 1.35 & - & 15.2 \\

    \bottomrule
  \end{tabular}